\documentclass[prd,twocolumn,nofootinbib,nobibnotes,showpacs,superscriptaddress]{revtex4-1}

\usepackage{graphicx}
\graphicspath{{../../data/plots/}}
\usepackage{float}
\usepackage{bm}
\usepackage[hidelinks]{hyperref}
\usepackage{color}

\usepackage[utf8]{inputenc}
\usepackage[T1]{fontenc}
\usepackage{slashed}
\newcommand{\spindersphere}{{}_s\slashed{\Delta}}
\newcommand{\edth}{\textnormal{\dh}}

\newcommand{\Edth}{\textnormal{\DH}}

\usepackage{fullpage}
\usepackage{enumitem}
\usepackage{mathalfa,amssymb,amsthm}
\usepackage{amsmath}

\theoremstyle{definition}

\newcommand*\oline[1]{%
   \vbox{%
     \hrule height 0.5pt
     \kern0.25ex
     \hbox{%
       \ifmmode#1\else\ensuremath{#1}\fi
     }
   }
}

\usepackage[caption=false]{subfig}

\begin{document}


\title{Numerical computation of second order
vacuum perturbations of Kerr black holes}
\author{Justin L. Ripley}
\email{lloydripley@gmail.com}
\affiliation{%
DAMTP,
Centre for Mathematical Sciences,
University of Cambridge,
Wilberforce Road, Cambridge CB3 0WA, UK.
}%
\author{Nicholas Loutrel}
\email{nloutrel@princeton.edu}
\affiliation{%
 Department of Physics, Princeton University, Princeton, New Jersey 08544, USA.
}%
\author{Elena Giorgi}
\email{egiorgi@princeton.edu}
\affiliation{%
 Department of Mathematics, Princeton University, Princeton, New Jersey 08544, USA.
}%
\affiliation{%
 Princeton Gravity Initiative, Princeton University, Princeton, New Jersey 08544, USA.
}%
\author{Frans Pretorius}
\email{fpretori@princeton.edu}
\affiliation{%
 Department of Physics, Princeton University, Princeton, New Jersey 08544, USA.
}%
\affiliation{%
 Princeton Gravity Initiative, Princeton University, Princeton, New Jersey 08544, USA.
}%

\date{\today}

\begin{abstract}
Motivated by the desire to understand the leading order nonlinear
gravitational wave interactions around arbitrarily rapidly rotating Kerr black
holes, we describe a numerical code designed to compute second order vacuum 
perturbations on such spacetimes.
A general discussion of the formalism we use is presented in 
\cite{Loutrel:2020wbw}; here we show how we numerically 
implement that formalism with a particular choice of coordinates
and tetrad conditions, and give example results for black holes with
dimensionless spin parameters $a=0.7$ and $a=0.998$. 
We first solve the Teukolsky equation for the linearly
perturbed Weyl scalar $\Psi_4^{(1)}$,
followed by direct reconstruction of the spacetime metric
from $\Psi_4^{(1)}$, and then solve for the dynamics of the second
order perturbed Weyl scalar $\Psi_4^{(2)}$.
This code is a first step toward a more general
purpose second order code, and we outline how 
our basic approach could
be further developed to address current questions of interest,
including extending the analysis of ringdown in black hole mergers
to before the linear regime, exploring
gravitational wave ``turbulence''
around near-extremal Kerr black holes, and studying
the physics of extreme mass ratio inspiral.
 
\end{abstract}

\maketitle

\allowdisplaybreaks
\section{Introduction}
   In this paper we initiate a numerical study of the dynamics
of second order vacuum perturbations of a Kerr black hole.
Linear black hole perturbation theory has played an important
role in the study of black holes, with diverse applications from mathematical 
physics to gravitational wave astrophysics
(for reviews see e.g. \cite{Sasaki:2003xr,Dafermos:2008en,Barack:2018yvs}).
Regarding the latter, it is used in interpreting the ringdown
phase of a binary black hole merger, and for extreme mass ratio
inspirals (EMRIs). For both of these physical regimes, it is presently
computationally intractable to full numerical solution without recourse
to perturbative methods.

For some applications it may be necessary to go beyond linear
perturbations. Here for brevity we
only mention a couple of key incentives
(a more thorough discussion that motivates this study
can be found in our companion paper~\cite{Loutrel:2020wbw}).
In order to extract 
subleading modes of the ringdown following comparable mass mergers,
it may be necessary to consider nonlinear effects.
The ``problem'' with such mergers is that the leading order
quadrupole mode is excited with such high amplitude relative
to subleading modes (see e.g.~\cite{2014PhRvD..90l4032L,2016PhRvD..94f9902L}),
that nonlinear mode coupling 
could produce features of similar strength to subleading modes; this is
particularly so
within the first few cycles of ringdown were most
of the observable signal, hence most opportunity for measurement, lies.
It will be important to quantitatively understand second order features
to reap the most out of ringdown analysis of future loud merger events.

Nonlinear physics may also play an important role
in the ringdown of near-extremal Kerr black holes\footnote{We note though
that comparable mass 
mergers cannot produce near-extremal remnants,
see e.g.~\cite{Hemberger:2013hsa,Lousto:2013wta,Healy:2014yta,Hofmann:2016yih},
and it is unknown how rapidly the typical
supermassive black holes in the universe,
of relevance to EMRIs,
rotate. Thus near-extremal ringdown may end up being more a question
of theoretical, rather than astrophysical, interest.
}. 
This can partly be motivated by the presence of a family of slowly
damped modes, whose damping timescale grows without bound
as the black hole spin approaches its extremal value
~\cite{Detweiler:1980gk,Hod:2008zz,Yang:2012pj}\footnote{Though 
there is some controversy about exactly what the 
spectrum of modes of extremal/near-extremal black holes is;
see e.g.~\cite{Hod:2015swa,Zimmerman:2015rua,Hod:2016aoe,Richartz:2015saa,Casals:2019vdb}.}. The slower damping of 
perturbations implies nonlinear effects could be more
pronounced; most intriguing in this regard is the suggestion
that mode coupling induces a turbulent energy cascade 
in near-extremal Kerr black holes~\cite{2015PhRvL.114h1101Y},
similar to that seen
in a few studies of black holes and black branes in
asymptotically Anti de-Sitter spacetime
\cite{Carrasco:2012nf,Green:2013zba,Adams:2013vsa}.
Nonlinear effects almost certainly play some role in the
physics of \emph{extremal} Kerr black holes, as those
have been shown to be linearly unstable 
\cite{Detweiler:1980gk,Aretakis:2012ei,Aretakis:2013dpa}
(the instability may be related
to the above mentioned slowly damped quasinormal modes,
that become ``zero damped''
in the extremal limit).

   Finally, we mention that second order black hole perturbation
theory plays an important role in computing the second order
self force of a small particle orbiting a black hole, which
is relevant to modeling EMRIs
(e.g. \cite{Barack:2009ux}). We note though that in this paper
we only consider the second order \emph{vacuum} perturbation of a Kerr
black hole, so our results are not directly applicable to modeling
EMRI physics. 

Following a brief summary of our formalism in
Sec.\ref{sec:summary_of_formalism}
(which is described in more detail in our companion paper
\cite{Loutrel:2020wbw}),
in the remainder of this paper we describe a numerical implementation
of the method and then analyze a few example runs from our code.
In the remainder of the introduction we give a general overview
of our numerical approach. 

Several steps are required to arrive at the desired second order
perturbation. First
is to solve for a linear gravitational wave perturbation 
characterized by the Weyl scalar $\Psi_4$ (or 
equivalently $\Psi_0$) using the Teukolsky equation
\cite{Teukolsky:1973ha}.
As described in Sec.~\ref{sec:coordinates}, with more details
given in Appendix~\ref{sec:derivation_coordinates}, 
we begin with the Kerr metric in Boyer-Lindquist form
and a rotated version of the Kinnersley tetrad~\cite{1969JMP....10.1195K},
then transform 
these to a hyperboloidal compactified, horizon penetrating coordinate system.
In these coordinates then we numerically solve the Teukolsky equation
in the time domain, starting with Cauchy data for $\Psi_4^{(1)}$.
(We note that many codes have been developed over the years
to solve the Teukolsky equation;
see, e.g.~\cite{
Krivan:1997hc,
Krivan:1997hc,
Baker:2001sf,
LopezAleman:2003ik,
Lousto:2005ip,
Burko:2006ua,
Sundararajan:2007jg,
Zenginoglu:2011zz,
Harms:2013ib}).

Campanelli and Lousto first showed that
the evolution of the second order perturbation of $\Psi_4$ 
is also governed by the Teukolsky equation, with a source
term that depends on all components of the first order metric perturbation
$h_{\mu\nu}$ \cite{Campanelli:1998jv}. 
The next step in our calculation the is therefore to
reconstruct this first order
metric correction from the first order perturbation of $\Psi_4$.
We \emph{directly} reconstruct the metric
by solving a set of nested null transport equations
in the outgoing radiation gauge~\cite{PhysRevD.11.2042}. 
Once this is complete, we numerically solve the Teukolsky equation
in the time domain
with the second order source term for the second order correction to $\Psi_4$.
This latter quantity is in general not invariant 
under first order gauge or tetrad transformations (see ~\cite{Campanelli:1998jv});
as discussed in Sec.~\ref{sec:gauge_invariance_measurements},
in our coordinate system we 
can avoid these issues if we measure the waves at future null
infinity in an asymptotically flat gauge, which is one reason why
we employ a hyperboloidal slicing and the outgoing
radiation gauge.

\begin{figure}
\includegraphics[width=2\linewidth,trim=0.5in 0.5in 0.0in 0.5in]{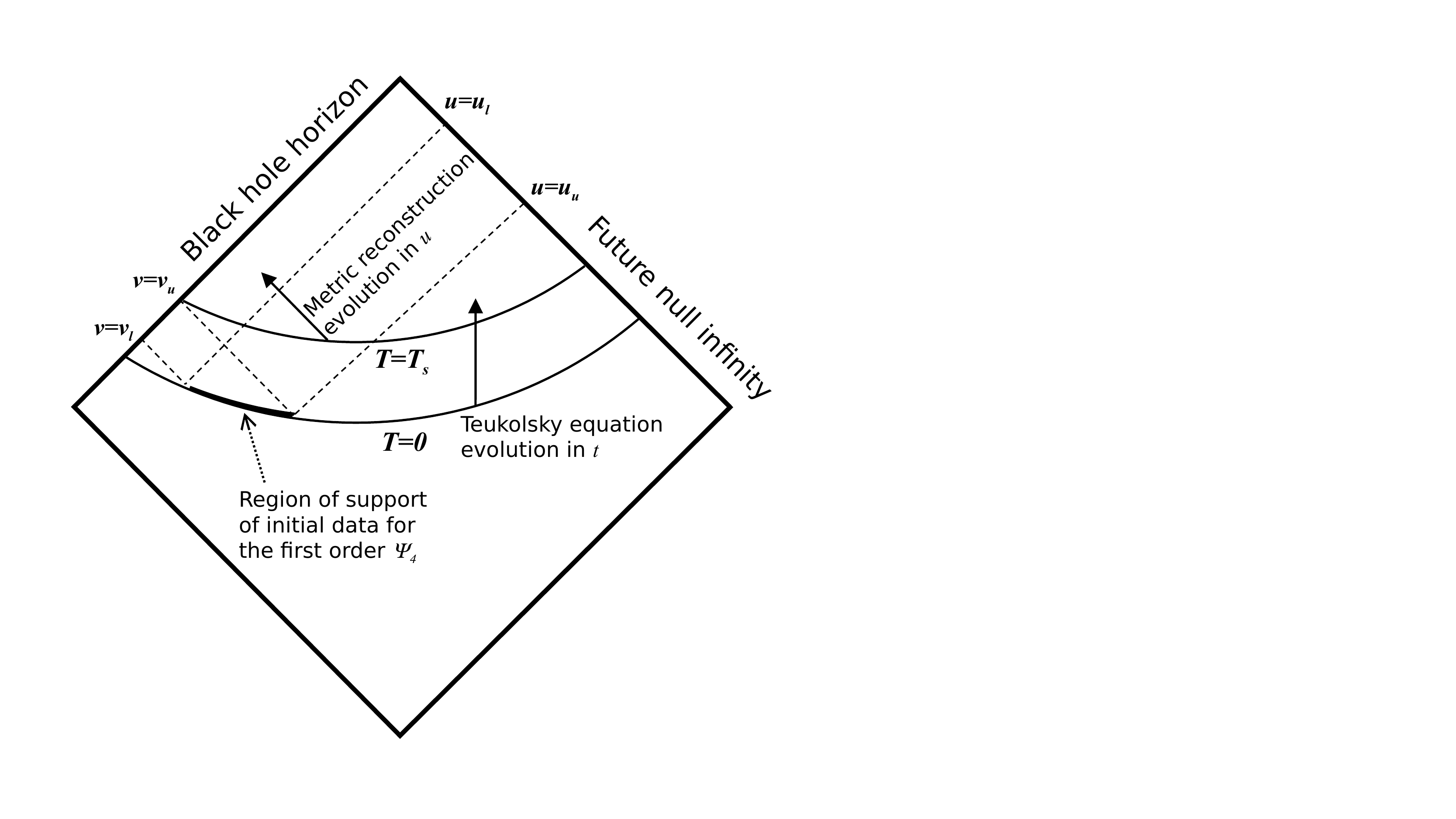}
\caption{
Schematic Penrose diagram demonstrating the domains of evolution
and metric reconstruction. The Teukolsky
equation is integrated using a Cauchy evolution scheme
along hypersurfaces of constant time $T$, whereas metric
reconstruction is carried out using null transport equations
in the $u$ direction, with characteristics tangent
to the ingoing Newman-Penrose vector $n^{\mu}$.
We provide initial data 
for the first order $\Psi_4^{(1)}$ at $T=0$, with
compact support in radial distance from the black hole in the range
$r=(r_l,r_u)$,
intersecting a range in advanced (retarded)
time of $u=(u_u,u_l)$ ($v=(v_l,v_u)$).
Thus only in the region $u>u_u, T>0,$ does the spacetime
differ from Kerr, which allows for trivial initial
data for the first order metric reconstruction
equations along $u=u_u$. For simplicity, as explained in the text,
we begin evolution of the second order perturbation
$\Psi_4^{(2)}$ at $T=T_s$;
in a sense then this is our ``true'' initial data surface
for the gravitational
wave perturbation of Kerr calculated to second order.
For technical reasons explained
in Sec.~\ref{sec:compact_initial_data},
this initial data setup only leads
to consistent metric reconstruction if $\Psi_4^{(1)}$ 
has azimuthal mode content $|m|\geq2$.}
\label{fig:metric_reconstruction_strategy_pictureb}
\end{figure}

One difficulty with using null transport equations in conjunction with
the (3+1) Cauchy evolution scheme we use for the Teukolsky equation
is that their domains of integration do not ``easily'' overlap. 
An implication of this is if we wanted to solve for the second
order perturbation over the entire domain of our Cauchy evolution
we would need to solve a set of first order constraint
equations on the initial $T=0$ slice to give
self-consistent initial data for the 
reconstruction equations. As explained more
in Sec.~\ref{sec:initial_data} and illustrated
in Fig.~\ref{fig:metric_reconstruction_strategy_pictureb},
we avoid this
issue here by choosing initial data for the first-order
$\Psi_4^{(1)}$ that has compact
support on the initial slice, and only solve for the first order metric
beginning on an ingoing null slice ($v_u$) intersecting
$T=0$ outside of this region (as we explain in
Fig.~\ref{fig:metric_reconstruction_strategy_pictureb},
our metric reconstruction equations are transport equations along
the ingoing null tetrad vector $n^{\mu}$).
Then, initial data for reconstruction only needs to be specified
along the outgoing null curves $u=u_u$, which is 
trivially the Kerr solution. In principle we could
immediately begin the second order evolution 
for $v\geq v_u$, though again to simplify
the Cauchy evolution we only start second order
evolution at a constant hyperboloidal time $T_s$ after
all the ingoing data from $v_l$ to $v_u$ has entered
the black hole horizon.  
The Cauchy evolution
prior to this is then, in a sense, simply providing
a map from initial data for the first order
$\Psi_4^{(1)}$ specified
on $T=0$ to the ``true'' initial time $T=T_s$. 
As implemented in 
the code, during each step of the Cauchy evolution we
also perform reconstruction (and second order integration
for $T>T_s$). The
reconstruction will therefore not be consistent until evolution
crosses $v=v_u$, but
for illustrative purposes we also show 
independent residuals of our reconstruction procedure
prior to that, to demonstrate 
that the inconsistency then has no effect on the consistency
of the solution afterward.

As described in Sec.\ref{sec:code_implementation}, we
use pseudo-spectral methods to solve both the 
Teukolsky and null transport equations; the code
can be downloaded from \cite{code_online}.
Example results are given 
in Sec.~\ref{sec:example_evolution_a0.7}
and Sec.~\ref{sec:example_evolution_a0.998} 
for a black hole with dimensionless spin $a=0.7$ and $a=0.998$ respectively.
Specifically, for the examples we consider the linear wave only contains
azimuthal $e^{i m \phi}$ angular dependence for $m=2$.
This linear field sources a second order $\Psi_4^{(2)}$ containing
modes with both $m=0$ and $m=4$.
We conclude in Sec.~\ref{sec:conclusion}, which includes
a discussion of how our methods and code will be extended 
to tackle the problems we are ultimately interested in addressing.

In Appendix \ref{sec:symbols} we describe the conventions
we use for the Fourier transform and normalization
of discrete quantities used to display some of our results.
In Appendix \ref{sec:derivation_metric_recon_source}
we provide a derivation of the metric
reconstruction equations and second order source term in
the specific gauge and coordinates used here, which is slightly different
from the analytical example case give in~\cite{Loutrel:2020wbw}.
We describe the transformation of the Kerr metric and Kinnersley tetrad
in Boyer-Lindquist coordinates
to the coordinates we use in the code in Appendix \ref{sec:derivation_coordinates}.
Finally, in Appendix \ref{sec:swaL} we collect some useful 
properties of spin-weighted spherical harmonics, which
are used in our pseudo-spectral code.

\subsection{Notation and conventions}
	We use geometric units ($8\pi G=1$, $c=1$)
and follow the definitions and sign conventions of Chandrasekhar
\cite{Chandrasekhar_bh_book}
(in particular the $+---$ signature for the metric), except we
use Greek letters $\mu,\nu...$ to denote spacetime indices.
We   use the non-standard symbols : $\varpi=3.14159...$ for the number `pi'
(to avoid confusion with the Newman-Penrose scalar $\pi$),
and the symbols $\mathcal{R}$ and $\mathcal{I}$ to respectively denote the 
real and imaginary parts of fields.
We use an overbar $\bar{f}$ to denote complex conjugation
of a quantity $f$.

We write the perturbed metric $g_{\mu\nu}$ about a background solution $g_{\mu\nu}^{(0)}$
as $g_{\mu\nu}=g_{\mu\nu}^{(0)} + h_{\mu\nu}$, where
$h_{\mu\nu}$ is the first order perturbation.
Other than the metric, we
denote an $n^{th}$ order
quantity in a perturbative expansion with a trailing superscript
$\ ^{(n)}$; e.g. the Newman-Penrose scalar $\Psi_2 = \Psi_2^{(0)} +
\Psi_2^{(1)}$ to linear order.
However, for brevity, in terms of expressions where
the correction to a background
quantity will lead to a higher order perturbation
than considered, we drop the $^{(0)}$ superscript
from the background quantity. For example,
in the Teukolsky equation, Eq.~\eqref{eq:Teuk_NP} below,
all symbols other than $\Psi_4^{(1)}$ are
background quantities.

\section{Description of the Scheme}
In this section we summarize details
of the reconstruction scheme and second
order perturbation equation described in ~\cite{Loutrel:2020wbw} 
that are particular to our numerical implementation.
\subsection{Use of the NP/GHP formalisms}
\label{sec:summary_of_formalism}
We make extensive use of the
Newman-Penrose (NP)\cite{Newman_Penrose_paper}
formalism. In the Appendix of the companion paper~\cite{Loutrel:2020wbw}
we gave a brief overview of the NP formalism,
and to avoid excessive repetition we do not reproduce that here.
However, we now mention the most salient definitions of the NP formalism
necessary to understand the main points of this paper.

The NP formalism decomposes the geometry and Einstein equations in
terms of an orthonormal basis of null vectors, $\{l^{\mu},n^{\mu},m^{\mu},\bar{m}^{\mu}\}$;
$l^{\mu}$ ($n^{\mu}$) is an outgoing (ingoing) real null
vector, and $m^{\mu}$ is a complex angular null vector
with $\bar{m}^{\mu}$ its complex conjugate.
These define the four directional derivative operators
\begin{eqnarray}\label{diff_op}
{D} &=& l^{\mu} \partial_{\mu}\,, \qquad {\Delta} = n^{\mu} \partial_{\mu}\,, \\
{\delta} &=& m^{\mu} \partial_{\mu}\,, \qquad \bar{{\delta}} = \bar{m}^{\mu} \partial_{\mu}\,.
\end{eqnarray}
A vacuum geometry is characterized
by the 5 complex scalars $\{\Psi_0,\Psi_1,\Psi_2,\Psi_3,\Psi_4\}$,
which are contractions of the Weyl tensor with various
products of the null tetrad. The complex spin coefficients
(essentially combinations of Ricci rotation coefficients,
the tetrad analogue of the connection in a metric formalism)
are $\{\alpha, \beta, \gamma, \epsilon, \rho, \lambda, \pi, \mu, \nu, \tau, \sigma, \kappa\}$.
We choose a null tetrad, the explicit form of which is given later,
such that for the background Kerr solution
$\Psi_0=\Psi_1=\Psi_3=\Psi_4=\sigma=\kappa=\nu=\lambda=0$
(which is always possible for a general type D background) and $\gamma=0$
(which is always possible when the background is Kerr).

In the NP formalism, perturbations of the Kerr spacetime lead to one master equation
for the linearly perturbed Weyl scalar $\Psi_{4}$
known as the Teukolsky equation~\cite{Teukolsky:1973ha}, specifically
\begin{eqnarray}
\label{eq:Teuk_NP}\nonumber
	\mathcal{T}\Psi_4^{(1)}
	\equiv
	\big[
	\left(\Delta+4\mu+\bar{\mu}\right)
	\left(D+4\epsilon-\rho\right) \\
-	\left(\edth^{\prime}+4\pi-\bar{\tau}\right)
	\left(\edth-\tau\right)
-	3\Psi_2
	\big]
	\Psi_4^{(1)}
	=
	0
	.
\end{eqnarray}
(Here we have made use of the GHP operators $\edth$ and $\edth^{\prime}$,
which we define in Eq.~\eqref{edth_def}.)
The first step of our procedure is to
solve the Teukolsky equation for $\Psi_4^{(1)}$.
We discuss how we numerically solve this
equation in Sec.~\ref{sec:teuk_in_coord_form}.

We also make some use of the Geroch-Held-Penrose (GHP) \cite{GHP_paper}
formalism; in particular we use the following GHP derivatives
acting on some NP scalar $\eta$
\begin{subequations}\label{edth_def}
\begin{align}
	\edth\eta
	\equiv
	\left(\delta-p\beta-q\bar{\alpha}\right)\eta
	,\\
	\edth^{\prime}\eta
	\equiv
	\left(\bar{\delta}-p\alpha-q\bar{\beta}\right)\eta
	,
\end{align}
\end{subequations}
	where $\{p,q\}$ are the (constant) weights of $\eta$
(related to its spin and boost weights).

\subsection{Choice of background coordinates and tetrad}
\label{sec:coordinates}
We choose a form for the background Kerr metric, with mass and spin parameters
$M$ and $a$ respectively,
that is horizon penetrating
and hyperboloidally compactified so that the constant time $T$
(spacelike) slices become asymptotically null, reaching future
null infinite at finite coordinate radius.
An outline of how we derive these coordinates is given
in Appendix \ref{sec:derivation_coordinates}; here we
simply summarize their most important qualities. 

We use a rotated version of the Kinnersley tetrad that is
regular at future null infinity;
in $(T,R,\vartheta,\phi)$ component form, the tetrad vectors read:
\begin{subequations}
\label{eq:tetrad_IEF_HC}
\begin{align}
	l^{\mu} 
	= &
	\frac{R^2}{L^4+a^2R^2\mathrm{cos}^2\vartheta}\bigg(
		2M\left(2M-\left(\frac{a}{L}\right)^2R\right),
        \nonumber\\&
	-	\frac{1}{2}\left(L^2-2MR+\left(\frac{a}{L}\right)^2R^2\right),
		0,
		a
	\bigg)
	, \\
	n^{\mu} 
	= &
	\left(
		2+\frac{4MR}{L^2},\frac{R^2}{L^2},0,0
	\right)
	, \\
	m^{\mu}
	= &
	\frac{R}{\sqrt{2}\left(L^2-iaR\mathrm{cos}\vartheta\right)}
	\left(
		-ia\mathrm{sin}\vartheta,
		0,
		-1,
		-\frac{i}{\mathrm{sin}\vartheta}
	\right)
	.
\end{align}
\end{subequations}
	Here $R$ is the compactified radial coordinate, 
related to the Boyer-Lindquist radial coordinate by
\begin{align}\label{R_def}
	r\equiv \frac{L^2}{R}
	,
\end{align}
	with $L$ a constant parameter ($R=0$ thus corresponds
to future null infinity). 

	With this tetrad and metric, the only nonzero Weyl scalar is
\begin{align}
	\Psi_2
	=
	-\frac{
		MR^3
	}{
		\left(L^2 - iaR\mathrm{cos}\left(\vartheta\right)\right)^3
	}
	,
\end{align}
	and the nonzero spin coefficients are
\begin{subequations}
\label{eq:NP_IEF_HC}
\begin{align}
	\rho
	= &
	-\frac{
		R \left(a^2 R^2+L^4-2 L^2 M R\right)
	}{
		2 \left(L^2-i a R \cos (\vartheta )\right)^2 
		\left(L^2+i a R \cos (\vartheta )\right)
	}
	, \\
	\mu
	= &
	\frac{R}{-L^2+i a R \cos (\vartheta )}
	, \\
	\tau
	= &
	\frac{
		i a R^2 \sin (\vartheta )
	}{
		\sqrt{2} \left(L^2-i a R \cos (\vartheta )\right)^2
	}
	, \\
	\pi
	= &
	-\frac{
		i a R^2 \sin (\vartheta )
	}{
		\sqrt{2} \left(a^2 R^2 \cos ^2(\vartheta )+L^4\right)
	}
	, \\
	\epsilon
	= &
	\frac{
		R^2 \left(a^2 (-R)-i a \cos (\vartheta ) 
		\left(L^2-M R\right)+L^2 M\right)
	}{
		2 \left(L^2-i a R \cos (\vartheta )\right)^2
		\left(L^2+i a R \cos (\vartheta )\right)
	}
	, \\
	\alpha
	= &
	\frac{
		R \cot (\vartheta )
	}{
		\sqrt{2} \left(2 L^2+2 i a R \cos (\vartheta )\right)
	}
	, \\
	\beta
	= &
	\frac{
		R \left(
			-L^2 \cot (\vartheta )
			+i a R \sin (\vartheta ) \left(\csc ^2(\vartheta )
			+1\right)
		\right)
	}{
		2 \sqrt{2} \left(L^2-i a R \cos (\vartheta )\right)^2
	}
	.
\end{align}
\end{subequations}
The coefficients $\alpha$ and $\beta$
are singular at the poles ($\vartheta=0,\varpi$), but when expanded
out in the equations of motion they only appear
with other terms that in combination are regular there.
Other than this, all spin coefficients are regular in the domain of interest,
namely on the black hole horizon and the region
exterior to it.
Notice that with the Kinnersley
tetrad $\epsilon=0$, but we have rotated to a tetrad where $\gamma=0$
instead.

	The quantities above are
sufficient to completely determine the Teukolsky
and metric reconstruction equations, and so for brevity we do not write out
the explicit form of the Kerr metric in these coordinates.

\subsection{Choice of linearized metric gauge and linearized tetrad}
	After computing $\Psi_4^{(1)}$, we need to specify
a gauge in which to reconstruct the first order metric;
we choose the outgoing radiation gauge, defined by the following conditions:
\begin{subequations}
\begin{align}\label{org_def}
	h_{\mu\nu}n^{\mu}
	=&0
	,\\
	g^{\mu\nu}h_{\mu\nu}
	=
	&0
	.	
\end{align}
\end{subequations}
        For type D background spacetimes one can always impose
the outgoing (or ingoing)
radiation gauge, despite the fact that this 
imposes five conditions on the linear metric \cite{Price:2006ke}.
The only nonzero tetrad projections of $h_{\mu\nu}$ in
outgoing radiation gauge are
\begin{subequations}
\begin{align}\label{h_projs}
	h_{ll}
	\equiv&
	h_{\mu\nu}l^{\mu}l^{\nu}
	,\\
	h_{lm}
	\equiv&
	h_{\mu\nu}l^{\mu}m^{\nu}
	,\\
	h_{mm}
	\equiv&
	h_{\mu\nu}m^{\mu}m^{\nu}
	,
\end{align}
\end{subequations}
and the complex conjugates of the last two,
i.e. $h_{l\bar{m}}\equiv h_{\mu\nu}l^{\mu}\bar{m}^{\nu}$ 
and $h_{\bar{m}\bar{m}}\equiv h_{\mu\nu}\bar{m}^{\mu}\bar{m}^{\nu}$.

	As detailed in 
Appendix \ref{sec:derivation_metric_recon_source}, 
in this gauge
we can choose the linearly perturbed tetrad vectors
so that the first order corrections to the
derivative operators are 
\begin{subequations}
\begin{align}\label{lin_ops}
	D^{(1)}
	=&
-	\frac{1}{2}h_{ll}\Delta
	,\\
	\Delta^{(1)}
	=&
	0
	,\\
	\label{eq:delta-1}
	\delta^{(1)}
	=&
-	h_{lm}\Delta
+	\frac{1}{2}h_{mm}\bar{\delta}
	.
\end{align}
\end{subequations}
\subsection{Metric reconstruction equations}
\label{sec:metric_reconstruction_equations}
	Our metric reconstruction procedure consists of solving a nested set
of transport equations that are derived by linearly expanding
some of the Bianchi and Ricci identities in the NP
formalism; see Appendix \ref{sec:derivation_metric_recon_source}.
Unlike metric reconstruction procedures that use ``Hertz potentials''
(see e.g. \cite{PhysRevD.11.2042}), our method \emph{directly} reconstructs
the metric from $\Psi_4^{(1)}$.
The basic idea of this metric reconstruction procedure was first described
by Chandrasekhar \cite{Chandrasekhar_bh_book}.
One advantage of our implementation of Chandrasekhar's idea
is that it does not require using any specific features of
a particular coordinate system beyond the gauge and tetrad choices
we have already stated; a similar approach has recently been
rigorously analyzed by
Andersson et. al. \cite{Andersson:2019dwi}. 
A disadvantage of our implementation though
is that outgoing radiation gauge is incompatible
with most forms of source term, including from 
matter with a stress energy tensor that is not trace-free,
or that coming from first
order vacuum perturbations\footnote{We note though that the general
approach of reconstructing the metric by solving a series of nested
transport equations does not require one use the radiation gauges;
indeed Chandrasekhar \cite{Chandrasekhar_bh_book}
does not use a radiation gauge.
For a brief review of recent works that directly reconstruct the metric:
Andersson et. al. \cite{Andersson:2019dwi} works in outgoing
radiation gauge for perturbations of Kerr,
Dafermos et. al. \cite{Dafermos:2016uzj} work in a double null gauge
for perturbations of Schwarzschild, and
Klainerman and Szeftel \cite{Klainerman:2017nrb} work in a Bondi
gauge for perturbations of Schwarzschild.
In a different gauge one could presumably
reconstruct the metric in the presence of linearized matter perturbations.
That being said, using a radiation
gauge greatly simplifies and reduces the number of metric reconstruction
equations that we need to solve, and is sufficient
for solving for the dynamics of the second order Weyl scalar
$\Psi_4^{(2)}$ about a Kerr background.}.
Therefore, we can compute the gravitational wave perturbation
$\Psi_4$ to second order, but the metric tensor only to first order.
Recently ~\cite{Green:2019nam}
proposed a method based (in part) on Hertz potentials
that does not seem to have such restrictions.
However, for our purposes
we believe our method is more straightforward to implement
(see our companion paper~\cite{Loutrel:2020wbw}
for more discussion on these different approaches to reconstruction).

Given a solution $\Psi_4^{(1)}$ to the Teukolsky
equation, below are the transport equations we solve
to reconstruct the first order metric:
\begin{subequations}
\label{eq:metric_reconstruction}
\begin{align}
\label{eq:psi3_recon}
-	\left(\Delta+4\mu\right)\Psi_3^{(1)}
+	\left(\edth-\tau\right)\Psi_4^{(1)}
	&=
	0
	,\\
\label{eq:la_recon}
-	\left(\Delta+\mu+\bar{\mu}\right)\lambda^{(1)}
-	\Psi_4^{(1)}
	&=
	0
	,\\
\label{eq:psi2_recon}
-	\left(\Delta+3\mu\right)\Psi_2^{(1)}
+	\left(\edth-2\tau\right)\Psi_3^{(1)}
	&=
	0
	,\\
\label{eq:hmbmb_recon}
-	\left(\Delta-\mu+\bar{\mu}\right)h_{\bar{m}\bar{m}}
-	2\lambda^{(1)}
	&=
	0
	,\\
\label{eq:pi_recon}
-	\Delta\pi^{(1)}
-	\Psi_3^{(1)}
-	\left(\bar{\pi}+\tau\right)\lambda^{(1)}
	&\nonumber\\
+	\frac{1}{2}\mu\left(\bar{\pi}+\tau\right)h_{\bar{m}\bar{m}}
	&=
	0
	,\\
\label{eq:hlmb_recon}
-	\left(\Delta+\bar{\mu}\right)h_{l\bar{m}}
-	2\pi^{(1)}
-	\tau h_{\bar{m}\bar{m}}
	&=
	0
	,\\	
\label{eq:muhll_recon}
-	\left(\Delta+\bar{\mu}\right)\left(\mu h_{ll}\right)
-	\mu\left(\edth+\bar{\pi}+2\tau\right)h_{l\bar{m}} &\nonumber\\
-	2\Psi_2^{(1)}
-	\pi\left(\edth^{\prime}-\pi\right)h_{mm}
	&\nonumber\\
+	\left(
		\mu\edth^{\prime}
	-	3\mu\pi
	+	2\bar{\mu}\pi
	-	2\mu\bar{\tau}
	\right)h_{lm}
	&\nonumber\\
-	2\left(\edth+\bar{\pi}\right)\pi^{(1)}
-	2\pi\bar{\pi}^{(1)}
	&=
	0
	.	
\end{align}
\end{subequations}
We solve the equations in the sequence listed,
in each step obtaining one of the following set
of unknowns: $\{
	\Psi_3^{(1)},\lambda^{(1)},\Psi_2^{(1)},
	h_{\bar{m}\bar{m}},
	\pi^{(1)},
	h_{l\bar{m}},
	h_{ll}
\}$. We set the initial values for all these
quantities to zero; as explained in the Introduction
and Sec.~\ref{sec:compact_initial_data}, this choice
is consistent from the ingoing slice $v=v_u$ shown
in Fig.\ref{fig:metric_reconstruction_strategy_pictureb} 
onward (i.e. for $v\geq v_u$),
as long as the initial data for $\Psi_4^{(1)}$ only contains
azimuthal modes $|m|\ge2$.

\subsection{Source term for the second order perturbation $\Psi_4^{(2)}$}
	Having computed the linearized metric, we can then solve for
the second order perturbation of the Weyl scalar, $\Psi_4^{(2)}$.
	As was first shown in \cite{Campanelli:1998jv},
the equation of motion for $\Psi_2^{(2)}$ can be written as
\begin{align}
\label{eq:second_order_teuk}
	\mathcal{T}\Psi_4^{(2)}
	=
	\mathcal{S}\left[h_{\mu\nu}\right]
	,
\end{align}
	where $\mathcal{T}$ is the Teukolsky operator
(Eq.~\eqref{eq:Teuk_NP}), and $\mathcal{S}$ is the ``source'' term
which depends on the first order perturbed metric.
The general expression for $\mathcal{S}$ is given in
\cite{Campanelli:1998jv,Loutrel:2020wbw}.
Here we write it down in outgoing radiation gauge and with our
background tetrad choice
(see Appendix \ref{sec:derivation_metric_recon_source} for a derivation):
\begin{align}
\label{eq:schematic_source}
	\mathcal{S}
	=
	\left(\Delta+4\mu+\bar{\mu}\right)\mathfrak{s}_d
+	\left(\edth^{\prime}+4\pi-\bar{\tau}\right)\mathfrak{s}_t
	,
\end{align} 
	where
\begin{subequations}
\label{eq:def_mathfrak_s}
\begin{align}
   \mathfrak{s}_d
   \equiv&
   \frac{1}{2}h_{ll}\left(\Delta+\mu\right)\Psi_4^{(1)}
   +  \Psi_4^{(1)}\bigg[
      \frac{1}{2}\left(
         \edth+\bar{\pi}+2\tau
      \right)h_{l\bar{m}}
      \nonumber\\&
   +  \left(\Delta-\mu+\bar{\mu}\right)h_{ll}
   -  \frac{1}{2}\left(
      \edth^{\prime}-5\pi-4\bar{\tau}
      \right)h_{lm}
   \bigg]
   \nonumber\\&
-  \frac{1}{2}\Psi_3^{(1)}\bigg[
      \left(\edth+\bar{\pi}+\tau\right)h_{\bar{m}\bar{m}}
   +  \left(
         \Delta
      -  2\mu
      +	 \bar{\mu}
      \right)h_{l\bar{m}}
   \bigg]
   \nonumber\\&
-  \left(
      h_{l\bar{m}}\Delta
   -  \frac{1}{2}h_{\bar{m}\bar{m}}\edth
   -  4\pi^{(1)}
   \right)\Psi_3^{(1)}
-  3\lambda^{(1)}\Psi_2^{(1)}
   ,\\
   \mathfrak{s}_t
   \equiv&
   -  h_{lm}\left(\Delta+\mu+2\bar{\mu}\right)\Psi_4^{(1)}
   +  \frac{1}{2}h_{mm}\edth^{\prime}\Psi_4^{(1)}
   \nonumber\\&
   +  \Psi_4^{(1)}\bigg[
   \bar{\pi}^{(1)}
   -  \Delta h_{lm}
   +  \left(
         \edth^{\prime}
      -  \frac{1}{2}\pi
      -	 \frac{1}{2}\bar{\tau}
      \right)h_{mm}
   \bigg]
   .
\end{align}
\end{subequations}
\section{Measurement of the gravitational wave
	at future null infinity
}
\label{sec:gauge_invariance_measurements}
	For outgoing radiation at future null infinity
in an asymptotically flat coordinate system
there is a simple relation between $\Psi_4^{(1)}$
and the linearized metric (e.g. \cite{Teukolsky:1973ha}):
\begin{align}\label{psi4_1_inf}
	\lim_{r\to\infty}\Psi_4^{(1)}
	=
-	\frac{1}{2}\left(
		\partial_t^2 h_{+}
	-	i\partial_t^2 h_{\times}
	\right)
	,
\end{align}
   where the $+$ and $\times$ subscripts refer to the ``plus''
and ``cross'' polarizations of the gravitational wave.
From this we can then also calculate other quantities of interest,
such as the energy
and angular momentum radiated to future null infinity.

	As is discussed in \cite{Campanelli:1998jv}
(see Sec III.C and Sec. IV therein), $\Psi_4^{(2)}$
is in general not invariant under gauge or tetrad
transformations that are first order in magnitude
(it is invariant under second order transformations).
This complicates the interpretation of $\Psi_4^{(2)}$,
unless the gauge/tetrad is fixed in an appropriate
way, or as outlined in \cite{Campanelli:1998jv},
a gauge invariant object is calculated that
by construction reduces to $\Psi_4^{(2)}$ in the desired
gauge at null infinity. Here, because we
use the outgoing radiation gauge in an asymptotically
flat representation of Kerr, and our first order
correction to the tetrad (\ref{tetrad_pert}) amounts to a Class II
transformation that leaves $\Psi_4$ invariant~\cite{Chandrasekhar_bh_book},
we can directly interpret $\Psi_4^{(2)}$ as we do
$\Psi_4^{(1)}$ in (\ref{psi4_1_inf}) at future
null infinity; i.e. we can interpret the real and imaginary
parts of $\Psi_4^{(2)}$ as the second time derivative of the
plus and cross gravitational wave polarizations, respectively.

Another way to understand the physical interpretation of the second
order gravational wave perturbation at future null infinity
is through the radiated energy and angular momentum \cite{Campanelli:1998jv}:
\begin{subequations}
\label{eq:radiated_quantities}
\begin{align}
\label{eq:radiated_energy}
   \frac{dE}{du}\Big|_{\mathcal{J}_+}
   =&
   \lim_{r\to\infty}
   \left\{
      \frac{r^2}{4\pi}\int_{\Omega}d\Omega 
         \left|\int_{-\infty}^u du^{\prime} \Psi_4\right|^2
   \right\}
   \\
\label{eq:radiated_angular_momenta}
   \frac{dJ_z}{du}\Big|_{\mathcal{J}_+} 
   =&
-  \lim_{r\to\infty}
   \mathcal{R}
   \Bigg\{
      \frac{r^2}{4\pi}\int_{\Omega}d\Omega
      \left(
         \partial_{\phi}\int_{-\infty}^udu^{\prime}\Psi_4
      \right)
      \nonumber\\ &\times
      \left(
         \int_{-\infty}^udu^{\prime}\int_{-\infty}^{u^{\prime}}du^{\prime\prime}\Psi_4
      \right)
   \Bigg\}
   .
\end{align}
\end{subequations}
In computing $\Psi_4^{(2)}$ at future null infinity using outgoing
radiation gauge to linear order, one can compute the linear and leading order
nonlinear contribution to the radiated energy and angular
momentum through future null infinity. 
\section{Initial data}\label{sec:initial_data}
\label{sec:compact_initial_data}
As discussed in the introduction and illustrated in
Fig.~\ref{fig:metric_reconstruction_strategy_pictureb} there,
on our $T=0$ initial data surface $\mathfrak{i}_0$ we set
$\Psi_4^{(1)}$ to be nonzero and smooth over a compact region
$\mathfrak{p}_0\subset\mathfrak{i}_0$,
and set the rest of our evolution variables 
(the metric reconstructed variables
$\{h_{ll},h_{l\bar{m}},h_{\bar{m}\bar{m}},
\Psi_3^{(1)},\Psi_2^{(1)},
\lambda^{(1)},\pi^{(1)}\}$, and $\Psi_4^{(2)}$)
to be zero everywhere
on $\mathfrak{i}_0$. 
The initial data is \emph{consistent} if it satisfies
all of the Einstein equations to the relevant order
in perturbation theory.
Our choice of initial data is in general only
consistent in the complement of $\mathfrak{p}_0$,
and then only, as discussed in the following subsection,
for angular components with $l \ge 2$. 
As the reconstructed metric variables are advected along
$n^{\mu}$ (the principal part of their corresponding transport equations is $\Delta$),
the constraint violating modes in our initial data will
also be advected along $n^{\mu}$, into the black hole.
As the constraint violating region is restricted to the
initial compact region $\mathfrak{p}_0$, within a finite
amount of evolution time all of the $l \ge 2$ constraint violating
modes will propagate off our computational domain.
 
To estimate how long we must wait until the constraint
violating modes are advected into the black hole,
we compute the travel time along $n^{\mu}$ from the outermost
point $R_{min}$ 
of the support of $\Psi_4^{(1)}$ on the initial data slice 
to the black hole horizon $R_H$ (recall that $R$ increases toward the horizon).
From
\begin{align}
	\Delta f
	=
	n^{\mu} \partial_{\mu}f
	=
	\left(2+\frac{4MR}{L^2}\right)\partial_Tf
+	\frac{R^2}{L^2}\partial_Rf
	,
\end{align}
	we see that along the characteristic we can write
\begin{align}
	\frac{dT}{dR}
	=
	\frac{L^2}{R^2}\left(2+\frac{4MR}{L^2}\right)
	.
\end{align}
	Thus the time we need to wait is
\begin{align}
	\frac{T_{mw}}{M}
	=&
	\int_{R_{min}}^{R_{H}}\frac{dR}{M}
	\frac{L^2}{R^2}\left(2+\frac{4MR}{L^2}\right)
	,\nonumber\\
	=&
	\frac{2L^2}{M}\left(\frac{1}{R_{min}}-\frac{1}{R_{H}}\right)
+	4\mathrm{ln}\left(\frac{R_{H}}{R_{min}}\right)
	.
\end{align}
	Using the relation $r\equiv L^2/R$ to convert
to Boyer-Lindquist $r$, with
$r_{max}\equiv L^2/R_{min}$, and for a conservative
estimate of the wait time setting $r_{H}\equiv L^2/R_H=M$,
Table ~\ref{table:wait_times_to_start_reconstruction}
gives several wait times for illustration.
\begin{table}
\centering
\begin{tabular}{ c|c }
\hline
 $r_{max}/M$ & $T_{mw}/M$ \\ 
\hline
 $3$  & $\sim 9$\\ 
 $5$  & $\sim 15$\\ 
 $10$ & $\sim 28$\\ 
 $50$ & $\sim 114$\\ 
\end{tabular}
\caption{Example minimum wait times, $T_{mw}$, before constraint violating
region exits computational the domain, and we begin evolving
$\Psi_4^{(2)}$.}
\label{table:wait_times_to_start_reconstruction}
\end{table}	
\subsection{Modes $|m|=0,1$}
\label{modes_m01}
   A field of spin weight $s$ and angular number $m$ has angular
support over modes $l\geq\mathrm{max}\left(|s|,|m|\right)$
(see Appendix \ref{sec:swaL}). 
Essentially because of this, and as is well known,
the $s=-2$ field $\Psi_4^{(1)}$
can not describe changes to the Kerr spacetime mass ($l=0$ modes)
and spin ($l=1$ modes), nor can it fix spurious
gauge modes with support over those angular numbers
\cite{doi:10.1063/1.1666203}.
Moreover, as the mass and spin modes do not propagate, we cannot
simply begin with a constraint violating region of compact support
and expect the constraint violating modes to propagate off our
domain in some finite amount of time (as they do for $l,|m|\geq2$
propagating modes). In order to obtain
fully consistent evolution we would need to add in consistent $l=0,1$ data
\emph{everywhere} on our initial data surface\footnote{Determining consistent
$l=0,1$ data is sometimes called ``completing the metric reconstruction''
procedure in the gravitational self-force literature and remains
only a partially solved problem in that field;
e.g. \cite{Merlin:2016boc,Dolan:2012jg} and references therein.}.

   We leave constructing such nontrivial
initial data for future research,
and content ourselves with metric reconstruction for $|m|\geq2$ modes. 
We note that while we can only reconstruct the metric over angular
modes $l,|m|\geq2$, we can still compute their contribution to the
evolution of $\Psi_4^{(2)}$ for $|m|=0,1$, as that field only has
support over angular numbers $l\geq 2$.
In particular, for the examples presented here we can still
consistently compute the contribution of the
$m=-2$ and $m=2$ metric reconstructed fields to the evolution of
the $m=0$ mode of $\Psi_4^{(2)}$.
\section{Code implementation details}
\label{sec:code_implementation}
	In this section we describe the details of our numerical implementation.
The code can be downloaded at \cite{code_online}.
A Mathematica notebook that contains our derivations of the equations
of motion in coordinate form can be downloaded at \cite{notebook_online}.
\subsection{Teukolsky \& Metric reconstruction equations in coordinate form}
\label{sec:teuk_in_coord_form}
One can economically write a master equation for both 
the spin $s=-2$ equation governing $\Psi_4$ (\ref{eq:Teuk_NP}),
and the spin $s=2$ equation governing $\Psi_0$ (see~\cite{Teukolsky:1973ha}), 
so we do that here, though the rest of the paper deals exclusively with $\Psi_4$.

	Following~\cite{Teukolsky:1973ha},
we define the functions $\psi_4^{(1)}$ and $\psi_0^{(1)}$ via
\begin{subequations}
\begin{align}
	\Psi_4^{(1)}
	\equiv&
	R \psi_4^{(1)} 
	,\\
	\Psi_0^{(1)}
	\equiv&
	R \left(\frac{\Psi_2}{M}\right)^{4/3}\psi_0^{(1)} 
	,
\end{align}
\end{subequations}
which are motivated by the 
``peeling theorem''\cite{Newman_Penrose_paper}: we expect
$\Psi_4^{(1)}\sim1/r=R$ and $\Psi_0^{(1)}\sim 1/r^5\sim R^5$.
We next multiply the NP form of the Teukolsky equation Eq.~\eqref{eq:Teuk_NP}
(and its analogue spin $2$ version)
by $2\Sigma_{BL}/R$  ($\Sigma_{BL}\equiv r^2+a^2\mathrm{cos}^2\vartheta$)
to make the leading order terms finite at $R=0$ (future null infinity).
These scalings allow one to directly
solve for and read off the gravitational waves at infinity 
as finite, non-zero fields.
The resultant spin $s=\pm2$ Teukolsky equation,
in terms of these variables in our coordinates and tetrad, is
\begin{widetext}
\begin{align}
\label{eq:Teukolsky_equation}
\left[
	8M\left(2M-\frac{a^2R}{L^2}\right)\left(1+\frac{2MR}{L^2}\right)
-	a^2\mathrm{sin}^2\vartheta
	\right]\partial_T^2\psi^{(1)}
-	2\left[
		L^2
	-	\left(8M^2-a^2\right)\frac{R^2}{L^2}
	+4	\frac{a^2M}{L}\frac{R^3}{L^3}
	\right]\partial_T\partial_R\psi^{(1)}
	\nonumber \\
-	\left(L^2-2MR+a^2\frac{R^2}{L^2}\right)
	\frac{R^2}{L^2}\partial_R^2\psi^{(1)}
-	\spindersphere\psi^{(1)}
 +	2a\left(
		1+\frac{4MR}{L^2}
	\right)\partial_T\partial_{\phi}\psi^{(1)}
+	2a\frac{R^2}{L^2}\partial_R\partial_{\phi}\psi^{(1)}
	\nonumber \\
+	2\bigg[
		2M\left(-s+(2+s)\frac{2MR}{L^2}-\frac{3a^2R^2}{L^4}\right)
	-	\frac{a^2R}{L^2}
	+	isa\mathrm{cos}\vartheta
	\bigg]\partial_T\psi^{(1)}
	\nonumber \\
+	2R\left[
		-(1+s)+(s+3)\frac{MR}{L^2}-\frac{2a^2R^2}{L^4}
	\right]\partial_R\psi^{(1)}
+	\frac{2aR}{L^2}\partial_{\phi}\psi^{(1)}
+	2\left[(1+s)\frac{MR}{L^2}-\frac{a^2R^2}{L^4}\right]\psi^{(1)}
	=
	0
	,
\end{align}
	where $s$ should be set to $-2$ ($2$) if $\psi^{(1)}=\psi_4^{(1)}$ ($\psi_0^{(1)}$),
and $\spindersphere$ is the spin-weight $s$ Laplace-Beltrami operator on
the unit two-sphere; see Appendix~\ref{sec:swaL}. 

	We rewrite Eq.~\eqref{eq:Teukolsky_equation} as a system of first
order partial differential equations by defining
\begin{subequations}
\label{eq:definitions_PQ}
\begin{align}
\label{eq:def_P}
	P^{(1)}
	\equiv &
	\bigg[
		8M\left(2M -\frac{a^2R}{L^2}\right)
		\left(1+\frac{2MR}{L^2}\right)
	-	a^2\mathrm{sin}^2\vartheta
	\bigg]\partial_T\psi^{(1)}
-	 2\left(
		L^2
	-	\left(8M^2-a^2\right)\frac{R^2}{L^2}
	+4	\frac{a^2M}{L}\frac{R^3}{L^3}
	\right)\partial_R\psi^{(1)}
	\nonumber \\ 
+	& 2a\left(
		1+\frac{4MR}{L^2}
	\right)\partial_{\phi}\psi^{(1)}
+	 2\bigg[
		2M\left(-s+(2+s)\frac{2MR}{L^2}-\frac{3a^2R^2}{L^4}\right)
	-	\frac{a^2R}{L^2}
	+	isa\mathrm{cos}\vartheta
	\bigg]\psi^{(1)}
	, \\
\label{eq:def_Q}
	Q^{(1)}
	\equiv &
	\partial_R\psi^{(1)}
	.
\end{align}
\end{subequations} 
\end{widetext}

We decompose the fields in terms of $e^{im\phi}$,
as the equations of motion are invariant under shifts in $\phi$.
Defining
\begin{align}
\label{eq:def_v}
	{\bf v}(T,R,\vartheta,\phi)
	\equiv
	\begin{pmatrix}
	P^{(1)}(T,R,\vartheta)\\
	Q^{(1)}(T,R,\vartheta)\\
	\psi^{(1)}(T,R,\vartheta) 
	\end{pmatrix}
	e^{i m\phi}
	,
\end{align}
	and factoring out the overall factor of $e^{i m\phi}$,
we can write the Teukolsky equation as
\begin{align}
\label{eq:evolution_teuk}
	\partial_T{\bf v}
	=
	\mathbb{A}\partial_R{\bf v}
+	\mathbb{B}\spindersphere{\bf v}
+	\mathbb{C}{\bf v}
	,
\end{align}
	where $\mathbb{A}$, $\mathbb{B}$, and $\mathbb{C}$ are matrices
that can be straightforwardly evaluated from
Eqs.~(\ref{eq:Teukolsky_equation}-\ref{eq:def_v}).
We empirically find for very rapidly rotating black holes ($a/M\gtrsim0.99$)
that the ``constraint'' $Q-\partial_R\psi_4=0$ is poorly maintained
by free evolution. To amend this, we evolve our runs by \emph{imposing}
the constraint $Q=\partial_R\psi_4$ at each intermediate step of
our time solver (fourth order Runge-Kutta scheme; see e.g. \cite{10.5555/1403886}),
and not freely evolving $Q$.
A test that our Teukolsky solver gives solutions that
converge to the continuum Teukolsky equation then comes from our check
that the late time behavior of $\Psi_4^{(1)}$ matches the behavior
of a mode that one would expect for a $s=-2$,
$l=\mathrm{max}\left[|s|,|m|\right]$ quasinormal mode (see Sec.~\ref{sec_ires}). 

	Using the coordinate forms of the tetrad
Eq.~\eqref{eq:tetrad_IEF_HC} and NP scalars
Eq.~\eqref{eq:NP_IEF_HC}, 
it is straightforward to write the metric reconstruction
equations \eqref{eq:metric_reconstruction} and directional
derivative operator $\Delta$ (\ref{diff_op}) in coordinate form;
the full expressions are not particularly illuminating,
so we do not give their explicit form here.
Their full form can be found in the Mathematica notebook \cite{notebook_online}.
We describe how we evaluate the GHP derivatives
$\edth$ and $\edth^{\prime}$ in Sec. \ref{sec:eval_ghp_der}.

\subsection{Pseudo-spectral evolution}
\label{sec:pseudo-spectral_evolution}
	We numerically solve the Teukolsky equation Eq.~\eqref{eq:evolution_teuk}
and the metric reconstruction equations Eq.~\eqref{eq:metric_reconstruction} using
pseudo-spectral methods.
Here we review the basic elements of pseudo-spectral methods that we 
implemented in our code;
see, e.g. \cite{fornberg_1996,Trefethen:2000,boyd2001chebyshev} for
a general discussion of these methods.
As mentioned, the equations of motion are invariant under shifts of $\phi$, so we first
decompose all variables in terms of definite
angular momentum number $m$
\begin{align}
\label{eq:decomposition_in_phi}
	\eta(T,R,\vartheta,\phi)
	\equiv
	\eta^{[ m]}(T,R,\vartheta)e^{i m\phi}
	.
\end{align} 
	For a given $m$ then, we have
to solve a $1+2$ ($T+(R,\vartheta)$) dimensional system of 
partial differential equations.

	We expand the fields as a sum
of Chebyshev polynomials and spin-weighted spherical harmonics. Writing
\begin{subequations}
\begin{align}
	R_{max}
	\equiv&
	\frac{L^2}{r_+}
	,\\
	x
	\equiv&
	2\frac{R}{R_{max}}-1
	,\\
	y
	\equiv&
	-\mathrm{cos}\vartheta
	,
\end{align}
\end{subequations}
	we have
\begin{align}
	\eta^{[ m]}(T,x,y)
	=
	\sum_{n,l}\eta_{nl}^{[ m]}(T)T_n(x){}_sP^m_l(y)
	,
\end{align}
	where $T_n$ is the $n^{th}$ Chebyshev polynomial,
\begin{align}
	T_n(x)
	\equiv
	\mathrm{cos}\left(n\;\mathrm{arccos}\left(x\right)\right)
	,
\end{align}
	and ${}_sP^m_l$ is a spin-weighted associated Legendre polynomial
(see Appendix \ref{sec:swaL}).
We use the spin weight $s$ of a quantity (related
to how it scales under certain tetrad transformations)
as introduced by GHP \cite{GHP_paper}.
All NP scalars except for $\{\alpha,\beta,\epsilon,\gamma\}$
have a definite spin weight, as do our first order metric
projections; see Table.~\eqref{table:spins_and_falloff}
for a listing of the spin weights and radial falloff of the
variables we solve for in our code.
Expanding each field with the matching spin-weighted spherical harmonic
 ${}_sP^m_l$ ensures 
 the fields automatically have the correct
regularity properties along the axis $\vartheta=0,\varpi$.

\begin{table}
\centering
\begin{tabular}{ c | c | c }
\hline
 variable  & spin weight & falloff \\ 
 \hline 
 $\Psi_4^{(1)},\lambda^{(1)},h_{\bar{m}\bar{m}}$ & -2 & $r^{-1}$ \\
 $\Psi_3^{(1)},\pi^{(1)},h_{l\bar{m}}$ & -1 & $r^{-2}$ \\
 $\Psi_2^{(1)}, h_{ll}$ &  0 & $r^{-3}$ \\
\hline
\end{tabular}
\caption{Spin weight and falloff of key variables. The falloff
   is derived by assuming $\Psi_4^{(1)}\sim 1/r$,
   and then considering the metric reconstruction
   equations \eqref{eq:metric_reconstruction};
   these falloffs are consistent with the
   ``peeling theorem'' \cite{Newman_Penrose_paper}
   and with what we observe in our code output.
   See \cite{Loutrel:2020wbw} for a more detailed discussion,
   in particular for a derivation of the radial falloff
   of $h_{ll}$, which depends on several cancellations
   in the equations of motion.
}
\label{table:spins_and_falloff}
\end{table}

	We evaluate the Chebyshev polynomials at Gauss-Lobatto
collocation points, and move to/from Chebyshev space
using Fast Cosine Transforms (FCT)\footnote{Specifically, we evaluate
the Fast Cosine Transforms using FFTW \cite{FFTW05}; see 
\cite{code_online}.}:
\begin{align}
	\eta(T,x,y)
	\mathrel{\mathop{\rightleftarrows}^{
		\mathrm{FCT}
	}_{\mathrm{FCT}}
	}
	\sum_n \eta_n(T,y) T_n(x)
	.
\end{align}
	We evaluate the spin weighted associated Legendre polynomials
at the roots of the $n^{th}$ Legendre
polynomial, and move to/from spin-weighted spherical harmonics using Gauss
quadrature and direct summation
\begin{align}
	\eta(T,x,y) 
	\mathrel{\mathop{\rightleftarrows}^{
		\mathrm{Gauss\;quadrature}
	}_{\mathrm{Summation}}
	}
	\sum_l \eta_l(T,x){}_sP^m_l(y)
	.
\end{align}
	We evaluate radial derivatives
by transforming to Chebyshev space, then recursively use the relation
\begin{align}
	\frac{1}{n+1}\frac{dT_{n+1}}{dx}
-	\frac{1}{n-1}\frac{dT_{n-1}}{dx}
	=
	2T_n
	,
\end{align}
	with the seed condition $T_{n_{max}+1}=0$ as
 we only expand out to
$n_{max}$ Chebyshev polynomials.
All the angular derivatives in our equations of
motion either appear in terms of the spin-weighted spherical Laplacian
$\spindersphere$, or in terms of the GHP covariant operators $\edth$
and $\edth^{\prime}$; we discuss how we evaluate
these in Sec.~\ref{sec:eval_ghp_der} below.

	We evolve the equations in time with the method of lines,
specifically using a fourth-order Runge-Kutta integrator
(see e.g. \cite{10.5555/1403886}). We use a 
time step $\Delta t$ 
of $9/\mathrm{max}\left(N_x^2,N_y^2\right)$, where $N_x (N_y)$
is the number of radial (angular) collocation points.
	After each time step we apply an exponential filter to all
the evolved variables in spectral space: 
\begin{align}
\label{eq:spectral_filter}
	c_{nl}
	\to
	\mathrm{exp}\left[
	- 	A \left\{
			\left(\frac{n}{n_{max}}\right)^p
		+	\left(\frac{l}{l_{max}}\right)^p
		\right\}
	\right]
	c_{nl}
	.
\end{align}
	For the results presented here we set $A=-40$ and $p=16$.
We use $A=-40$ as $e^{-40}\sim 10^{-18}$ is roughly the relative
precision of the double precision floating point arithmetic 
we used.
We set $p=16$ so that spectral coefficients
of low $n$ and $l$ are largely unaffected by the filter.
Note that the filter converges away with increased resolution (i.e.
larger $n_{max},l_{max}$). 
We found using a \emph{smooth} spectral filter such as
Eq.~\eqref{eq:spectral_filter} (as opposed to simply
zeroing $c_{nl}$ above a certain $(n,l)$) was crucial to achieve stable
evolution for high spin ($a\gtrsim0.99$) black holes.

We evaluate the source term, Eq.~\eqref{eq:schematic_source}
in two steps. We first compute $\mathfrak{s}_d$
and $\mathfrak{s}_t$ (Eqs.~\eqref{eq:def_mathfrak_s}),
and then Eq.~\eqref{eq:schematic_source}.
We can rewrite time derivatives in $\mathfrak{s}_d$
and $\mathfrak{s}_t$ in terms of spatial derivatives using the
evolution equations Eqs.~\eqref{eq:metric_reconstruction},
which can then be evaluated using pseudo-spectral methods
(we use $P$ to evaluate $\partial_t\Psi_4^{(1)}$).
We compute the time derivative for, e.g.
$\left(\Delta+4\mu+\bar{\mu}\right)\mathfrak{s}_d$ by saving several
time steps for $\mathfrak{s}_d$ and evaluating $\partial/\partial T$ with 
a fourth order backward difference stencil 
(again spatial derivatives are computed using pseudo-spectral methods).
\subsection{Evaluation of the GHP $\edth$ and $\edth^{\prime}$ operators}
\label{sec:eval_ghp_der}
	We can straightforwardly evaluate the background NP 
scalars at each collocation point using Eqs.~\eqref{eq:NP_IEF_HC}. 
Using the expressions for the tetrad \eqref{eq:tetrad_IEF_HC}, we can also straightforwardly
evaluate the NP derivatives in Eq.~\eqref{diff_op}.
The only potential difficulty comes from
$\{\alpha,\beta,\delta,\bar{\delta}\}$, as they all contain components
that go as $\sim 1/\mathrm{sin}\vartheta$; i.e. they blow up on the
coordinate axis $\vartheta=0,\varpi$.
To obtain regular answers using $\{\alpha,\beta,\delta,\bar{\delta}\}$,
we use these terms in combinations that have definite spin weight.
In particular, these terms only appear in combinations that make up the GHP
derivative operators $\{\edth,\edth^{\prime}\}$, which do have
definite spin weight when acting on scalar fields of
definite spin weight\footnote{We have already substituted
$\{\edth,\edth^{\prime}\}$ for $\{\alpha,\beta,\delta,\bar{\delta}\}$
in the metric reconstruction equations (\ref{eq:metric_reconstruction})
and source terms (\ref{eq:schematic_source},\ref{eq:def_mathfrak_s}).}.
In our coordinate system, these operators evaluate to 
\begin{subequations}
\begin{align}
	\edth\eta
	=&
	\frac{R}{\sqrt{2}}
	\frac{
		1
	}{
		\left(L^2-iaR\mathrm{cos}\vartheta\right)
	}
	\left(
	-	ia\mathrm{sin}\vartheta\partial_T
	+	\Edth
	\right)
	\eta
        \nonumber\\&
-	\frac{ip}{\sqrt{2}}
	\frac{
		aR^2\mathrm{sin}\vartheta
	}{
		\left(L^2-iaR\mathrm{cos}\vartheta\right)^2
	}
	\eta
	,\\
	\edth^{\prime}\eta
	=&
	\frac{R}{\sqrt{2}}
	\frac{
		1
	}{
		\left(L^2+iaR\mathrm{cos}\vartheta\right)
	}
	\left(
		ia\mathrm{sin}\vartheta\partial_T
	+	\Edth^{\prime}
	\right)
	\eta
        \nonumber\\&
+	\frac{iq}{\sqrt{2}}
	\frac{
		aR^2\mathrm{sin}\vartheta
	}{
		\left(L^2+iaR\mathrm{cos}\vartheta\right)^2
	}
	\eta
	,
\end{align} 
\end{subequations}
	where $\{\Edth,\Edth^{\prime}\}$ are the raising and
lowering operators for spin weighted spherical harmonics;
(see Appendix~\ref{sec:swaL}), and $\{p,q\}$ are the weights of
the NP field in question (see \cite{GHP_paper}).
Note that we evaluate $\{\Edth,\Edth^{\prime}\}$
in spectral space using the relations (\ref{Dplus}) and (\ref{Dminus}).
Written this way, the GHP derivatives are clearly regular at
$\vartheta=0,\varpi$
(as they should be, as they are GHP-covariant quantities).
\subsection{Boundary conditions}
	We place the radial boundaries of our domain at the black
hole horizon and at future null infinity, which is possible as our coordinates
are hyperboloidally compactified and horizon penetrating
(for more of a discussion on hyperboloidal
compactifications, see e.g. \cite{Zenginoglu_2008}).
At these locations
none of the field characteristics point into our computational domain,
so we do not need to impose boundary conditions at those boundaries.

The polar boundaries of the computational domain $\vartheta=\{0,\varpi\}$ 
are not boundaries of the physical domain, and often in such
situations regularity conditions need to be applied there.
However, as we have rewritten all the equations so 
they are regular at the poles, in
particular in that we calculate angular derivatives using
the GHP $\edth$ and $\edth^{\prime}$ operators applied
to the correct spin weighted harmonic decomposition of each variable,
regularity is ensured at ${\vartheta=0,\varpi}$ without any additional
conditions.
\subsection{Second order equation and radial rescaling}
	For the second order perturbation, the corresponding
Teukolsky equation we solve is
\begin{align}
	\left(2\Sigma_{BL}\frac{1}{R}\right)
	\mathcal{T}\left(R\psi_4^{(2)}\right)
	=
	\left(2\Sigma_{BL}\frac{1}{R}\right)\mathcal{S}
	,
\end{align}
        where $\mathcal{T}$ is the same spin weight $-2$
Teukolsky operator in Eq.~\eqref{eq:Teuk_NP} as acts on the first order perturbation,
hence the coordinate form of the left hand side is the same as in 
Eq.~\eqref{eq:Teukolsky_equation} with $s=-2$, but with $\psi_4^{(1)}$
replaced with $\psi_4^{(2)}$.

        The different radial falloff behavior of different NP scalars
and first order metric fields can make it challenging to accurately evaluate
the source term $\mathcal{S}$ using double precision arithmetic. To alleviate some of this,
in the code we use versions of these quantities rescaled by their assumed falloff,
as summarized in Table~\ref{table:spins_and_falloff}. 
We use a circumflex to denote the rescaled form of a variable; for example
$\Psi_4^{(1)}\equiv R\hat{\Psi}_4^{(1)}$,
$\rho\equiv R\hat{\rho}$,
$h_{l\bar{m}}\equiv R^2\hat{h}_{l\bar{m}}$, etc.
Note the radial derivative acting on a rescaled field is 
\begin{align}
   \partial_Rf
   =
   R^{n-1}\left(n + R\partial_R\right)\hat{f}
   .
\end{align}
\subsection{Evolution of different $ m$ modes}
	As the Kerr background is invariant under rotations in $\phi$,
to linear order in perturbation theory each $ m$ mode
is preserved. To second order in perturbations there is mode
mixing. In particular, from the form of the source term,
Eq.~\eqref{eq:schematic_source}, and given at present we only evolve
a single magnitude $|m|$ mode of $\Psi_4^{(1)}$ in our code,
we will have mixing of the form
\begin{align}
	& \{
		\Psi_4^{(1)[m]},
		\Psi_4^{(1)[-m]}
	\} 
        \rightarrow
	\{
		\Psi_4^{(2)[2m]},
		\Psi_4^{(2)[0]},
		\Psi_4^{(2)[-2m]}
	\}
	.
\end{align}
	For any given run then we simultaneously evolve 
first order perturbative modes with angular numbers
$\pm m$, and second order perturbations with angular
numbers  $\{0,\pm2 m\}$. 

	In astrophysical scenarios, we expect all $m$ modes to be excited,
which would lead to more complicated mode mixing: from the source term
we see any pair of first order modes $m_1,m_2$ will in general 
produce the four second order modes $\pm m_1 \pm m_2$.
While our code can handle such cases, in this paper we only consider
mode mixing of the form $[m]\to\{[0],[\pm 2m]\}$.

\subsection{Functional form of our initial data}
\label{sec:functional_form_initial_data}
	Here we present the specific functional form of initial data
for $\Psi_4^{(1)}$, in terms 
of the evolved fields
$\{\psi_4,Q,P\}$ (as defined in Sec.~\ref{sec:teuk_in_coord_form} above).

	As discussed in the introduction, we choose initial data for $\Psi_4^{(1)}$ 
that has compact support in $r$, to simplify the initial conditions for
the first order reconstruction within the part of the domain where we eventually solve
for the second order perturbation $\Psi_4^{(2)}$.
	For $\psi_4^{(1)}$
($\equiv r\Psi_4^{(1)}$), we choose the following rescaled ``bump function''
\begin{align}
\label{eq:functional_form_initial_data}
	&\psi_4^{(1)[m]}\big|_{T=0}
	=
                 \\&
	\begin{cases}
	a_0
			\left(\frac{r-r_l}{r_u-r_l}\right)^2
			\left(\frac{r_u-r}{r_u-r_l}\right)^2
        \nonumber\\
	\ \ \times	\mathrm{exp}\left[
				-\frac{1}{r-r_l}-\frac{2}{r_u-r}
			\right] 
			{}_sP^m_{l_0}\left(\vartheta,\phi\right)
                        ,& r_l<r<r_u
	\\
	0, & \mathrm{otherwise} 
	\end{cases}
\end{align}
	where $r_u>r_l$, $a_0$, $l_0$ and $m$ are constants,
and ${}_sP^m_l$ is a spin-weighted associated Legendre polynomial 
(see Appendix \ref{sec:swaL}).
We set $Q^{(1)}=\partial_R\psi^{(1)}_4$ as per its definition Eq.~\eqref{eq:def_Q}.
We solve the following equation for $P^{(1)}$ (Eq.~\eqref{eq:def_P}) at $T=0$
so that the initial gravitational wave pulse is initially radially ingoing:
\begin{align}
	n^T\partial_T\psi^{(1)}_4+n^R\partial_R\psi^{(1)}_4
	=
	0
	.
\end{align}
	The reason for this choice is
to minimize the ``prompt'' response at future null infinity
from an outgoing pulse that would largely be a reflection of the initial data,
thus more quickly being able to measure the ringdown response of the
black hole to the perturbation.
\subsection{Independent residuals and code tests}\label{sec_ires}
	Our metric reconstruction procedure does not use all of the Bianchi
and Ricci identities; we can thus use some of these 
``extra'' equations as independent residual checks of our numerical computation.
We directly evaluate the following Bianchi identity
(see Eq. (1.321.d) in \cite{Chandrasekhar_bh_book}):
\begin{eqnarray}
\label{eq:def_B3}
	\mathcal{B}_3
	\equiv
	\left(\edth^{\prime}+4\pi\right)\Psi_3^{(1)}
	\nonumber\\
+	\left(-D-4\epsilon+\rho\right)\Psi_4^{(1)}
-	3\lambda^{(1)}\Psi_2
	=
	0
	.
\end{eqnarray} 
Beginning from 
(Eq. (1.321.c) in \cite{Chandrasekhar_bh_book}):
\begin{subequations}
\begin{align}
\label{eq:def_B2_a}
	\left(
	-	\edth^{\prime}-3\pi
	\right)
	\Psi_2^{(1)}
+	\left(
	-	\edth^{\prime}-3\pi
	\right)^{(1)}
	\Psi_2
	\nonumber\\
+	\left(D+2\epsilon-2\rho\right)\Psi_3^{(1)}
	=&
	0
	,
\end{align}
\end{subequations} 
	using the first order perturbed equations for
$\bar{\delta}^{(1)}$ in Eq.~\eqref{eq:delta-1} and $\alpha^{(1)}$
(see ~\cite{Loutrel:2020wbw}),
and the type D equations for $\Psi_2$:
\begin{subequations}
\begin{align}
	\Delta\Psi_2
	=&
-	3\mu\Psi_2
	,\\
	\edth\Psi_2
	=&
	3\tau\Psi_2
	,
\end{align}
\end{subequations}
	we obtain
\begin{align}
\label{eq:def_B2}
	\mathcal{B}_2
	\equiv
	\left(
	-	3\mu h_{l\bar{m}} 
	- 	\frac{3}{2}\tau h_{\bar{m}\bar{m}}
	-	3\pi^{(1)}
	\right)\Psi_2
	\nonumber\\
-	\left(\edth^{\prime}+3\pi\right)\Psi_2^{(1)}
+	\left(D+2\epsilon-2\rho\right)\Psi_3^{(1)}
	=&
	0
	.	
\end{align}

	Another nontrivial test of
our computation is to check that $h_{ll}$ 
converges to a real function. The reason it is not
manifestly real in our code is because we factor
out definite harmonic angular
$\phi$ dependence from all variables via the complex function $e^{i m\phi}$.
It turns out
that inconsistent
initial data (as we have prior 
to $v=v_u$ in Fig.~\ref{fig:metric_reconstruction_strategy_pictureb}),
as well as truncation error, 
introduces an imaginary component to $h_{ll}$
after we reassemble it from the rescaled code variables.
Specifically then, we check the following residuals
\begin{subequations}
\label{eq:def_H}
\begin{align}
	\mathfrak{R}\mathcal{H}
	\equiv
	\mathcal{R}\left(h_{ll}^{[  m]}\right) 
-	\mathcal{R}\left(h_{ll}^{[- m]}\right) 
	=&
	0
	,\\
	\mathfrak{I}\mathcal{H}
	\equiv
	\mathcal{I}\left(h_{ll}^{[  m]}\right)
+	\mathcal{I}\left(h_{ll}^{[- m]}\right)
	=&
	0
	,
\end{align}
\end{subequations}
where the superscript $^{[m]}$ denotes the corresponding
variable excluding an $e^{i m\phi}$ piece (see Eq.~\eqref{eq:decomposition_in_phi}).

Finally, we have also tested our Teukolsky solver 
by evolving initial data with several different
azimuthal numbers $m$, and various black hole spins, and 
confirmed that the late time quasinormal mode 
decay (before power law decay sets in) at null infinity 
is consistent, to within estimated truncation error,
with known parameters of the dominant
$l=m$ mode\footnote{We take these quasinormal ringdown frequencies from
\cite{2009CQGra..26p3001B}, who computed them using Leaver's method.}.

We have not implemented an independent residual check for our source
term $\mathcal{S}$ in Eq.~\eqref{eq:schematic_source}. We are not
aware of, and have not been able to devise, a method that can do so without
knowledge of the full second order metric. In the future
we plan to check the result with a full numerical relativity code,
though that will require some non-trivial work in providing initial
data for the latter consistent to second order with our perturbative solution.
\section{Numerical results}
In this section we present two example scenarios,
first for a perturbation of a Kerr black hole with spin $a=0.7$,
then for one with spin $a=0.998$. In both cases we choose $m=2$
for the first order perturbations' azimuthal dependence,
and show the $m=0$ and $m=4$ second order $\Psi_4^{(2)}$
this produces.
\subsection{Example evolution with black hole spin $a=0.7$}
\label{sec:example_evolution_a0.7}
	Here we consider a perturbation of a black hole with
spin $a=0.7$, which is close to the value found after the merger of two
initially slowly-spinning, near equal mass black holes
(see e.g. \cite{Centrella:2010mx}).
The simulation parameters are listed
in Table.~\ref{table:sim_params_a0.7}.

\begin{table}
\centering
\begin{tabular}{ c | c }
\hline
mass & $0.5$ \\
spin & $0.35$ ($a=0.7$) \\
low  resolution & $N_x=160$, $N_l=28$ \\
med  resolution & $N_x=176$, $N_l=32$ \\
high resolution & $N_x=192$, $N_l=36$ \\
$T_w$ & $2\times T_{mw}\approx17.6M$ \\
$m$   & $2$ \\
$l_0$ & $2$\\
$a_0$ & $0.1$\\
$r_l$ & $1.1\times r_H$\\
$r_u$ & $2.5\times r_H$\\
\hline
\end{tabular}
\caption{Parameters for spin $a=0.7$ black hole evolution
(unless stated otherwise in the figure captions).
$T_w$ is the ``wait'' time before starting the evolution
of $\Psi_4^{(2)}$, which we choose to be
twice the ``minimum'' wait time $T_{mw}$ for the initial
data we choose; see Sec.~\eqref{sec:compact_initial_data}.}
\label{table:sim_params_a0.7}
\end{table}

\begin{figure*}
  \subfloat[$\mathcal{R}\psi_4^{(1)[2]}$, $\mathcal{R}\psi_4^{(2)[4]}$]{{\includegraphics[width=0.5\textwidth]{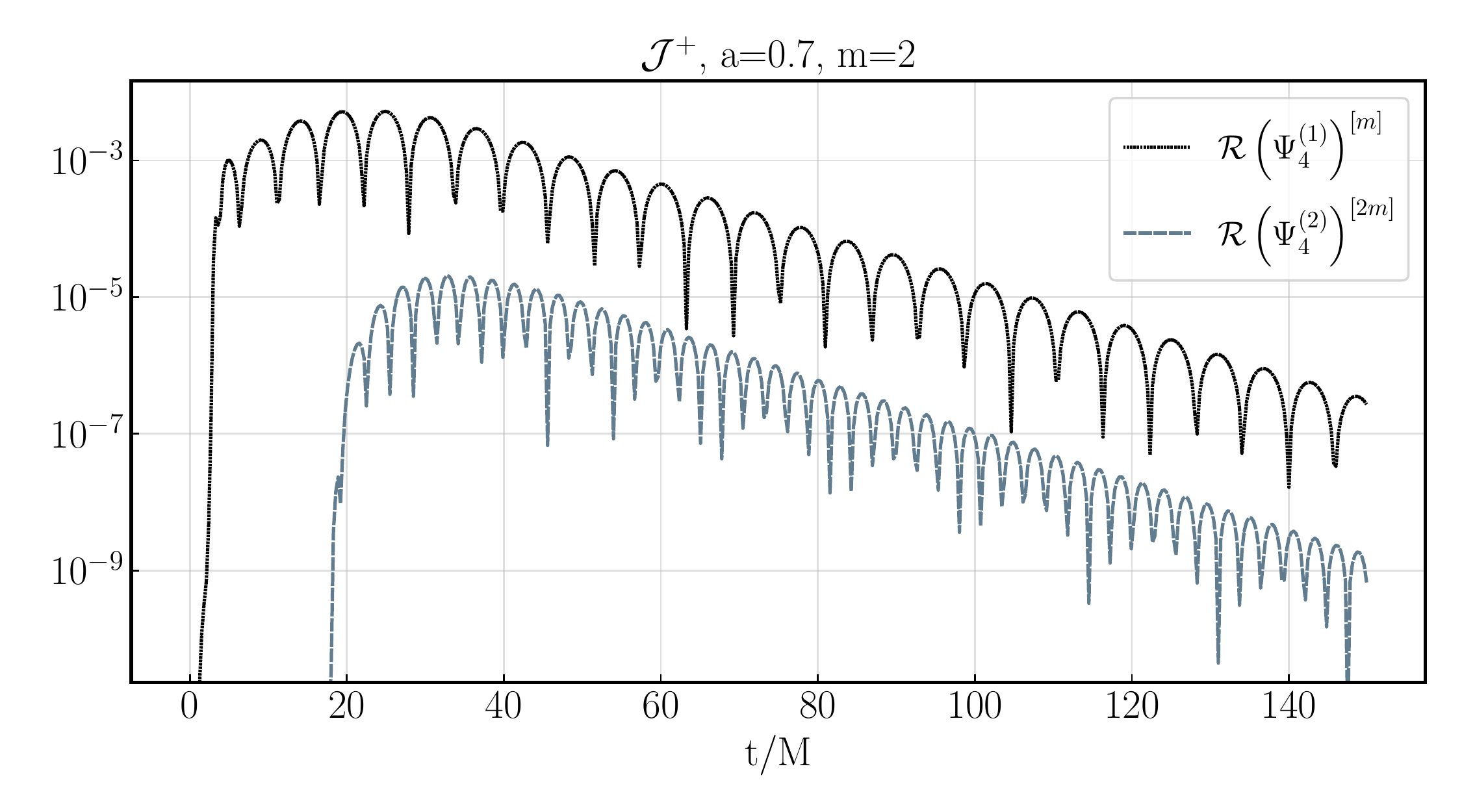}}}%
  \hfill
  \centering
  \subfloat[$\mathcal{I}\psi_4^{(1)[2]}$, $\mathcal{I}\psi_4^{(2)[4]}$]{{\includegraphics[width=0.5\textwidth]{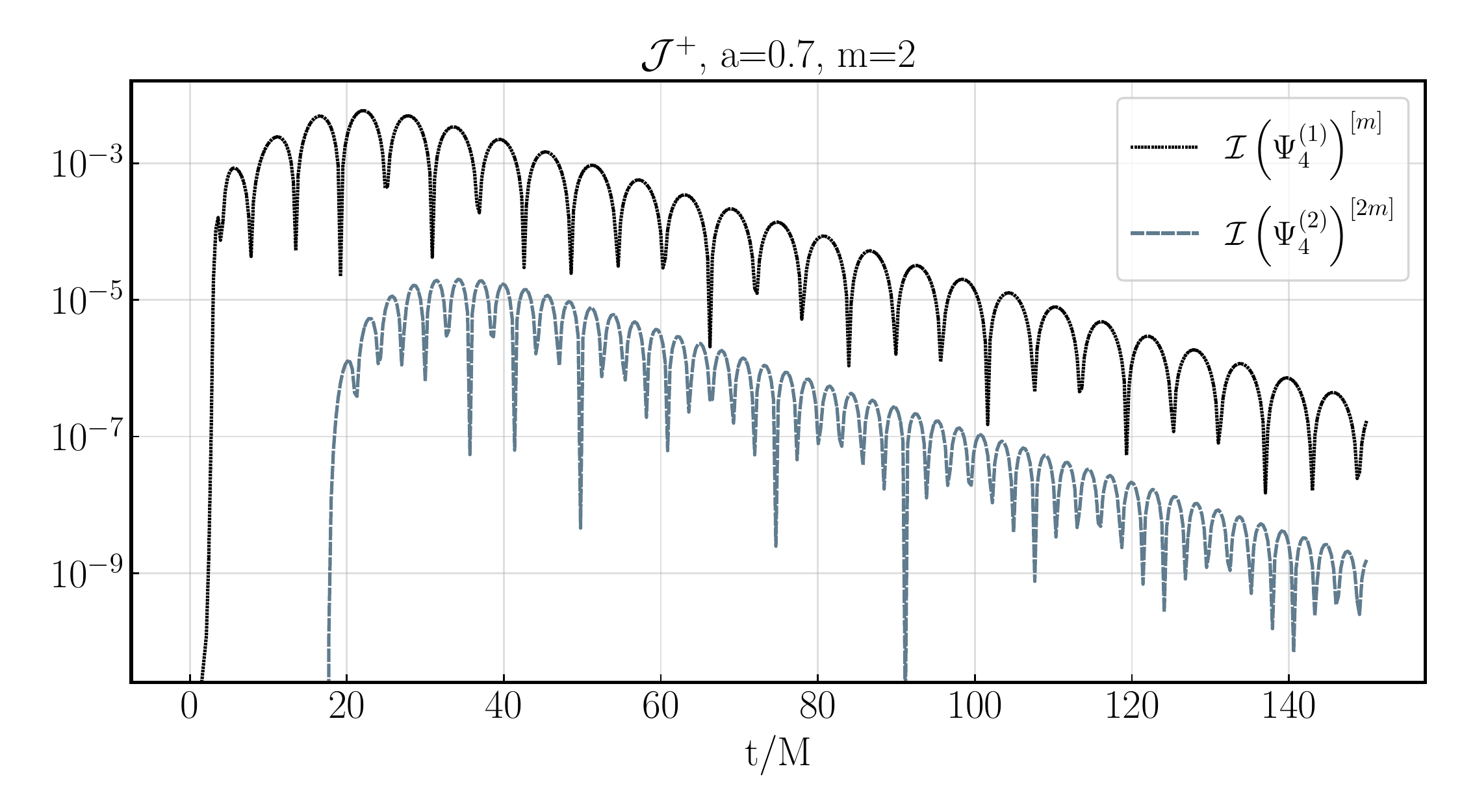}}}%
  \hfill
  \subfloat[$\mathcal{R}\psi_4^{(1)[2]}$, $\mathcal{R}\psi_4^{(2)[0]}$]{{\includegraphics[width=0.5\textwidth]{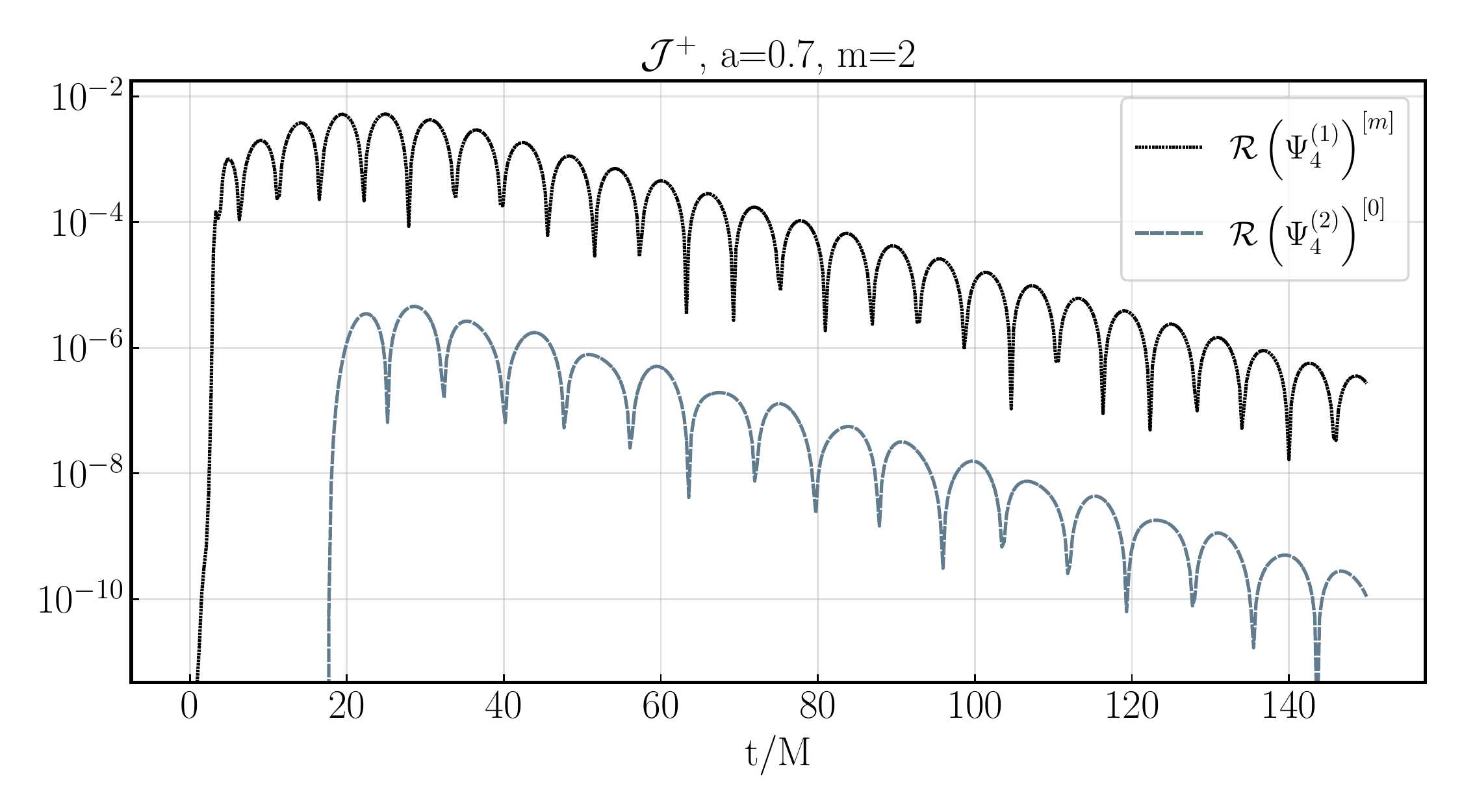}}}%
  \hfill
  \centering
  \subfloat[$\mathcal{I}\psi_4^{(1)[2]}$, $\mathcal{I}\psi_4^{(2)[0]}$]{{\includegraphics[width=0.5\textwidth]{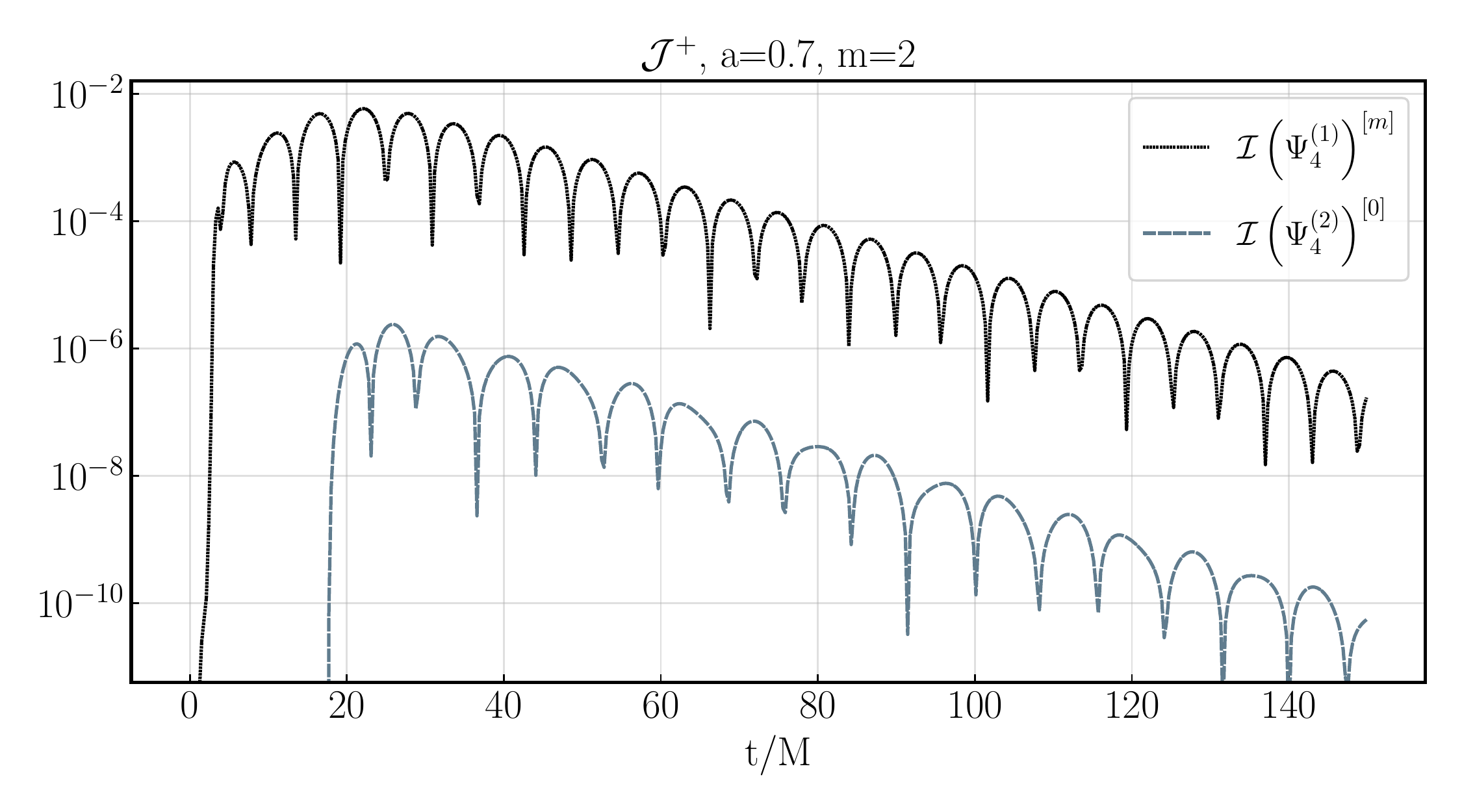}}}%
  \hfill
\caption{
Behavior of the real (left) and
imaginary (right) parts of $r\times \Psi_4^{(1),[m]}$ (here $m=2$)
at future null infinity ($\mathcal{J}^+$), compared with
$r\times\Psi_4^{(2),[2m]}$ (top) and $r\times\Psi_4^{(2),[0]}$ (bottom),
for the $a=0.7$ case (see Table.~\ref{table:sim_params_a0.7} for simulation parameters).
For reference we show the same $\Psi_4^{(1)}$ data in the top and bottom
panel for each case, though notice the different vertical scales.
$\Psi_4^{(1)}$ is initially zero as the data is compactly
supported near the origin, and $\Psi_4^{(2)}$ is zero until
we begin its evolution at $T_w=17.6M$; see Fig.~\ref{fig:a07_m2_startStudy}
for results with this turn-on time delayed to $2 T_w$ and $3 T_w$.
The data is from the `high' resolution run, and 
the truncation error estimate for all these functions remains
$\lesssim 1\%$ throughout the evolution.
}
\label{fig:spin_0.7_psi4}
\end{figure*}

\begin{figure*}
  \centering
  \subfloat[$\mathcal{R}\psi_4^{(1)[2]}$, $\mathcal{R}\mathcal{S}^{[4]}$]{{\includegraphics[width=0.5\textwidth]{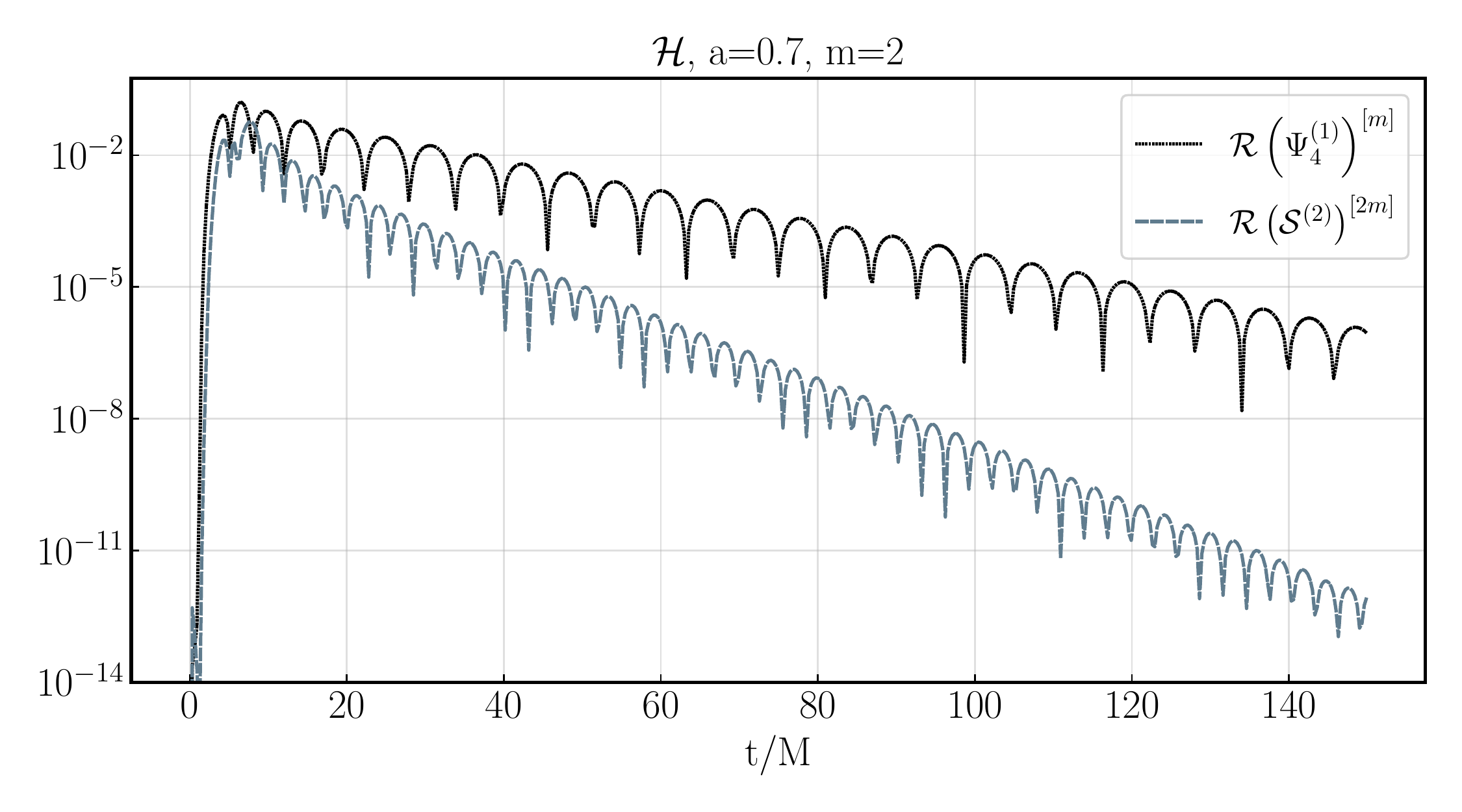}}}%
  \hfill
  \centering
  \subfloat[$\mathcal{I}\psi_4^{(1)[2]}$, $\mathcal{I}\mathcal{S}^{[4]}$]{{\includegraphics[width=0.5\textwidth]{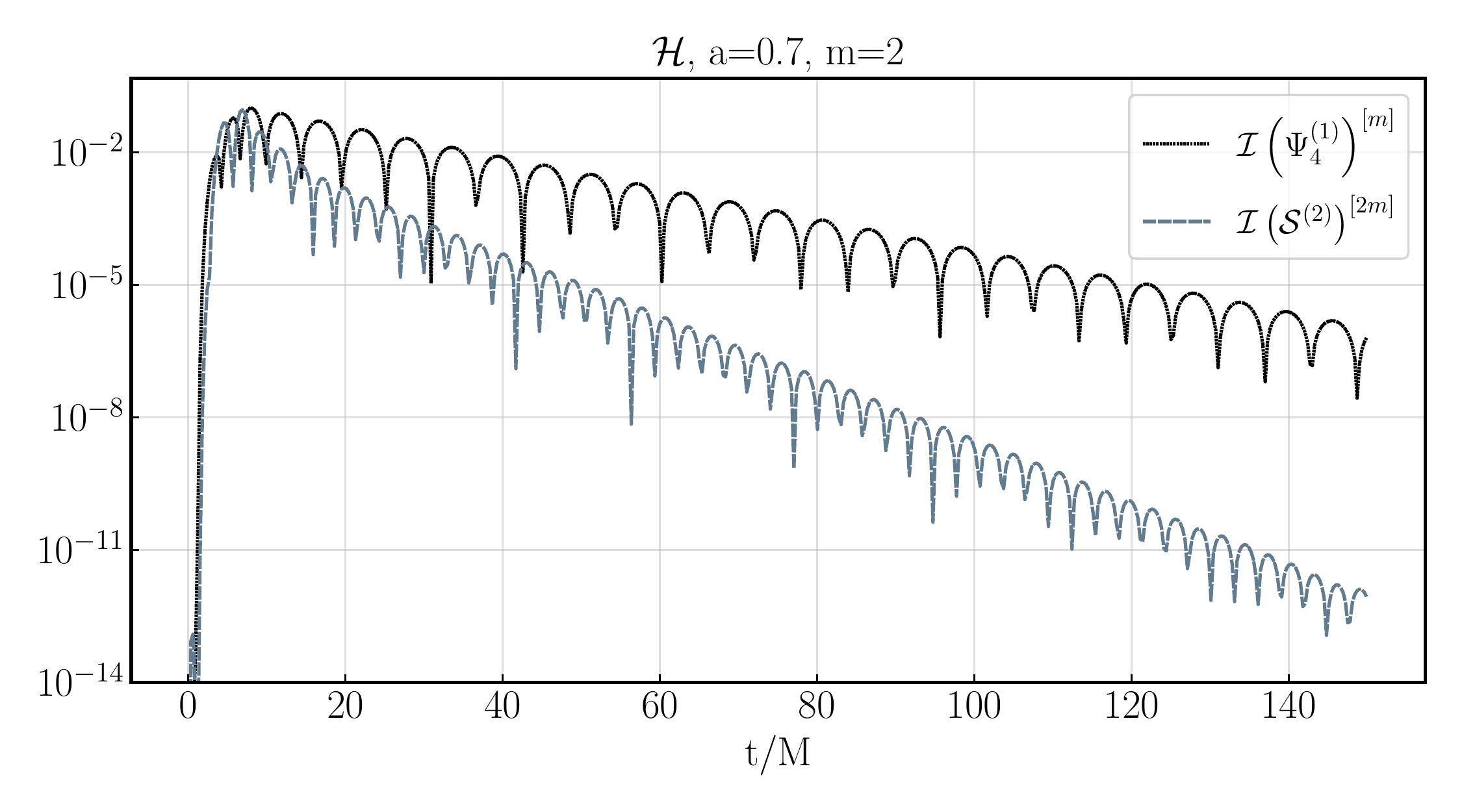}}}%
  \hfill
  \centering
  \subfloat[$\mathcal{R}\psi_4^{(1)[2]}$, $\mathcal{R}\mathcal{S}^{[0]}$]{{\includegraphics[width=0.5\textwidth]{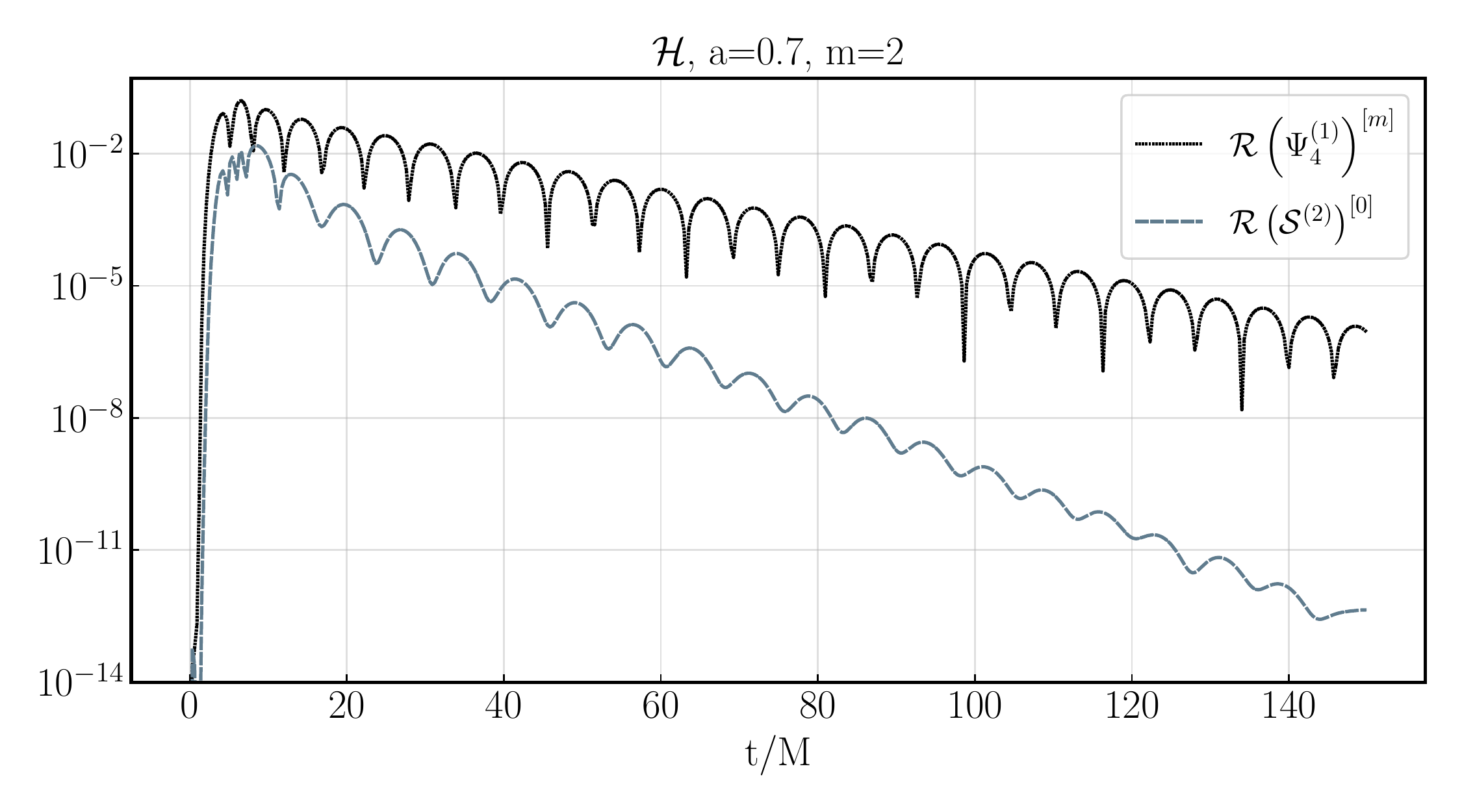}}}%
  \hfill
  \centering
  \subfloat[$\mathcal{I}\psi_4^{(1)[2]}$, $\mathcal{I}\mathcal{S}^{[0]}$]{{\includegraphics[width=0.5\textwidth]{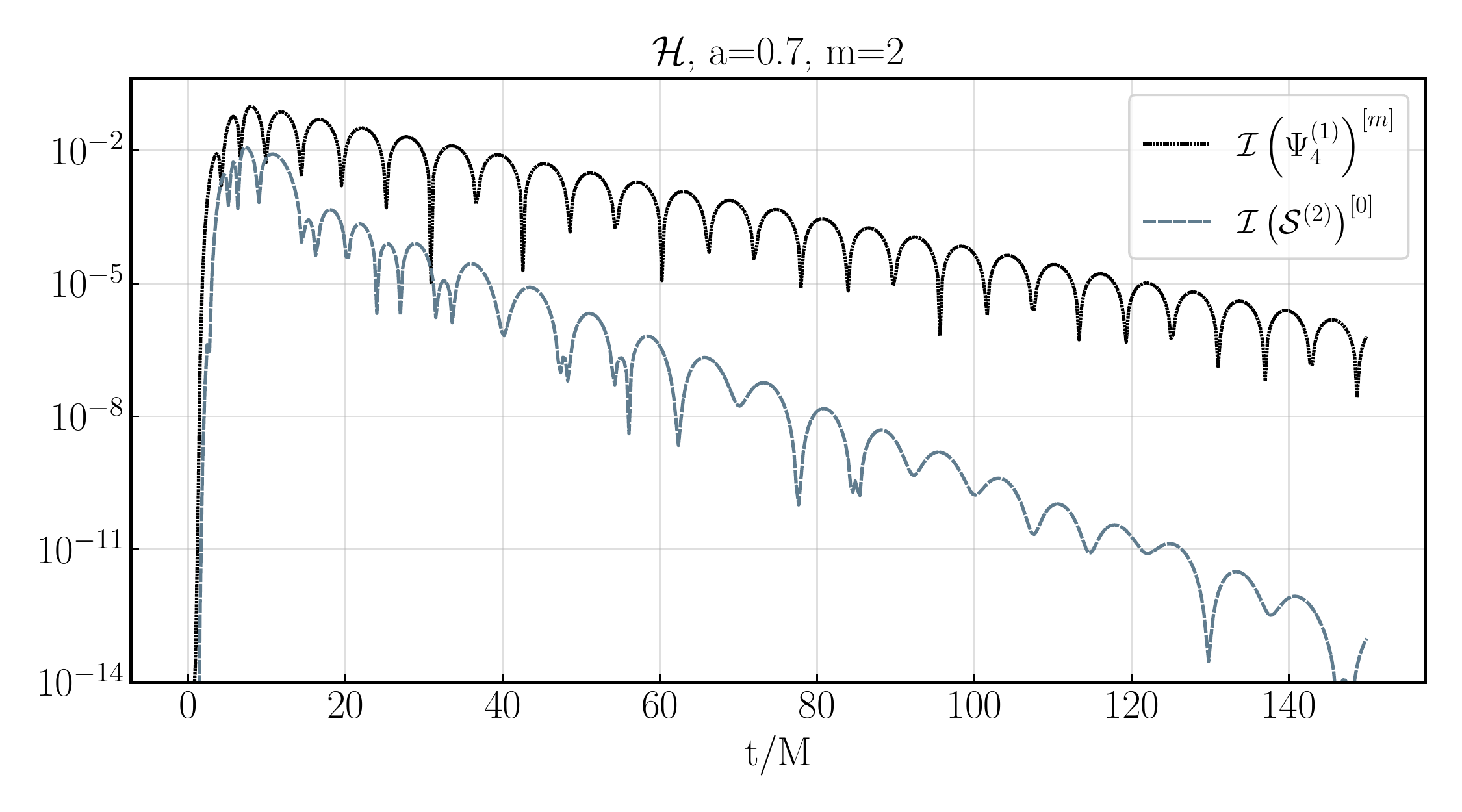}}}%
  \hfill
\caption{Comparison of the magnitude of the real (left) and imaginary (right) components of $\Psi_4^{(1)}$ with the 
corresponding components of the second order source terms
for $\mathcal{S}^{(2),[4]}$ (top) and $\mathcal{S}^{(2),[0]}$ (bottom),
at the black hole horizon, for the $a=0.7$ case (Table ~\ref{table:sim_params_a0.7}).
For reference we show the same $\Psi_4^{(1)}$ data in the top and bottom
panel for each case.
}
\label{fig:a07_m2_horizon_psi4_source}
\end{figure*}

\begin{figure*}
  \centering
  \subfloat[$\mathcal{R}\mathcal{S}^{[4]}$]{{\includegraphics[width=0.5\textwidth]{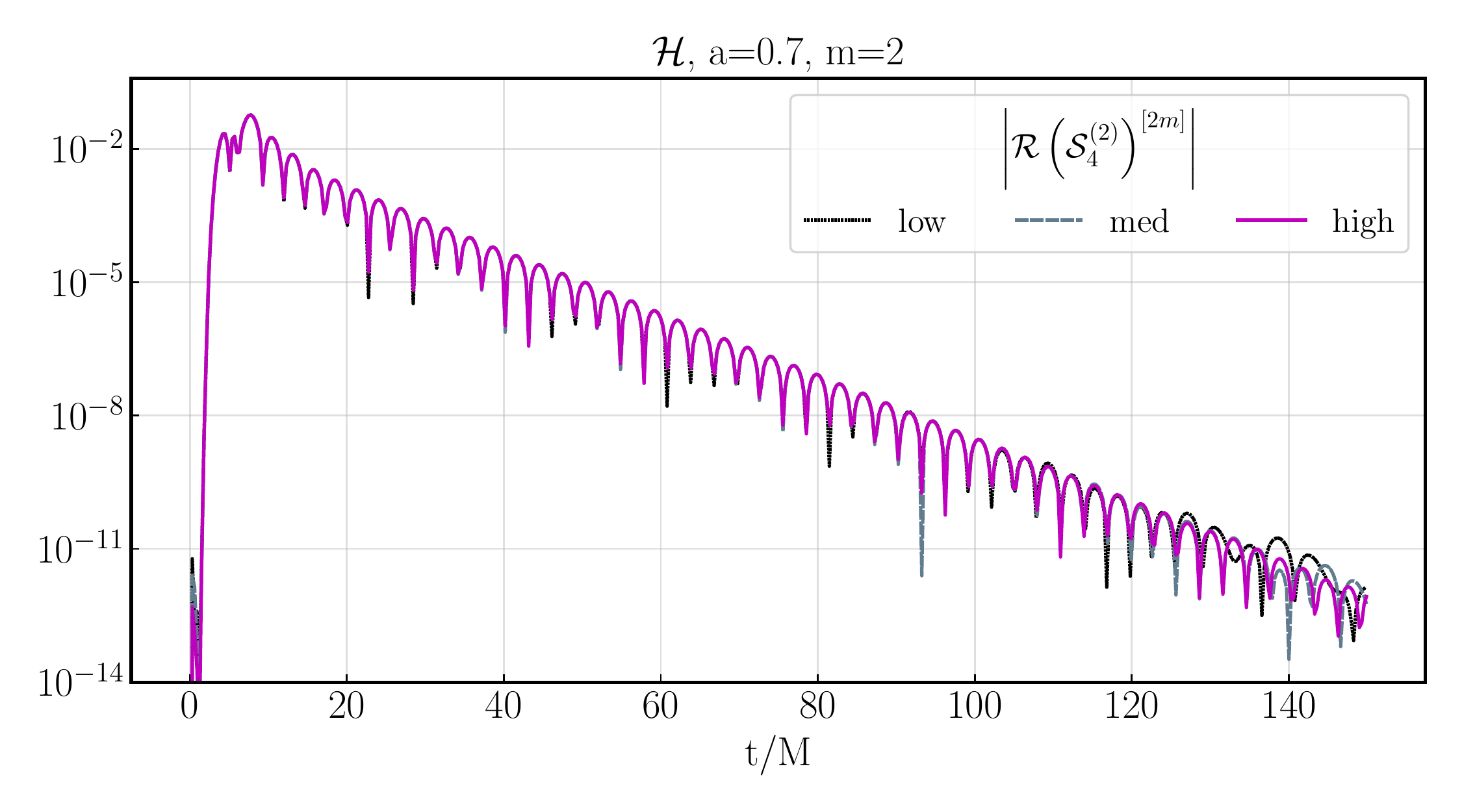}}}%
  \hfill
  \centering
  \subfloat[$\mathcal{I}\mathcal{S}^{[4]}$]{{\includegraphics[width=0.5\textwidth]{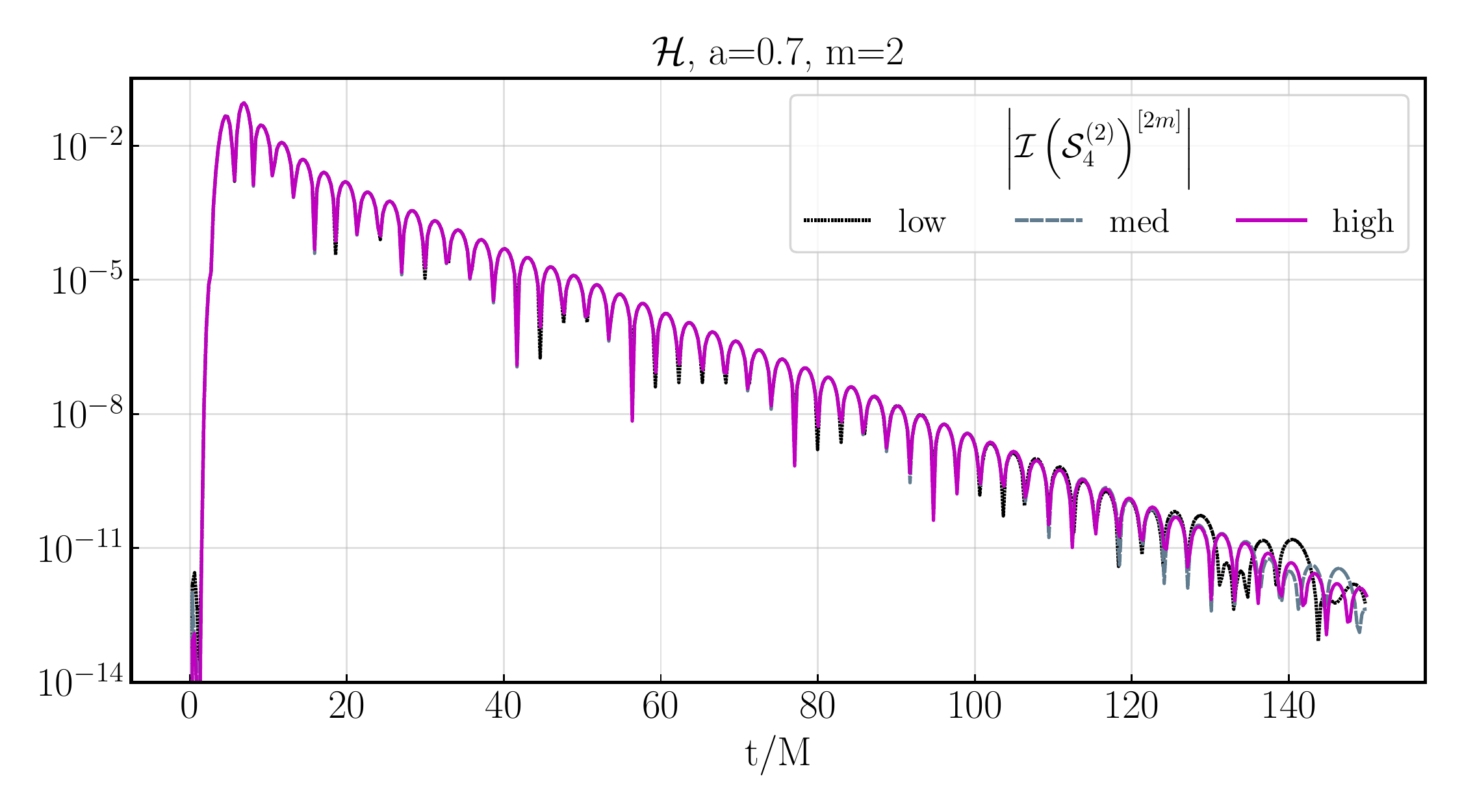}}}%
  \hfill
  \centering
  \subfloat[$\mathcal{R}\mathcal{S}^{[0]}$]{{\includegraphics[width=0.5\textwidth]{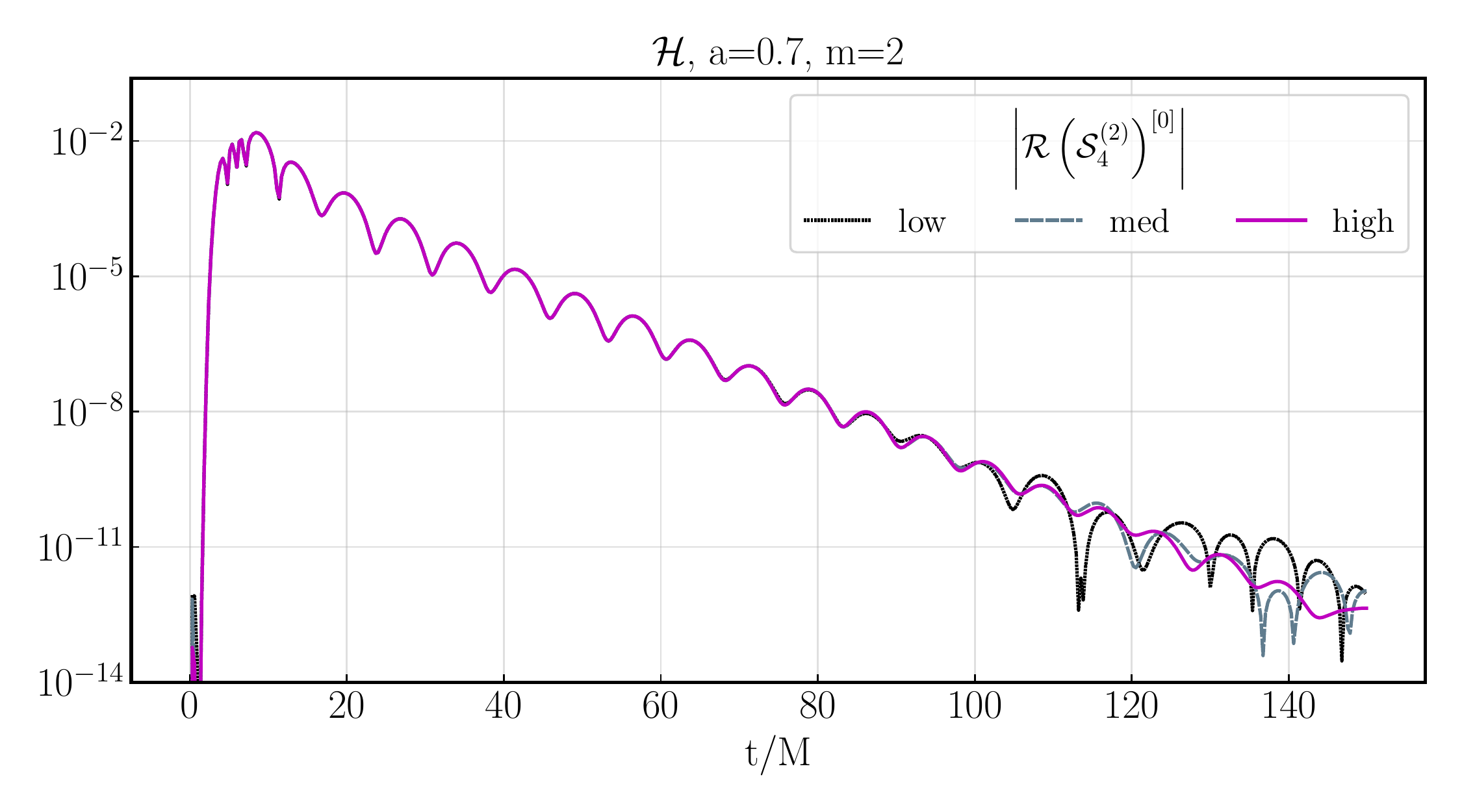}}}%
  \hfill
  \centering
  \subfloat[$\mathcal{I}\mathcal{S}^{[0]}$]{{\includegraphics[width=0.5\textwidth]{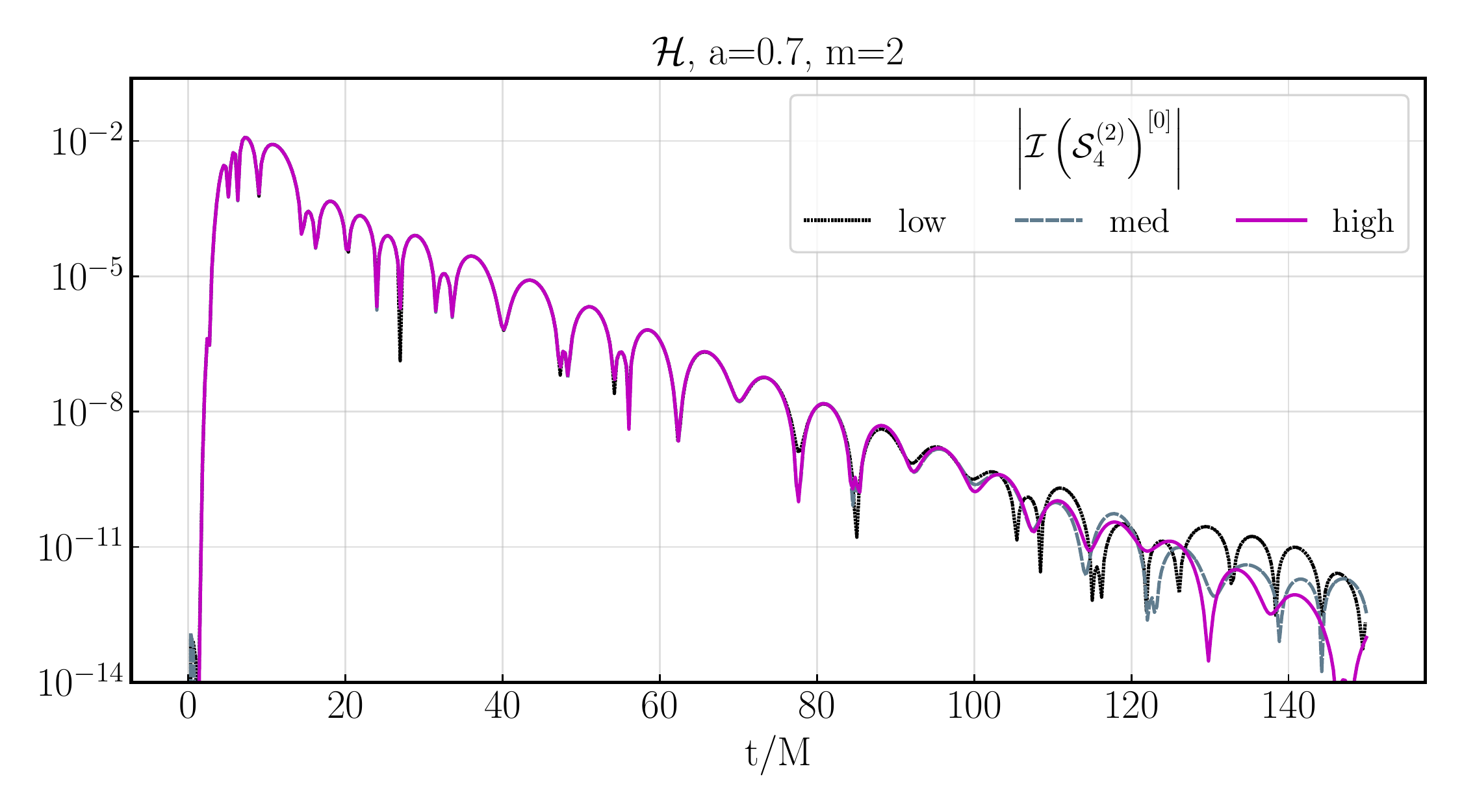}}}%
  \hfill
\caption{A resolution study of
the real (left) and imaginary (right) parts of the second order source terms
$\mathcal{S}^{(2)[4]}$ (top) and $\mathcal{S}^{(2)[0]}$ (bottom) at the
black hole horizon, for the $a=0.7$ case (Table ~\ref{table:sim_params_a0.7}).
This demonstrates that we are resolving 
the source terms until relatively late times ($t/M\sim120$ at
high resolution).
}
\label{fig:a07_m2_horizon_source2_source0_resolution}
\end{figure*}

\begin{figure*}
  \centering
  \subfloat[$\left|\mathcal{B}_2\right|_2$]{{\includegraphics[width=0.5\textwidth]{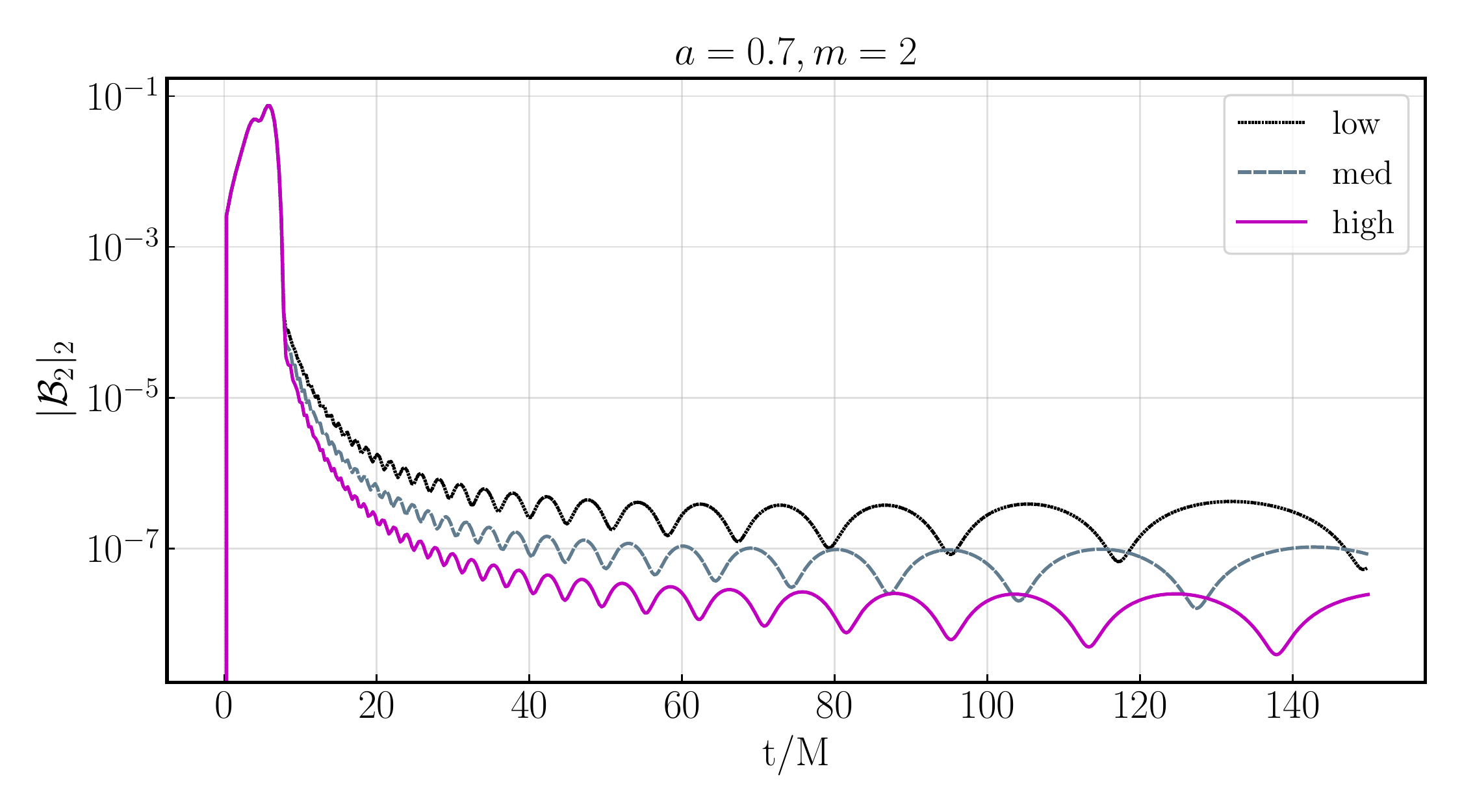}}}%
  \subfloat[$\left|\mathcal{B}_3\right|_2$]{{\includegraphics[width=0.5\textwidth]{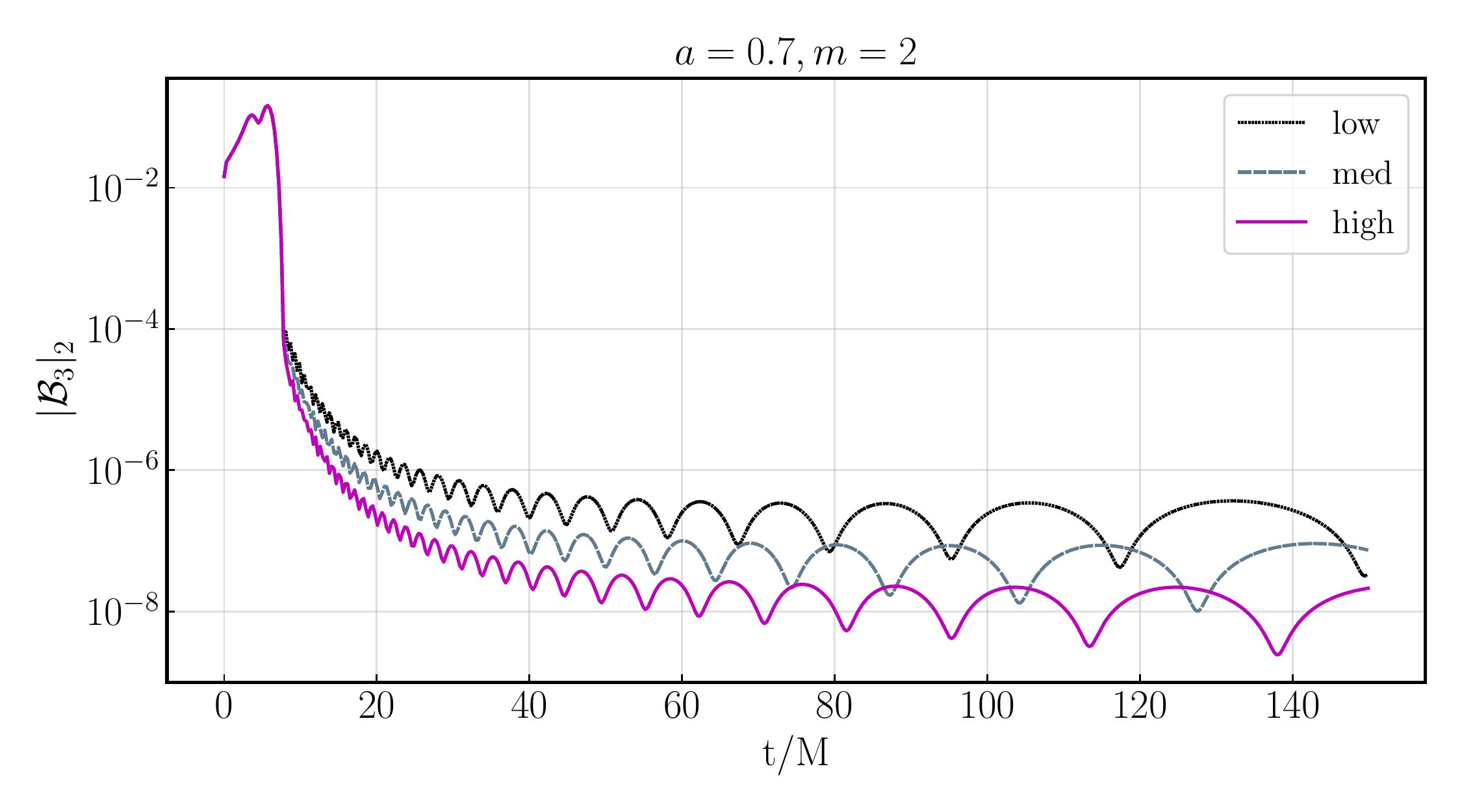}}}%
  \hfill
  \centering
  \subfloat[$\left|\mathcal{H}\right|_2$]{{\includegraphics[width=0.5\textwidth]{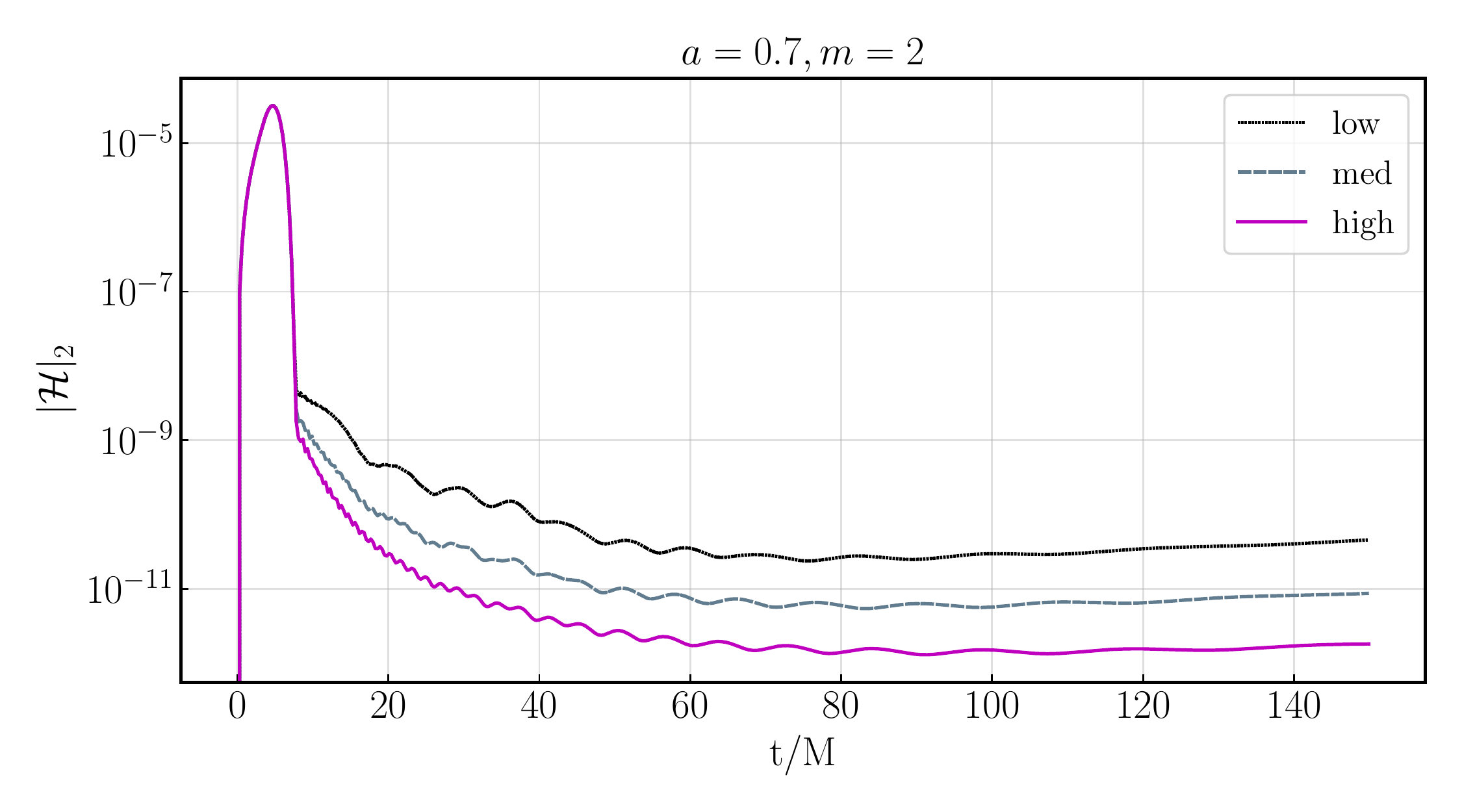}}}%
  \hfill
\caption{
The discrete two norm (see Eq.~\eqref{eq:def_two_norm})
of independent residuals $\mathcal{B}_3$, $\mathcal{B}_2$,
and $\mathcal{H}$ for metric reconstruction 
(see respectively Eq.~\eqref{eq:def_B3}, Eq.~\eqref{eq:def_B2},
and Eq.~\eqref{eq:def_H}), for the spin $a=0.7$ case,
as a function of time for three different resolutions
(Table. \ref{table:sim_params_a0.7}).
We only begin to obtain convergence to zero once the region
with inconsistent initial data has left our computational domain (around $t/M\sim10$).
}
\label{fig:convergence_indep_res_a07}
\end{figure*}

	In Fig.~\ref{fig:spin_0.7_psi4}
we plot the absolute value of the real and imaginary
parts of $\Psi_4^{(1),[m]}$, along with
$\Psi_4^{(2),[2m]}$ and $\Psi_4^{(2),[0]}$, measured at future null infinity.
The time offset between the start of the first
and second order components of the waveform is
due to the delayed integration start time $T_w$ of the latter compared
to the former; $T=T_w$ is twice the earliest time we can begin the second
order evolution with a consistent source term. 
   In Fig.~\ref{fig:a07_m2_horizon_psi4_source}
we plot the absolute value of the real and imaginary
parts of $\Psi_4^{(1)}$ and $\mathcal{S}^{(2)}$ on the black
hole horizon; Fig.~\ref{fig:a07_m2_horizon_source2_source0_resolution}
shows a resolution study of the latter. The region near the horizon is where the
source term is most significant (it decays faster than $1/r$ 
going to null infinity), and as expected 
$\mathcal{S}^{(2)}\sim \left(\Psi_4^{(1)}\right)^2$ there.
In Fig.~\ref{fig:convergence_indep_res_a07}
we plot norms of the metric reconstruction independent residuals 
discussed in Sec.\ref{sec_ires}, at three different resolutions.
After the constraint violating portions have left the domain,
we find roughly exponential convergence to zero,
in agreement with what one
would expect from a pseudo-spectral code with a sufficiently small time
step so that the time integration truncation error is subdominant.


\begin{figure*}
  \centering
  \subfloat[$\mathcal{R}\psi_4^{(2)[4]}$]{{\includegraphics[width=0.5\textwidth]{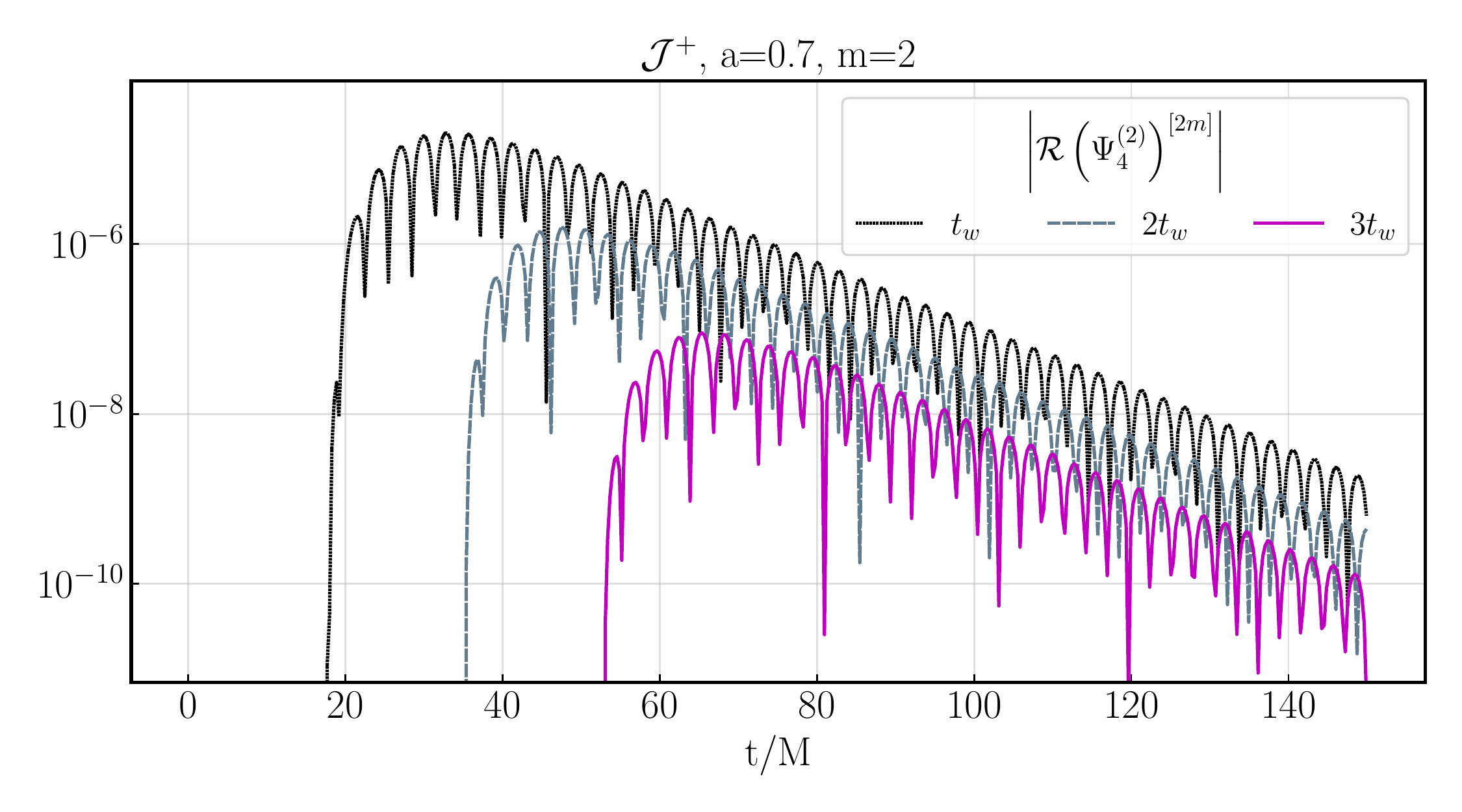}}}%
  \hfill
  \centering
  \subfloat[$\mathcal{I}\psi_4^{(2)[4]}$]{{\includegraphics[width=0.5\textwidth]{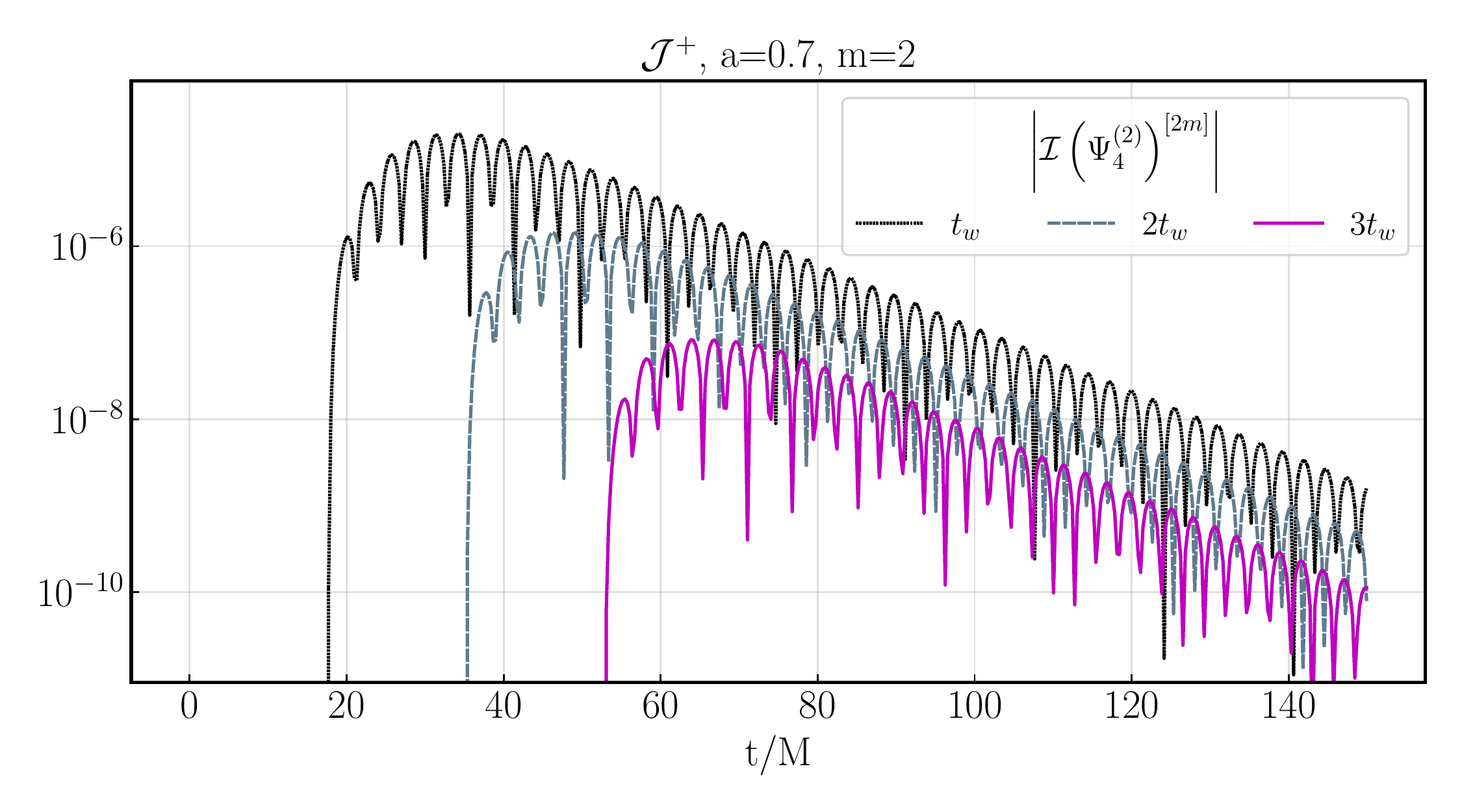}}}%
  \hfill
  \centering
  \subfloat[$\mathcal{R}\psi_4^{(2)[0]}$]{{\includegraphics[width=0.5\textwidth]{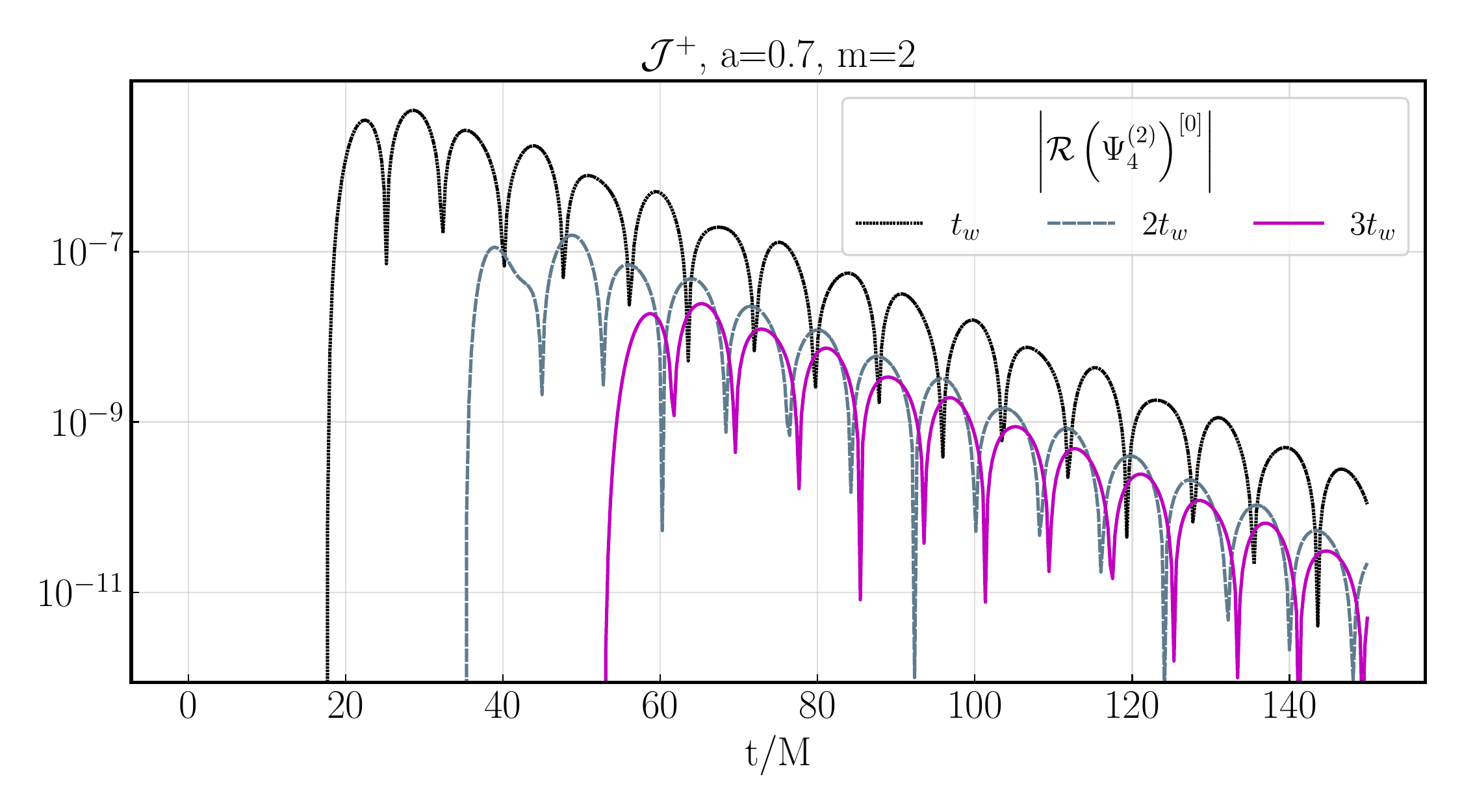}}}%
  \hfill
  \centering
  \subfloat[$\mathcal{I}\psi_4^{(2)[0]}$]{{\includegraphics[width=0.5\textwidth]{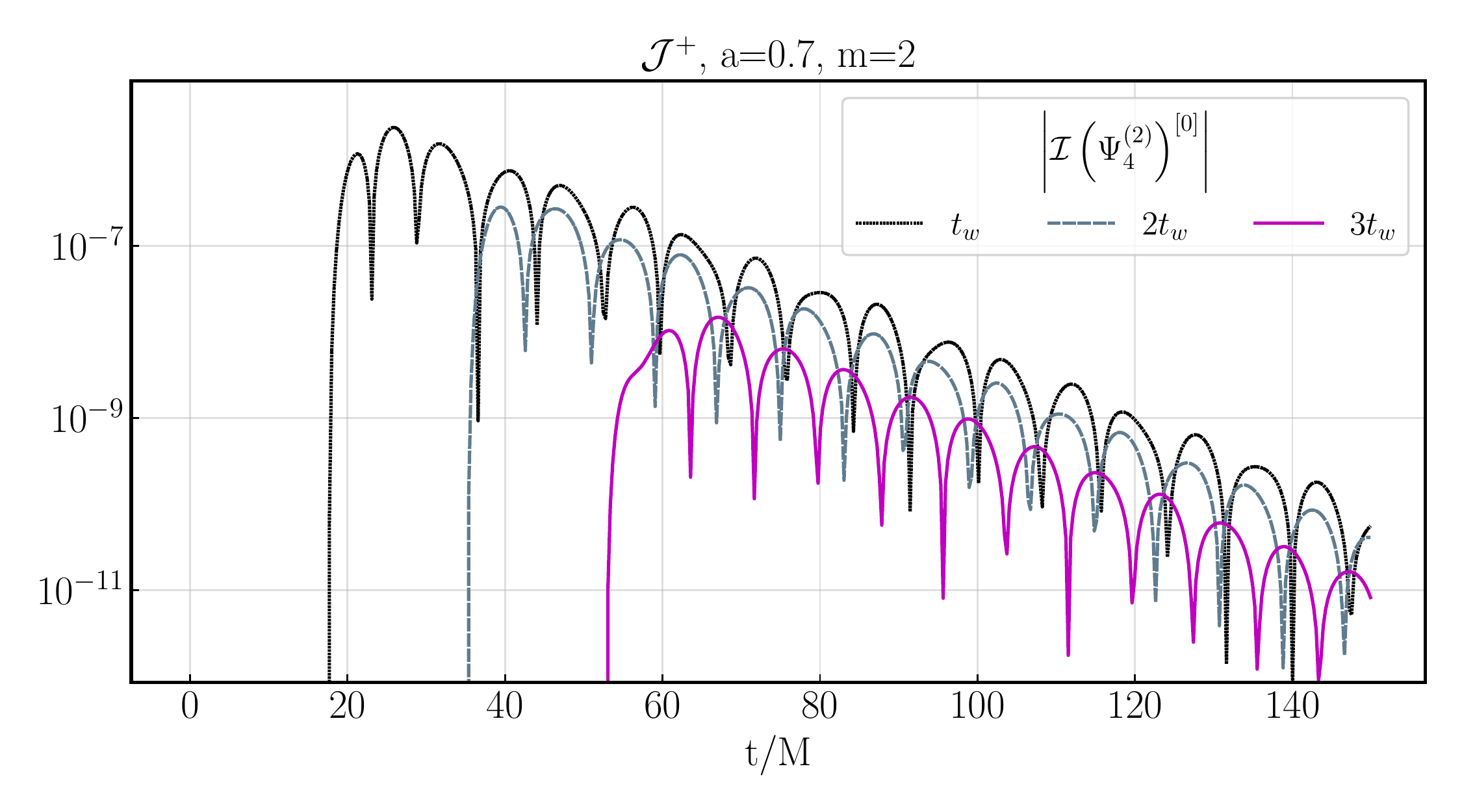}}}%
  \hfill
\caption{Comparison of the real (left) and imaginary (right)
components of the second order $\Psi_4^{(2),[4]}$ (top)
and $\Psi_4^{(2),[0]}$ (bottom) fields, from the same $a=0.7$ first
order perturbation depicted in Fig.~\ref{fig:spin_0.7_psi4}, as a function of when we begin
evolving the second order field. Three cases are shown, including for reference the $T_w$
case also shown in Fig.~\ref{fig:spin_0.7_psi4}.
}
\label{fig:a07_m2_startStudy}
\end{figure*}

\begin{figure*}
   \centering
   \subfloat[$\mathcal{F}\mathcal{R}$, window $(53M,150M)$]{{\includegraphics[width=0.5\textwidth]{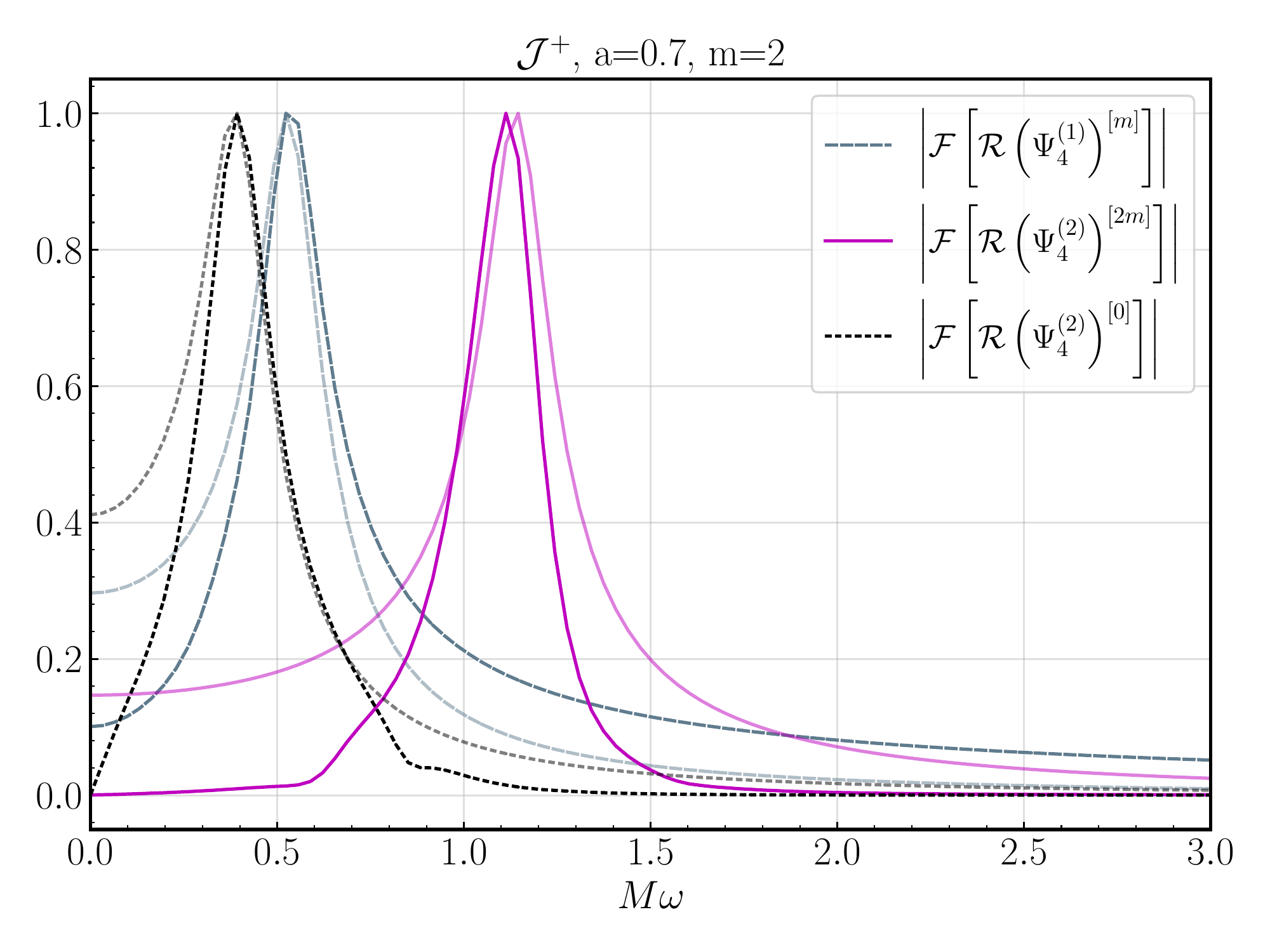}}}%
   \hfill
   \centering
   \subfloat[$\mathcal{F}\mathcal{I}$, window $(53M,150M)$]{{\includegraphics[width=0.5\textwidth]{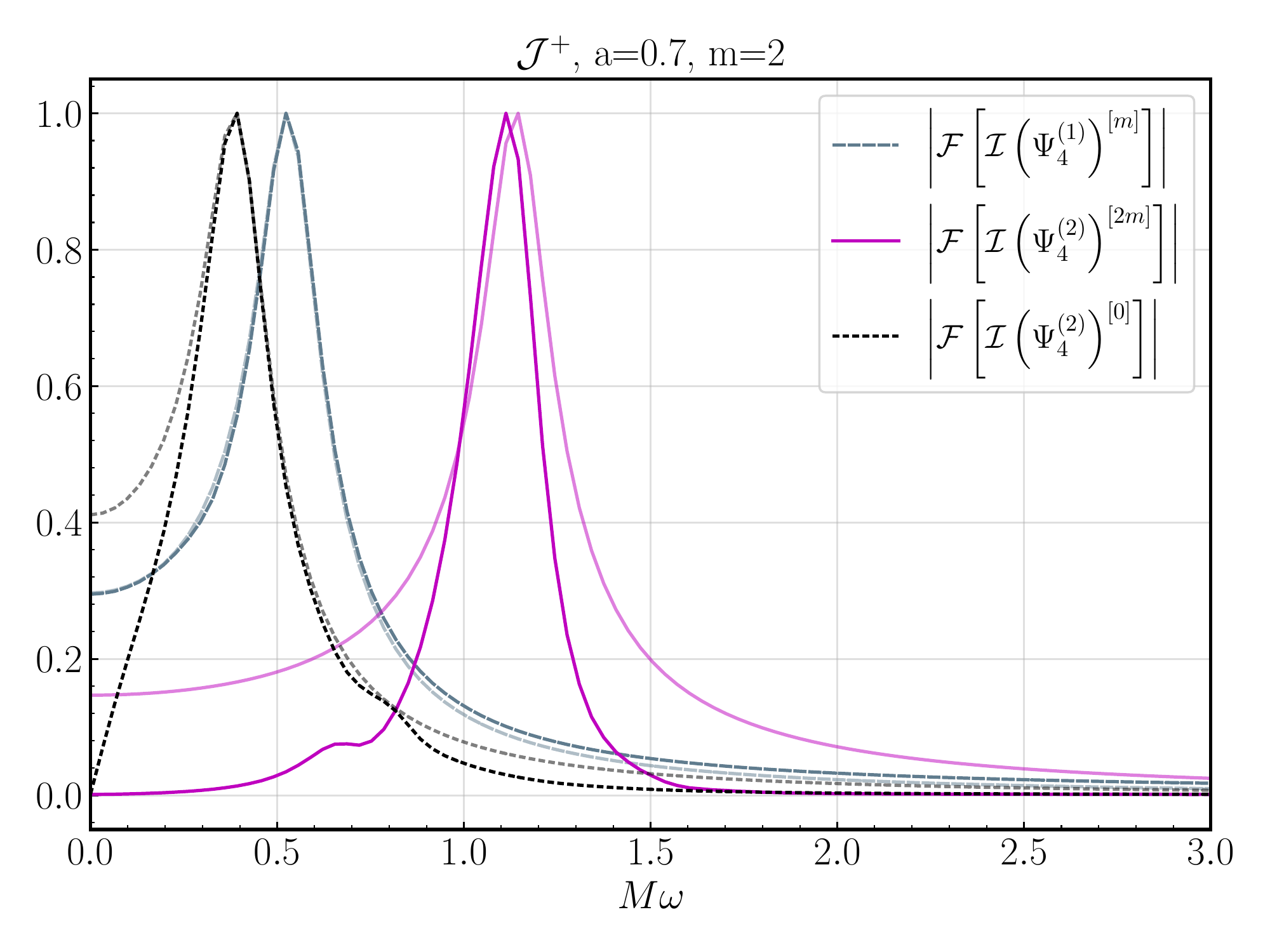}}}%
   \hfill
   \centering
   \subfloat[$\mathcal{F}\mathcal{R}$, window $(88M,150M)$]{{\includegraphics[width=0.5\textwidth]{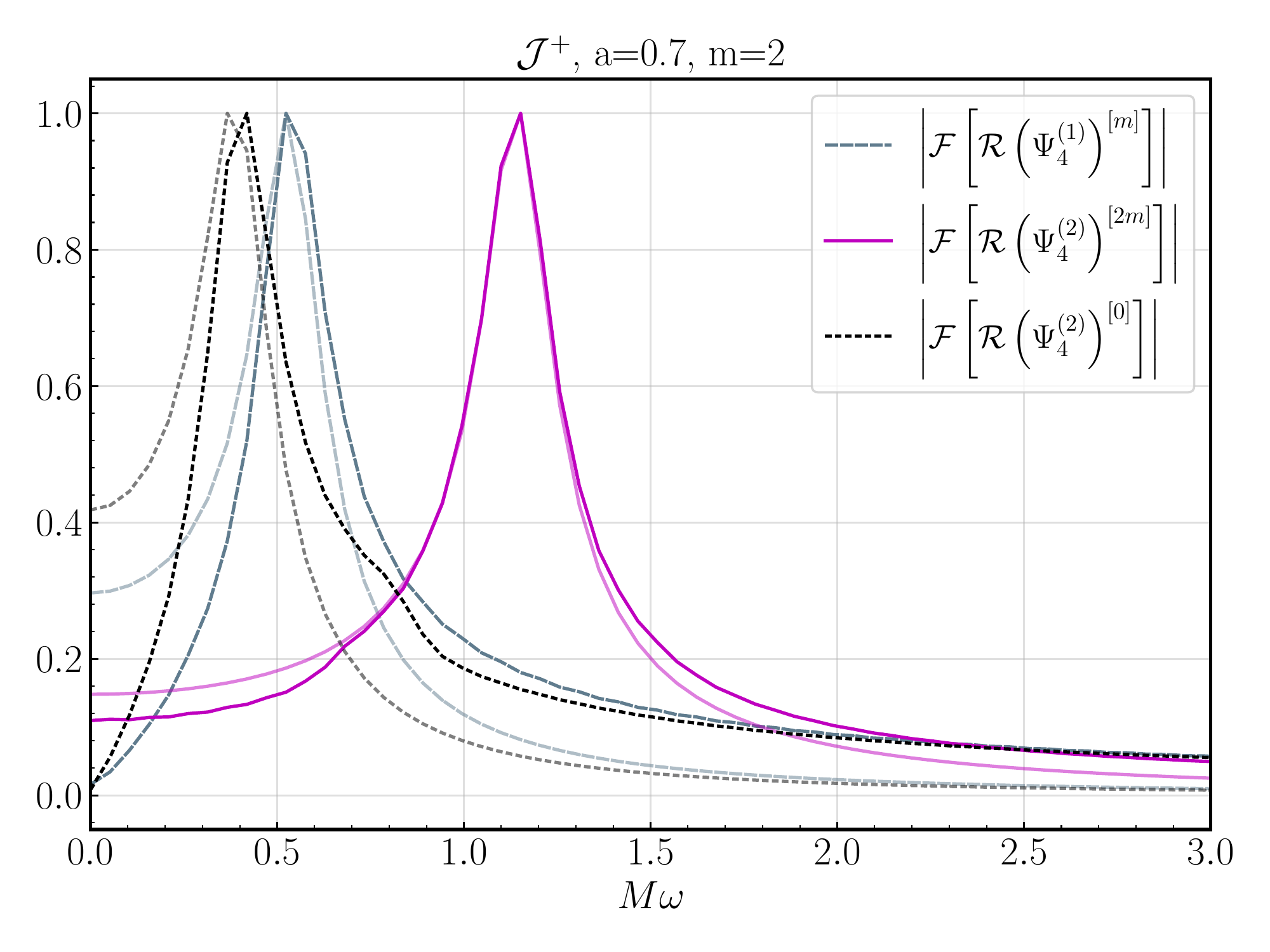}}}%
   \hfill
   \centering
   \subfloat[$\mathcal{F}\mathcal{I}$, window $(88M,150M)$]{{\includegraphics[width=0.5\textwidth]{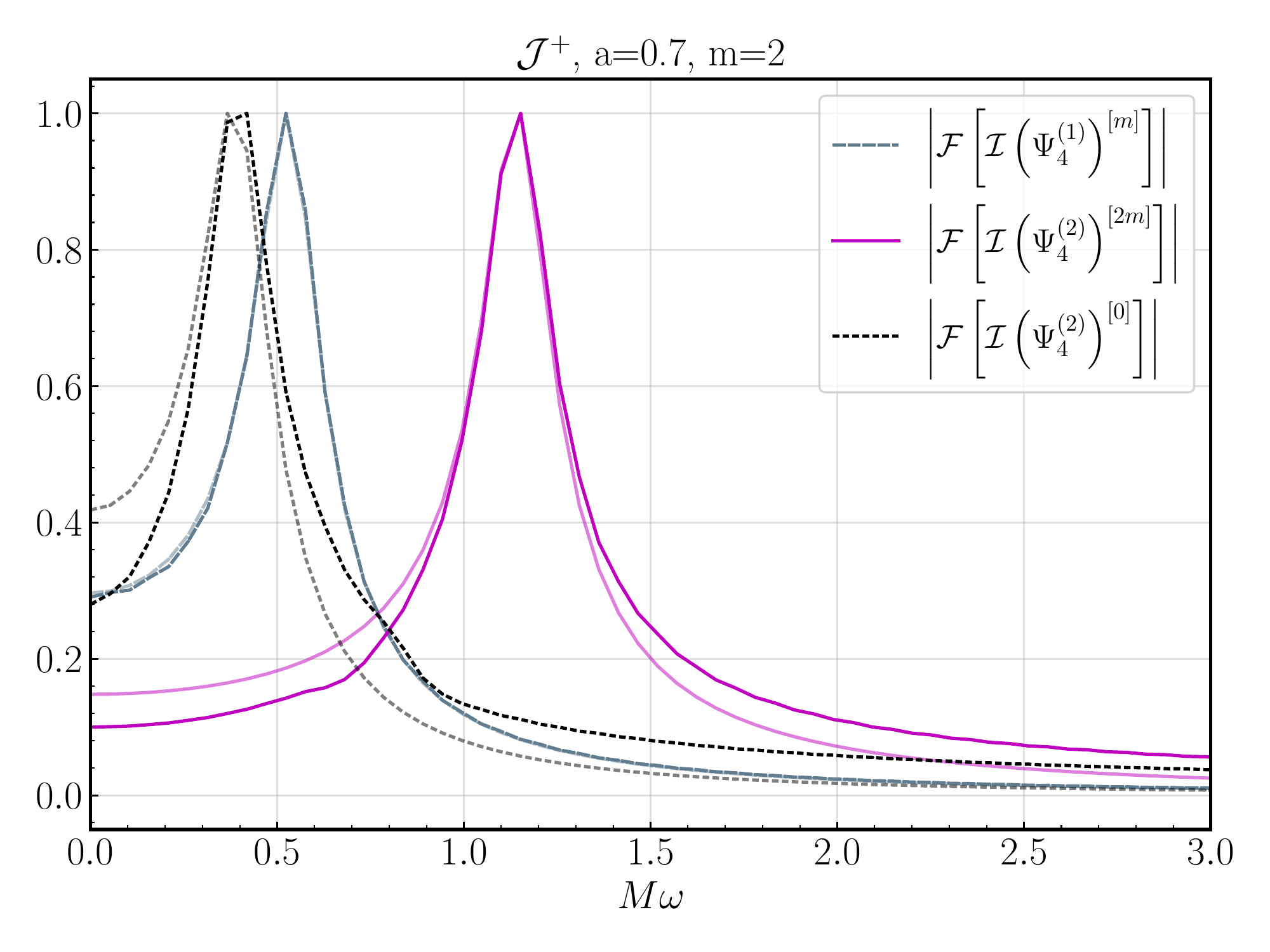}}}%
   \hfill
\caption{
Normalized absolute value of the Fourier Transform (\ref{norm_ft}) of
the real (left) and imaginary (right) parts of
$\Psi_4^{(1),[2]}$, $\Psi_4^{(2),[4]}$, and $\Psi_4^{(2),[0]}$,
taken over two different windows, for the $a=0.7$ case (see Table~\ref{table:sim_params_a0.7}).
The data for the second order
components come from the $3 T_w$ start time (see Fig.\ref{fig:a07_m2_startStudy}).
The window for the top panels is from $[3 T_w,150M]$, thus including the early
time behavior of the response, while for the bottom panels is $[5 T_w,150M]$ to focus
on the late time response. The darker plotted lines are from the numerical output,
while the lighter plotted lines are the Fourier
transform of $e^{-\omega_I t}\mathrm{sin}\left(\omega_Rt\right)$
with damping time $1/\omega_I$ and frequency $\omega_R$
of the $l=m$ (e.g. $l=2,m=2$) quasinormal mode for an $a=0.7$ spin black hole computed
via Leaver's method (taken from \cite{2009CQGra..26p3001B}).
}
\label{fig:spin_0pt7_FT_comparison}
\end{figure*}

Though this initial data is more to illustrate
our solution scheme and is not astrophysically accurate,
it is useful to begin to understand the non-linear response
when the black hole is excited by the fundamental $l=m=2$
quasinormal mode, in particular if we wait a sufficiently
long time for overtones present in the initial data
to decay\footnote{In a Kerr spacetime, setting initial
data (\ref{eq:functional_form_initial_data}) with a single $l_0$ mode of
the spin-weighted Legendre polynomials $_s P^m_{l_0}$ will
excite a spectrum of different
$l$ quasinormal modes measured at infinity, unless the black
hole spin $a=0$.}. Fig.~\ref{fig:spin_0.7_psi4}
suggests $T=T_w$ might not be early enough, as $\Psi_4^{(1)}$
has just started to enter its decaying phase. In Fig.~\ref{fig:a07_m2_startStudy}
then we show results for the second order modes with the
evolution begun at $2 T_w$ and $3 T_w$, in addition
to $T_w$ depicted in Fig.~\ref{fig:spin_0.7_psi4}.
The later start times show qualitatively similar behavior,
except the amplitude is lower by a factor close to the square
of the decay in the amplitude of $\Psi_4^{(1)}$ over the relevant
delay time. To help interpret the results further, in 
Fig.~\ref{fig:spin_0pt7_FT_comparison} we plot the normalized
absolute value of the Fourier transform of $\Psi_4$,
taken with two different windows: an earlier time window 
to capture the prompt second order response (but still
sufficiently past $T=0$ that the first order source is 
dominated by the single decaying quasinormal mode), and a later time window
to show the late time behavior once second order transient effects have decayed.
Also shown for reference are Fourier transforms of pure damped
sinusoids corresponding to the dominant fundamental quasinormal modes
expected for each $m$.

These plots illustrate a couple of interesting aspects of the second
order piece of a quasinormal mode perturbation of an $a=0.7$ spin Kerr black hole.
First, beginning with zero initial data for $\Psi_4^{(2)}$ on our
$T={\rm const.}$ slice, the response at future null infinity
builds up to a maximum over 1-2 local dynamical times scales,
before settling down to a quasinormal mode-like decay.
This is in part because where the source term is most significant is spread out
over a region a few Schwarzschild radii
about the horizon, and in part because of our prompt start of the second
order evolution.
Second, the source
term clearly excites the fundamental $m=0$ and $m=4$ quasinormal modes 
(i.e. solutions one would obtain from the {\em source-free} Teukolsky
equation), and these dominate the late time response due to their slower decay.
Or said another way, suppose 
the late time response was purely a driven mode, then (following the behavior
of the source term in Fig.~\ref{fig:a07_m2_horizon_psi4_source}) one would expect
the slope to be twice that of the first order mode, and 
the amplitude of the second order piece at a given late-time should not depend
on the start time, unlike what is shown in
Figs.~\ref{fig:spin_0.7_psi4} and~\ref{fig:a07_m2_startStudy}.\footnote{All this behavior
can qualitatively be captured by a driven, damped harmonic
oscillator model, $d^2 y(t)/dt^2 + \lambda\ dy(t)/dt + \omega^2\ y(t) = f(t)$,
where the source $f(t)$ is zero before being turned
on at time $t_0$. In addition to the driven
(particular solution) response to $f(t)$, demanding continuity in $y$
and $dy/dt$ at $t=t_0$ will generically require that
the fundamental modes (homogeneous solutions) of the oscillator are also 
excited then.} From the perspective of the Fourier transforms
in Fig.~\ref{fig:spin_0pt7_FT_comparison}, for $m=4$
the presence of this early time behavior
can be inferred by the narrower shape of the numerical data curve compared to 
that of the fundamental quasinormal mode: the driven and fundamental
modes have essentially the same frequency to within the resolution of the
Fourier transform here, and despite the more rapid decay of the former,
the initial growth phase (Fig.~\ref{fig:a07_m2_startStudy})
makes the transform of their sum look slightly closer to that of an
undamped sinusoid (a delta function).
The interpretation of the $m=0$ mode in this sense is less clear.

An implication of the above for ringdown
studies are (caveats about the physical accuracy of our
initial data aside):
if an $l=m=4$ component is searched for following a comparable
mass merger, given this mode's low amplitude relative
to the $l=m=2$ mode, in the first few cycles of ringdown non-linear
energy transfer from the $l=m=2$ to the $l=m=4$ mode could be
observable and should be accounted for. Furthermore,
once past this and in the linear regime, the amplitude
and phase of the $l=m=4$ mode that may be measured then
would differ from the linear evolution of what one could consider as the 
``initial'' amplitude and phase of this mode excited by merger. 
With proposals to coherently stack
multiple detected events to search for common subdominant modes
that rely on knowledge of predicted amplitudes and phases~\cite{Yang:2017zxs},
this implies non-linear effects need to be accounted for, 
either by incorporating them in the models, or using the ``final''
amplitudes and phases if only the linear portion of the waveforms
are included.

\subsection{Example evolution with black hole spin $a=0.998$}
\label{sec:example_evolution_a0.998}
	Here we show results
from a simulation of
the perturbation of a black hole with a spin near the
``Thorne limit'' \cite{1974ApJ...191..507T} $a\sim0.998$,
which is expected to be the maximum black hole spin that can 
be achieved within a class of thin-disk accretion models.
	Our simulation parameters are listed in
Table.~\ref{table:sim_params_a0.998}.
Note that the relevant dynamical timescale for
a near extremal black hole is $T_d\sim M/\sqrt{1-a}$.
For $a=0.998$, $T_d\sim22M$, so evolving for $T\sim150M$ corresponds
to $T\sim 7\times T_d$, a considerably
shorter time in terms of $T_d$ than for $a=0.7$.
Given that it is computationally intensive to evolve the $a=0.998$ case
for a comparable number of dynamical timescales with our present code,
we leave investigating late time effects to future work.

\begin{table}
\centering
\begin{tabular}{ c | c }
\hline
mass & $0.5$ \\
spin & $0.499$ ($a=0.998$) \\
low  resolution & $N_x=176$, $N_l=32$ \\
med  resolution & $N_x=192$, $N_l=36$ \\
high resolution & $N_x=208$, $N_l=40$ \\
$T_w$ & $2\times T_{mw}\approx13.6M$  \\
$m$   & $-2$\\
$l_0$ & $2$\\
$a_0$ & $0.1$\\
$r_u$ & $1.1\times r_H$\\
$r_l$ & $2.5\times r_H$\\
\hline
\end{tabular}
\caption{Parameters for spin $a=0.998$ black hole evolution
(unless stated otherwise in the figure captions).
$T_w$ is the ``wait'' time before starting the evolution
of $\Psi_4^{(2)}$, which we choose to be
twice the ``minimum'' wait time $T_{mw}$ for the initial
data we choose; see Sec.~\eqref{sec:compact_initial_data}.}
\label{table:sim_params_a0.998}
\end{table}

\begin{figure*}
  \subfloat[$\mathcal{R}\psi_4^{(1)[2]}$, $\mathcal{R}\psi_4^{(2)[4]}$]{{\includegraphics[width=0.5\textwidth]{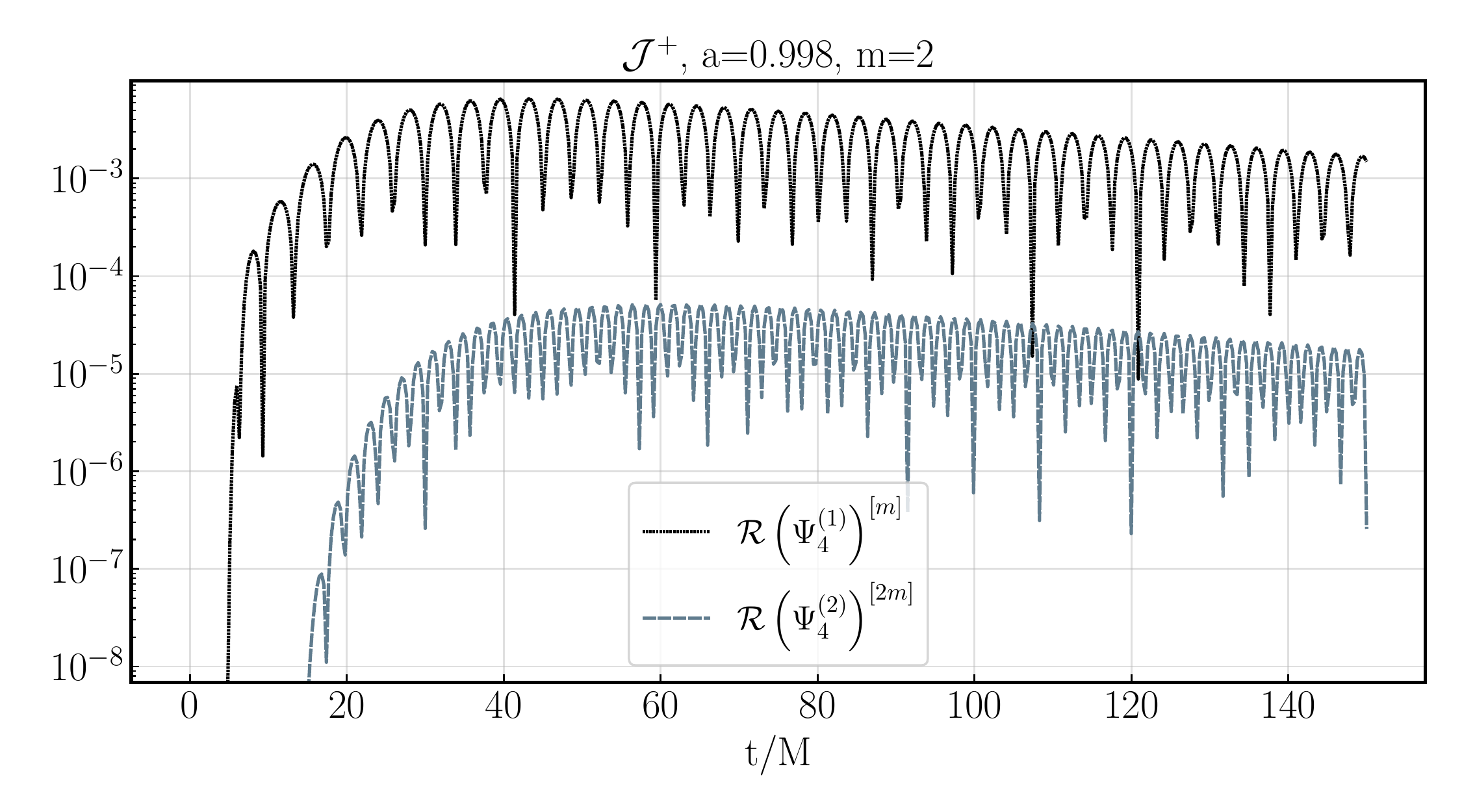}}}%
  \hfill
  \centering
  \subfloat[$\mathcal{I}\psi_4^{(1)[2]}$, $\mathcal{I}\psi_4^{(2)[4]}$]{{\includegraphics[width=0.5\textwidth]{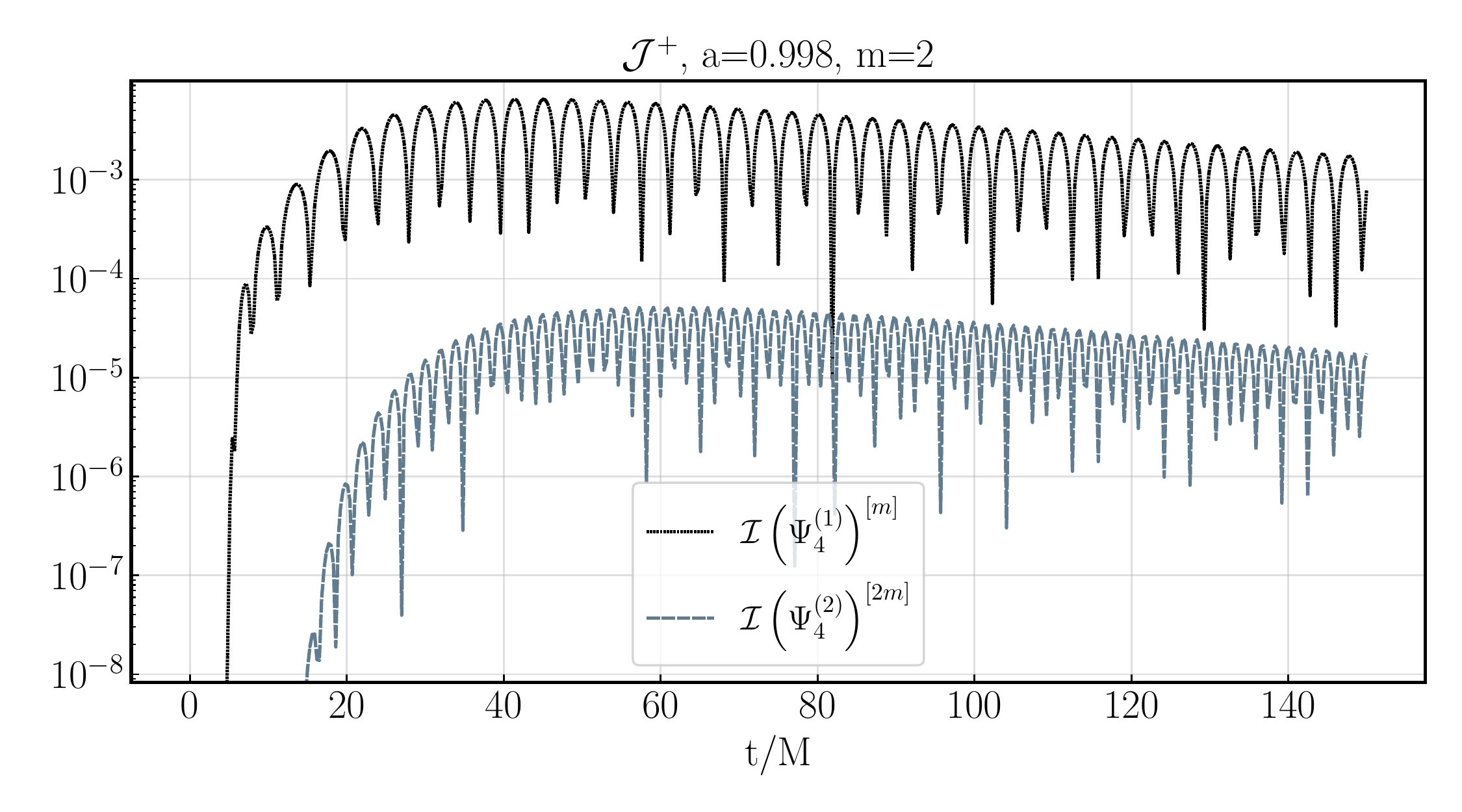}}}%
  \hfill
  \subfloat[$\mathcal{R}\psi_4^{(1)[2]}$, $\mathcal{R}\psi_4^{(2)[0]}$]{{\includegraphics[width=0.5\textwidth]{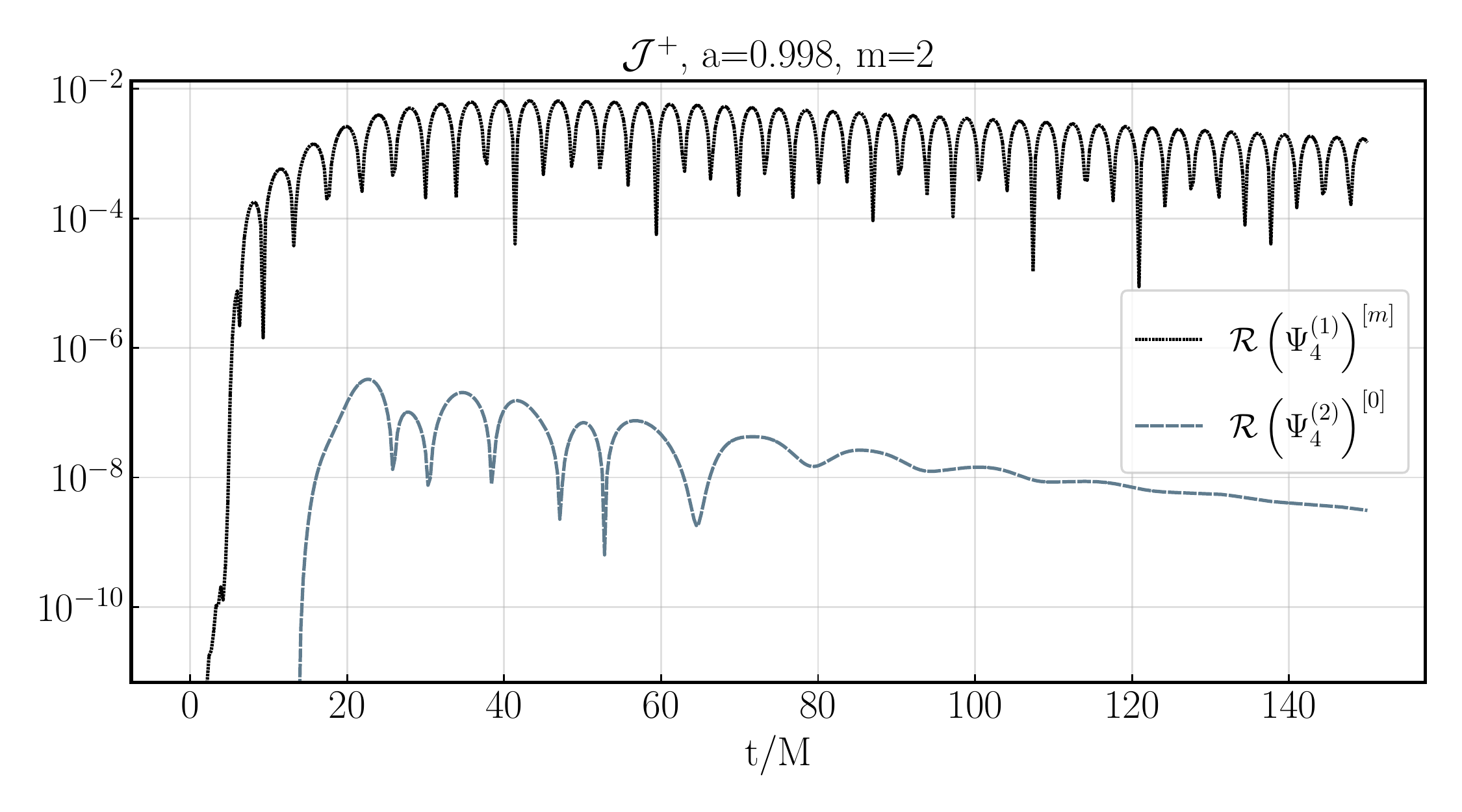}}}%
  \hfill
  \centering
  \subfloat[$\mathcal{I}\psi_4^{(1)[2]}$, $\mathcal{I}\psi_4^{(2)[0]}$]{{\includegraphics[width=0.5\textwidth]{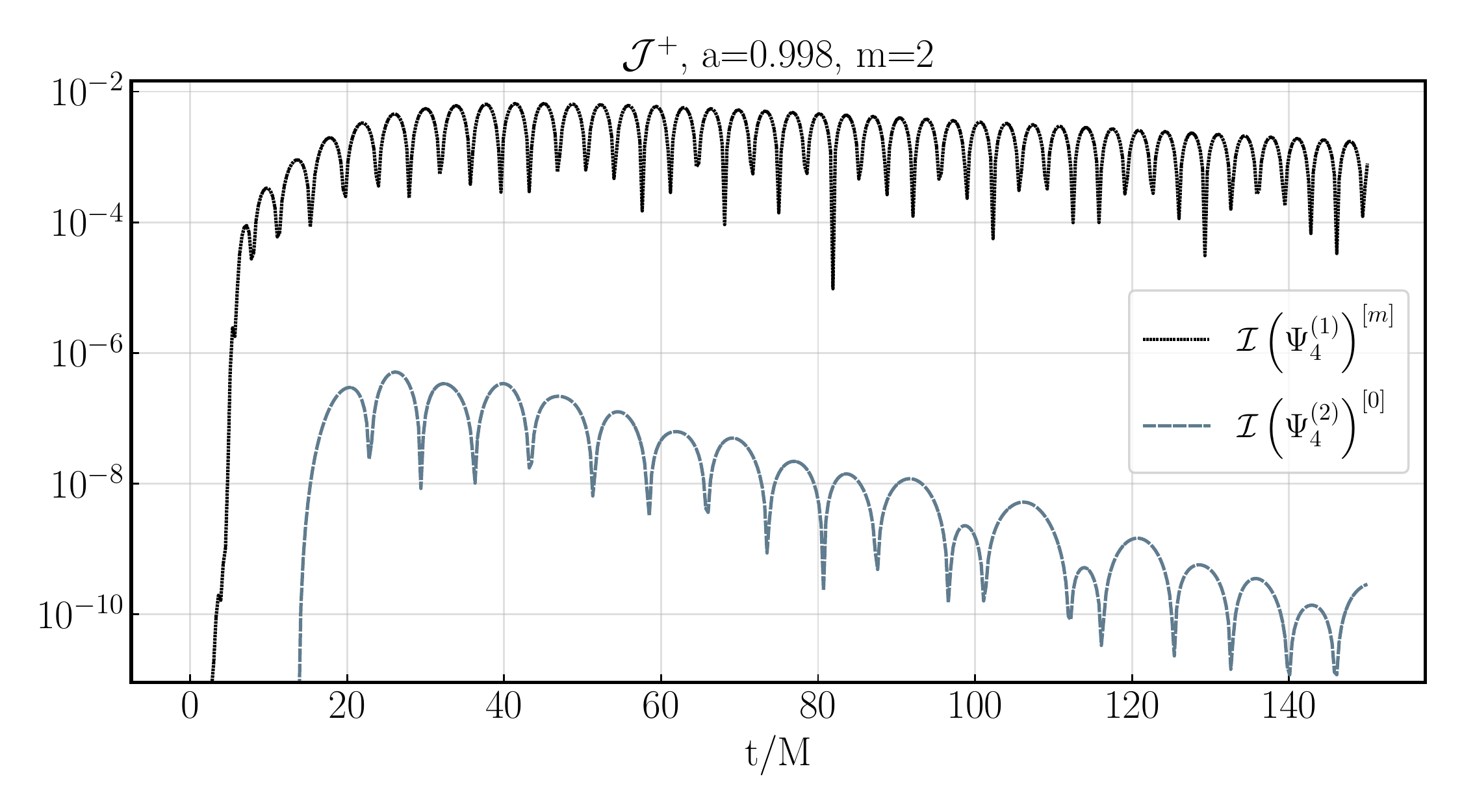}}}%
  \hfill
\caption{Behavior of the real (left) and
imaginary (right) parts of $r\times \Psi_4^{(1),[m]}$ (here $m=2$)
at future null infinity ($\mathcal{J}^+$), compared with
$r\times\Psi_4^{(2),[2m]}$ (top) and $r\times\Psi_4^{(2),[0]}$ (bottom),
for the $a=0.998$ case (see Table.~\ref{table:sim_params_a0.998} for simulation parameters).
As with the data shown for the
$a=0.7$ case in Fig.\ref{fig:spin_0.7_psi4}, the truncation error estimates
are less than $\sim 1\%$ throughout.
}
\label{fig:spin_0.998_psi4}
\end{figure*}

\begin{figure*}
  \centering
  \subfloat[$\mathcal{R}\psi_4^{(1)[2]}$, $\mathcal{R}\mathcal{S}^{[4]}$]{{\includegraphics[width=0.5\textwidth]{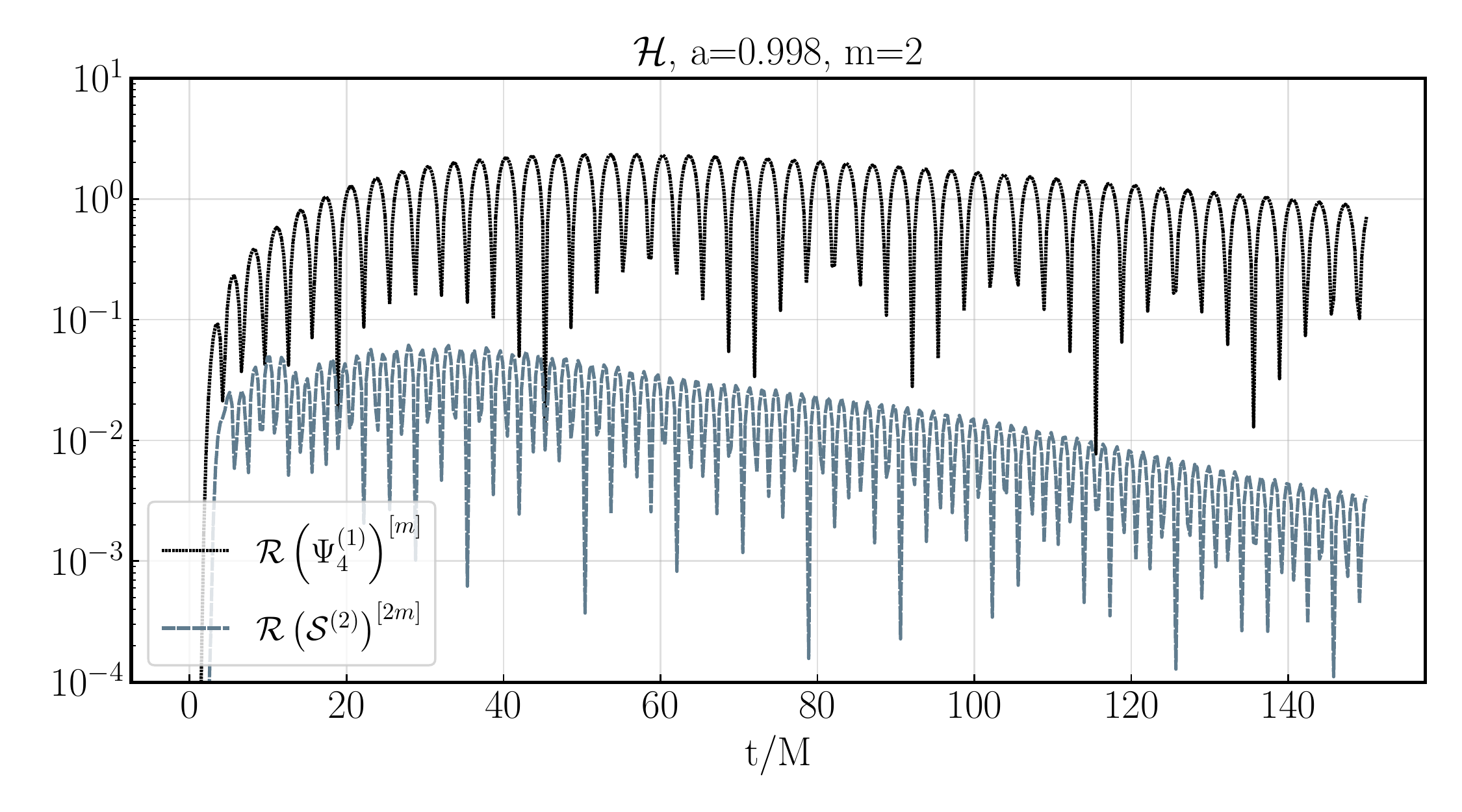}}}%
  \hfill
  \centering
  \subfloat[$\mathcal{I}\psi_4^{(1)[2]}$, $\mathcal{I}\mathcal{S}^{[4]}$]{{\includegraphics[width=0.5\textwidth]{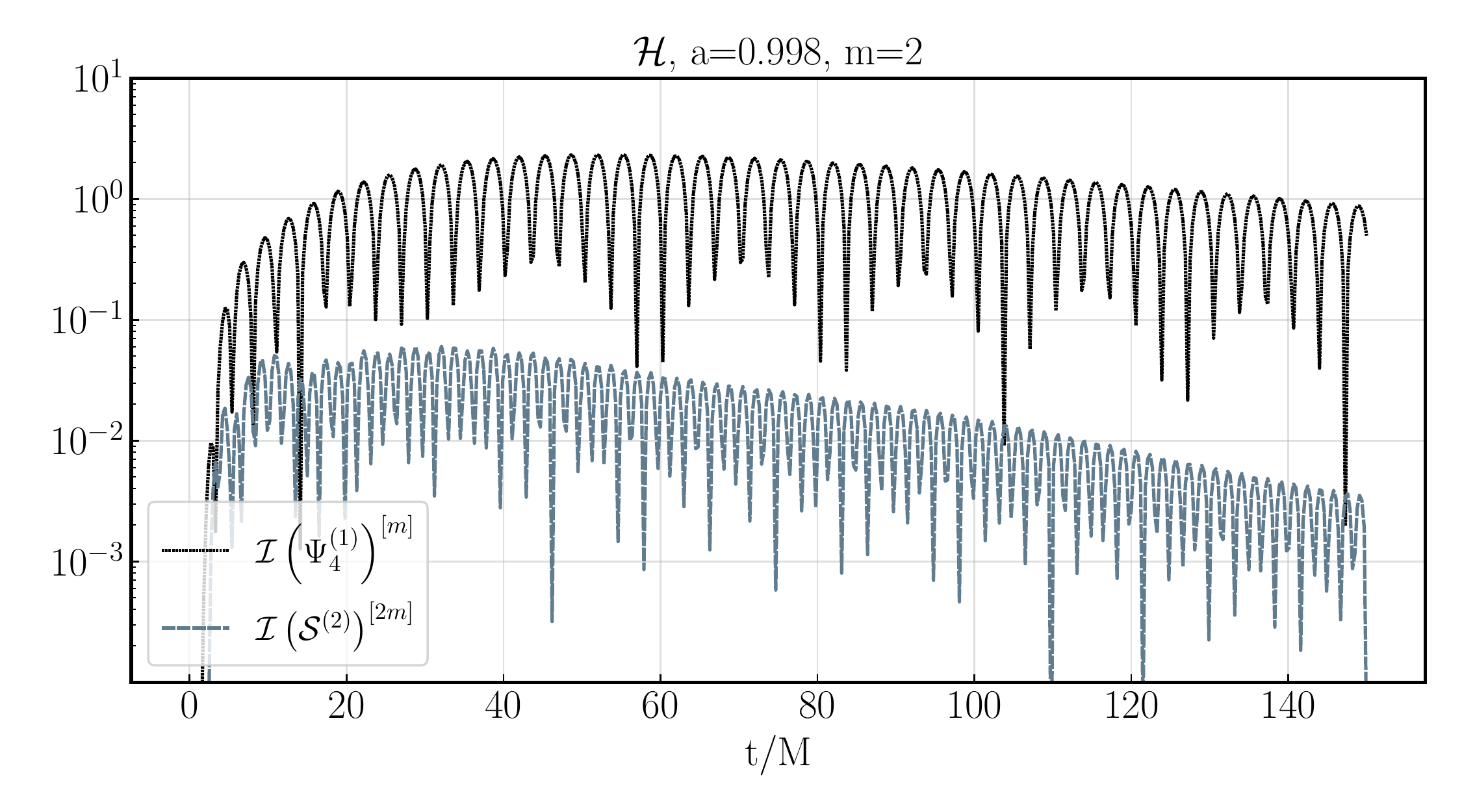}}}%
  \hfill
  \centering
  \subfloat[$\mathcal{R}\psi_4^{(1)[2]}$, $\mathcal{R}\mathcal{S}^{[0]}$]{{\includegraphics[width=0.5\textwidth]{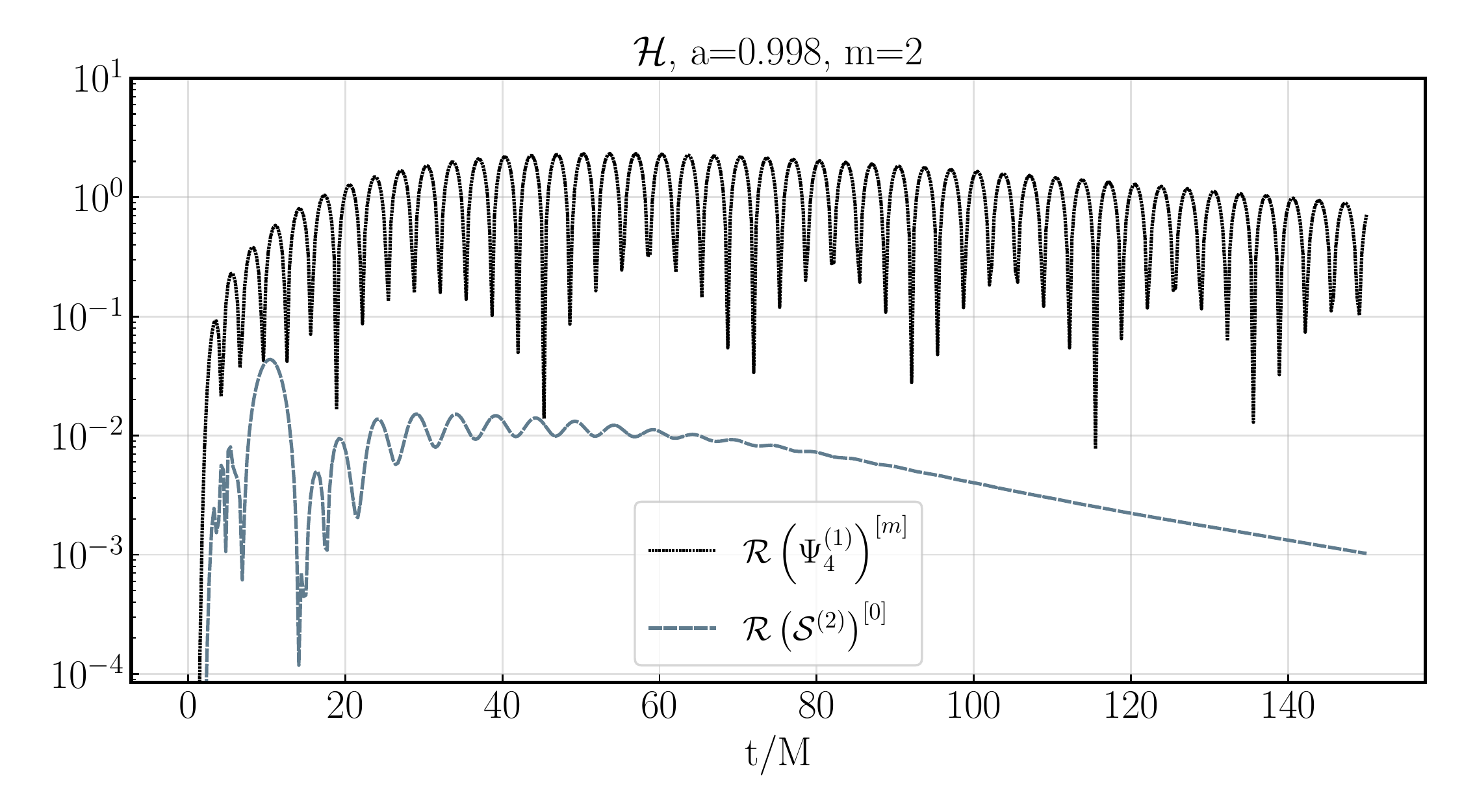}}}%
  \hfill
  \centering
  \subfloat[$\mathcal{I}\psi_4^{(1)[2]}$, $\mathcal{I}\mathcal{S}^{[0]}$]{{\includegraphics[width=0.5\textwidth]{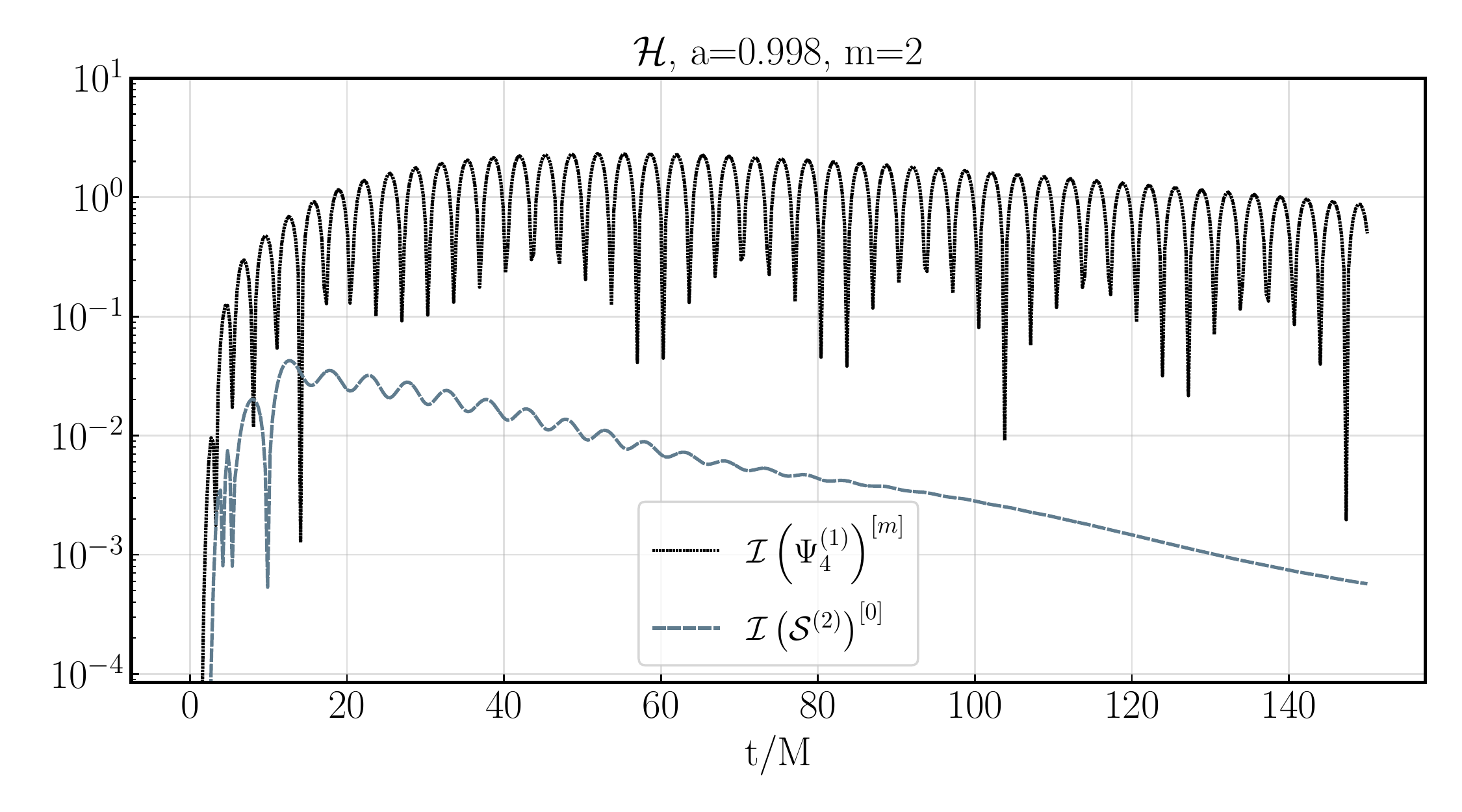}}}%
  \hfill
\caption{
Comparison of the magnitude of the real (left) and imaginary (right) components of $\Psi_4^{(1)}$ with the 
corresponding components of the second order source terms
for $\mathcal{S}^{(2),[4]}$ (top) and $\mathcal{S}^{(2),[0]}$ (bottom),
at the black hole horizon, for the $a=0.998$ case (Table ~\ref{table:sim_params_a0.7}).
}
\label{fig:a0998_m2_horizon_psi4_source}
\end{figure*}

\begin{figure*}
  \centering
  \subfloat[$\mathcal{R}\mathcal{S}^{[4]}$]{{\includegraphics[width=0.5\textwidth]{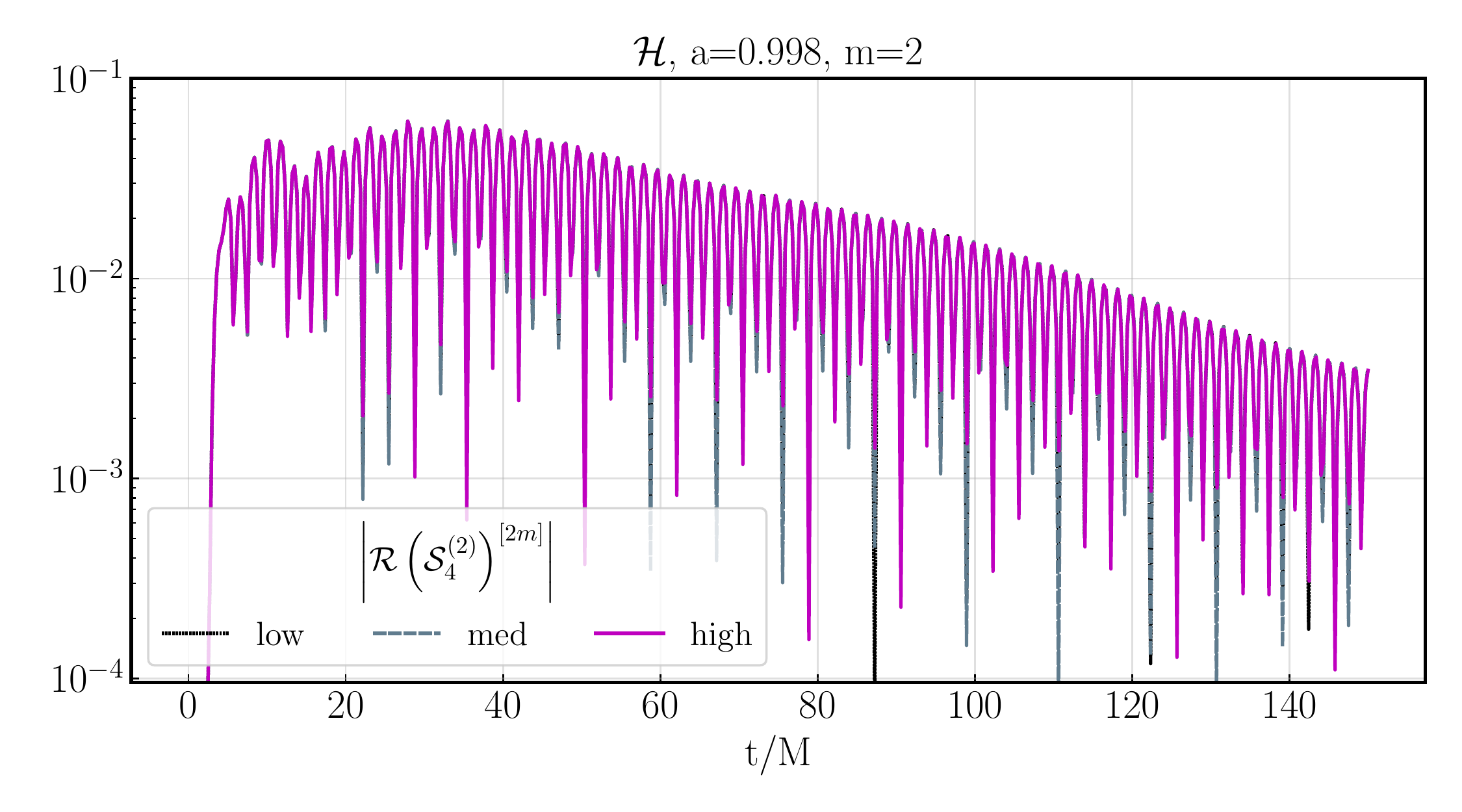}}}%
  \hfill
  \centering
  \subfloat[$\mathcal{I}\mathcal{S}^{[4]}$]{{\includegraphics[width=0.5\textwidth]{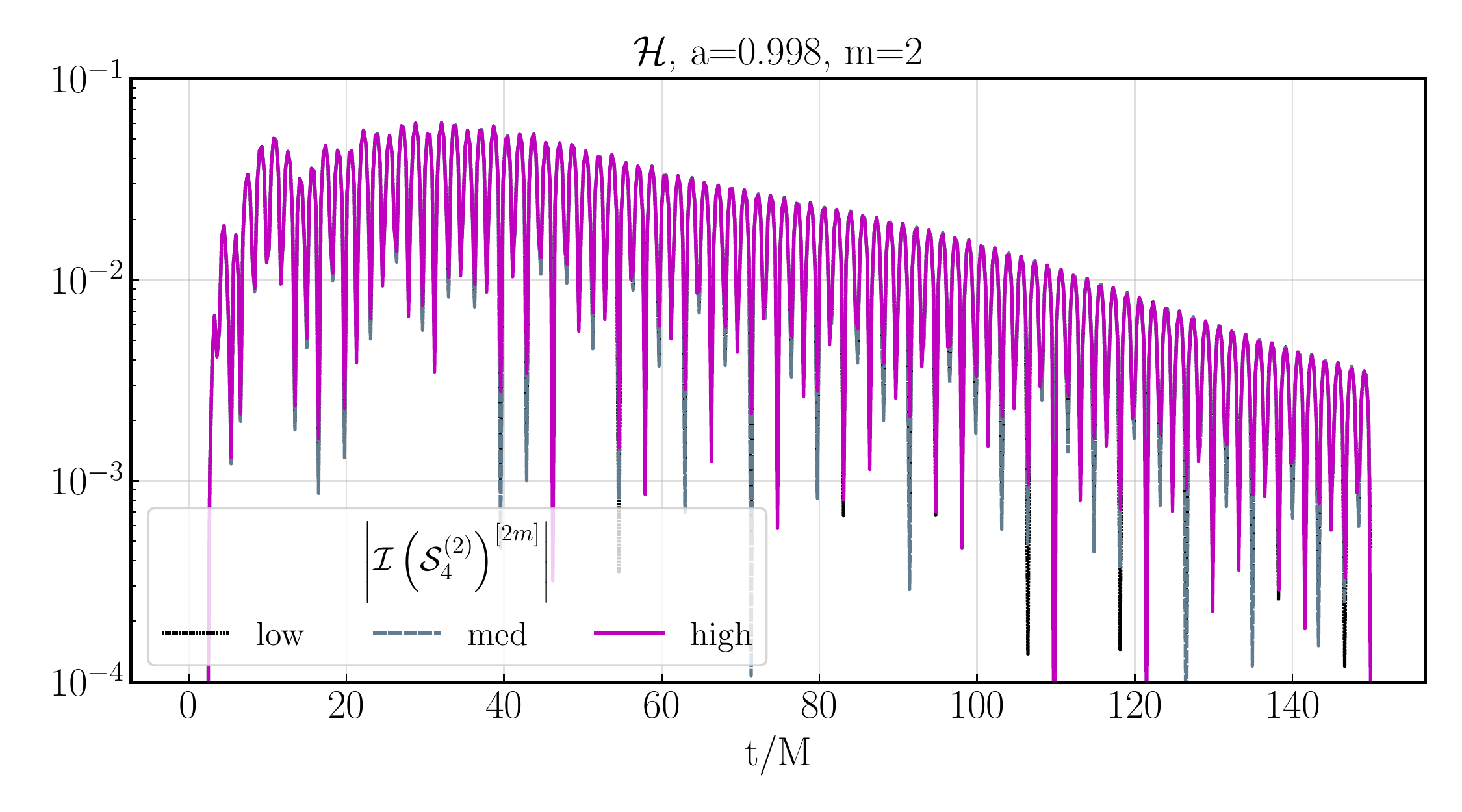}}}%
  \hfill
  \centering
  \subfloat[$\mathcal{R}\mathcal{S}^{[0]}$]{{\includegraphics[width=0.5\textwidth]{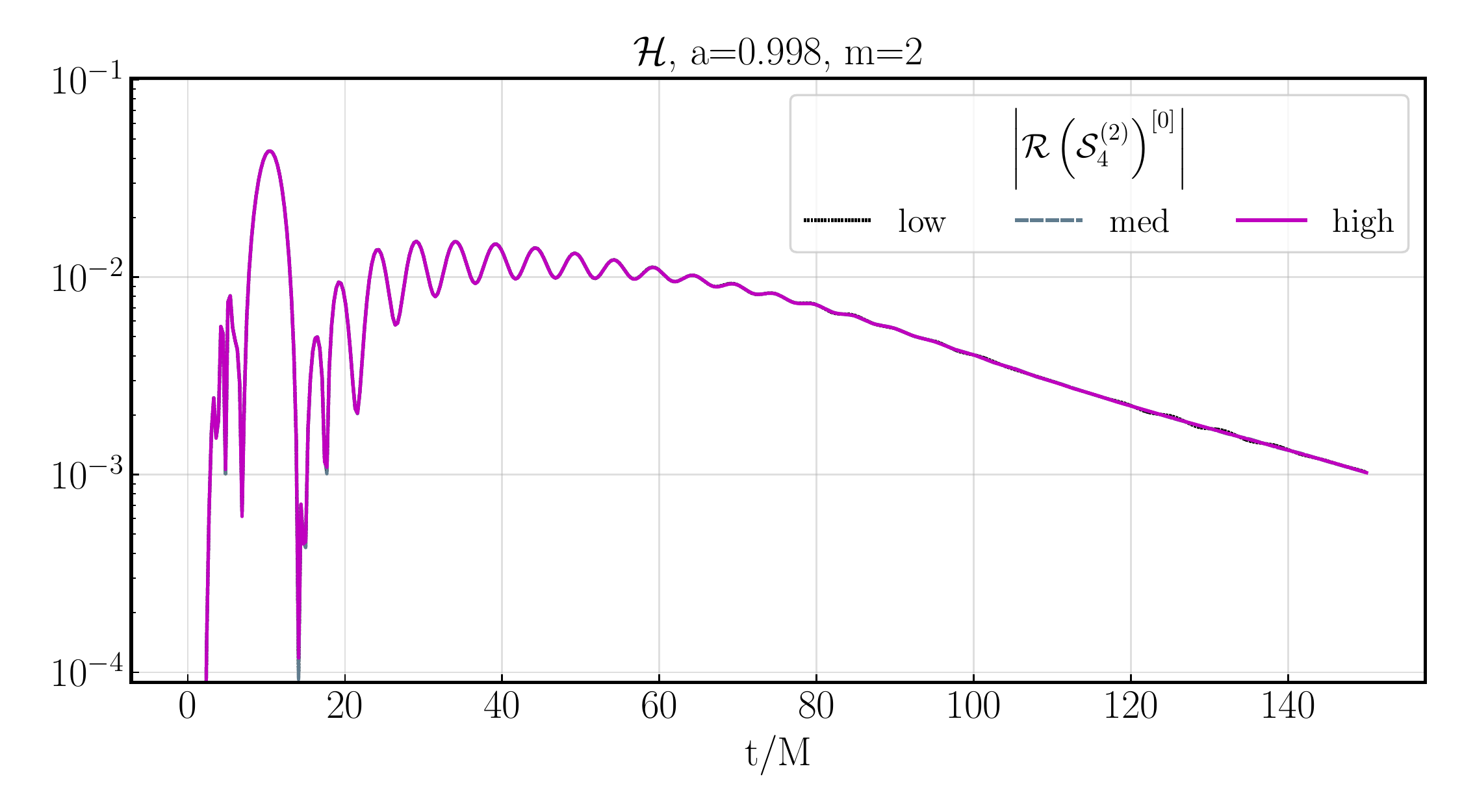}}}%
  \hfill
  \centering
  \subfloat[$\mathcal{I}\mathcal{S}^{[0]}$]{{\includegraphics[width=0.5\textwidth]{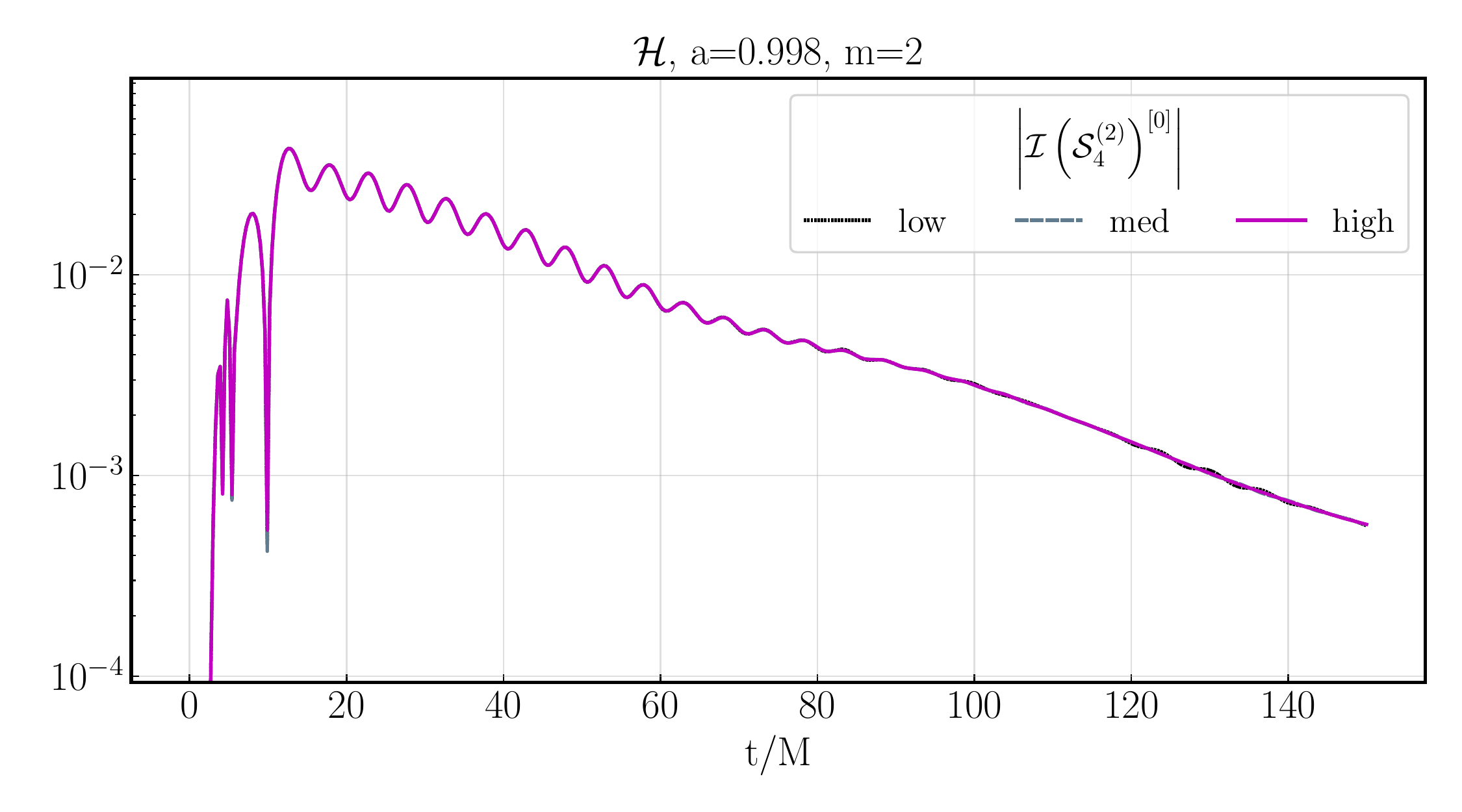}}}%
  \hfill
\caption{A resolution study of
the real (left) and imaginary (right) parts of the second order source terms
$\mathcal{S}^{(2)[4]}$ (top) and $\mathcal{S}^{(2)[0]}$ (bottom) at the
black hole horizon, for the $a=0.998$ case (Table ~\ref{table:sim_params_a0.998}).
This demonstrates that we are resolving 
the source terms over the entire integration
time ($T=0,150M$).
}
\label{fig:a0998_m2_horizon_source2_source0_resolution}
\end{figure*}

\begin{figure*}
   \subfloat[$\left|\mathcal{B}_2\right|_2$]{{\includegraphics[width=0.5\textwidth]{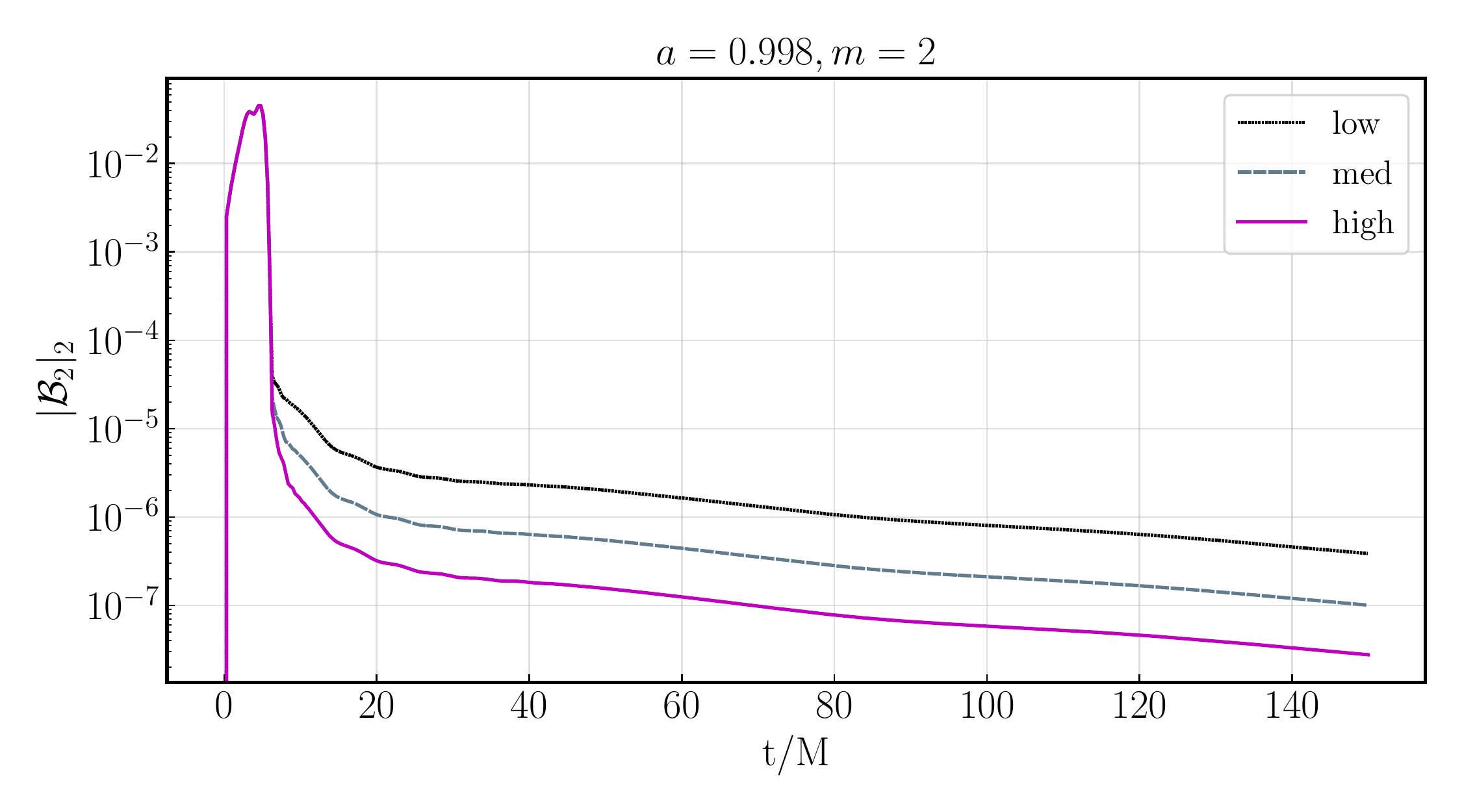}}}%
   \subfloat[$\left|\mathcal{B}_3\right|_2$]{{\includegraphics[width=0.5\textwidth]{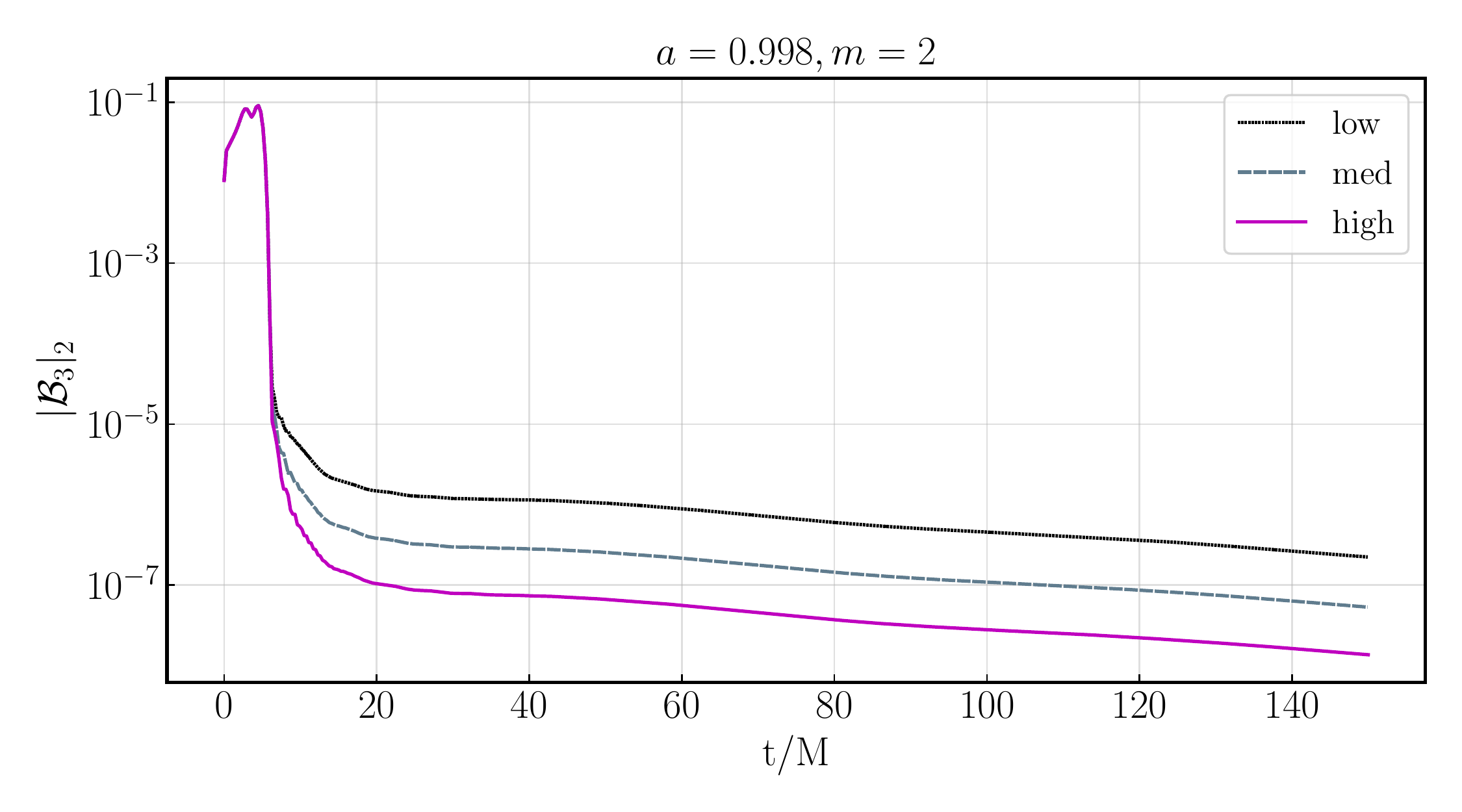}}}%
   \hfill
   \centering
   \subfloat[$\left|\mathcal{H}\right|_2$]{{\includegraphics[width=0.5\textwidth]{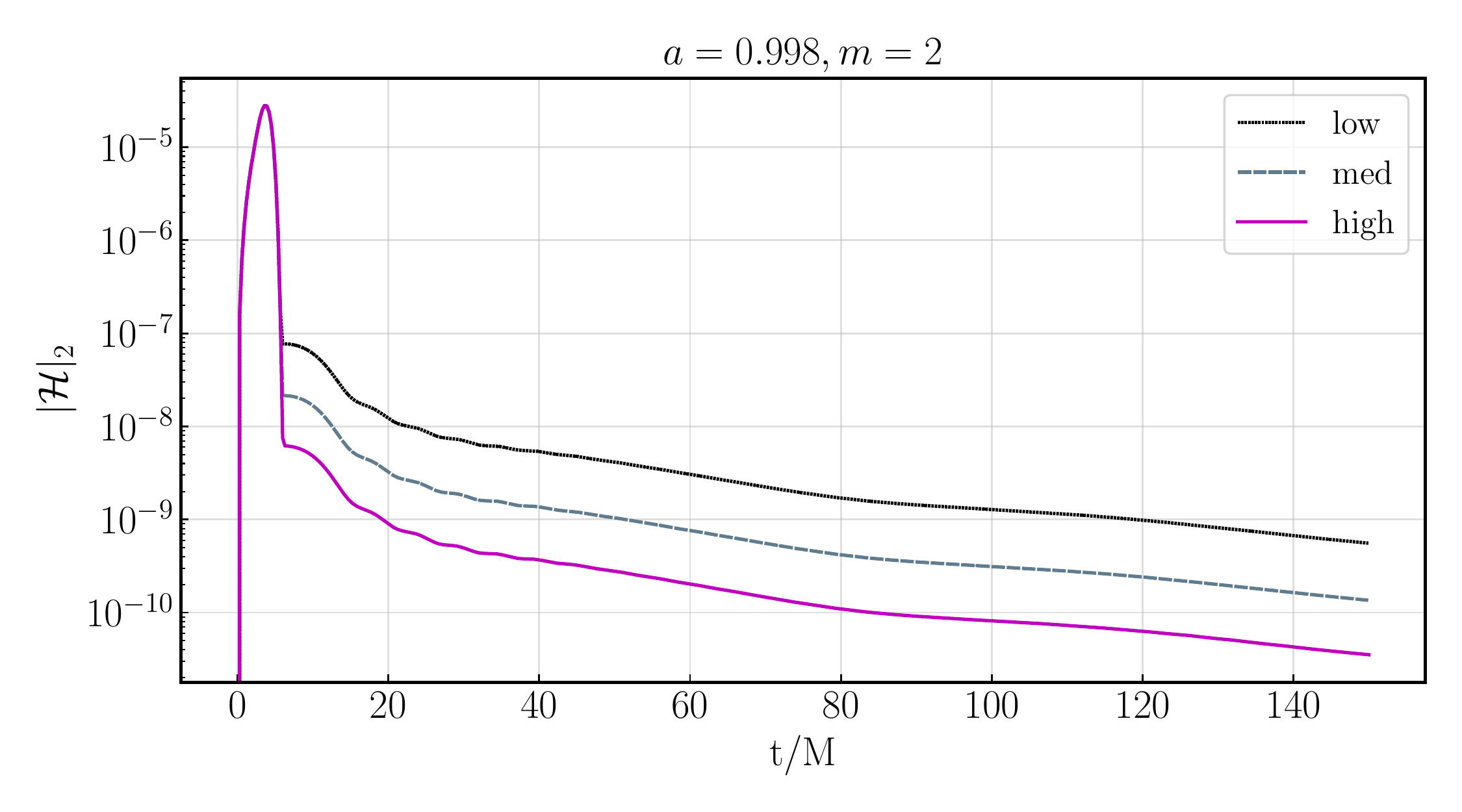}}}%
   \hfill
\caption{
The discrete two norm (see Eq.~\eqref{eq:def_two_norm})
of independent residuals $\mathcal{B}_3$, $\mathcal{B}_2$,
and $\mathcal{H}$ for metric reconstruction
(see respectively Eq.~\eqref{eq:def_B3}, Eq.~\eqref{eq:def_B2},
and Eq.~\eqref{eq:def_H}), for the spin $a=0.998$ case,
as a function of time for three different resolutions
(Table. \ref{table:sim_params_a0.998}).
We only begin to obtain convergence to zero once the region
with inconsistent initial data has left our computational domain (around $t/M\sim5$).
}
\label{fig:convergence_indep_res_a0998}
\end{figure*}

\begin{figure*}
  \centering
  \subfloat[$\mathcal{R}\psi_4^{(2)[4]}$]{{\includegraphics[width=0.5\textwidth]{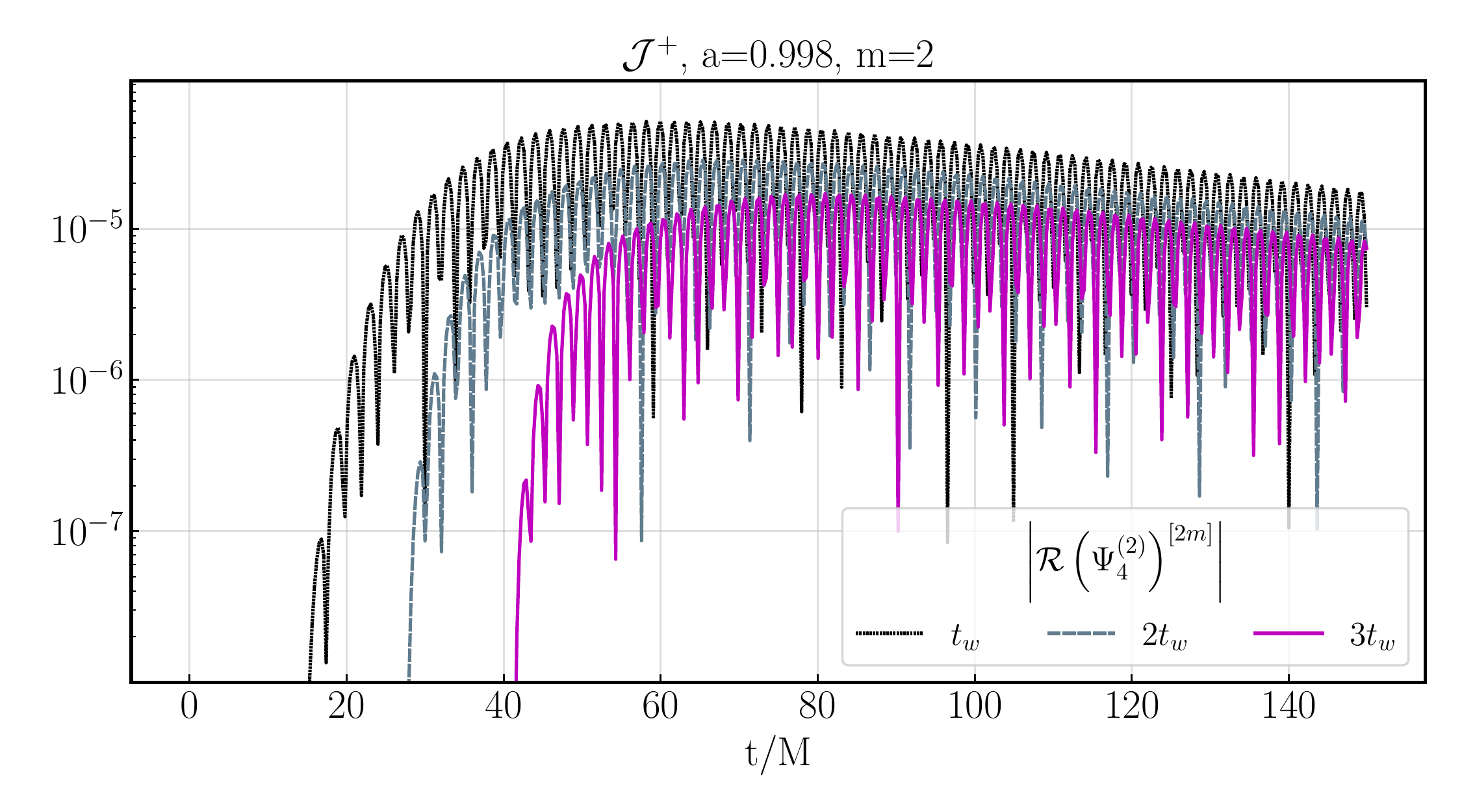}}}%
  \hfill
  \centering
  \subfloat[$\mathcal{I}\psi_4^{(2)[4]}$]{{\includegraphics[width=0.5\textwidth]{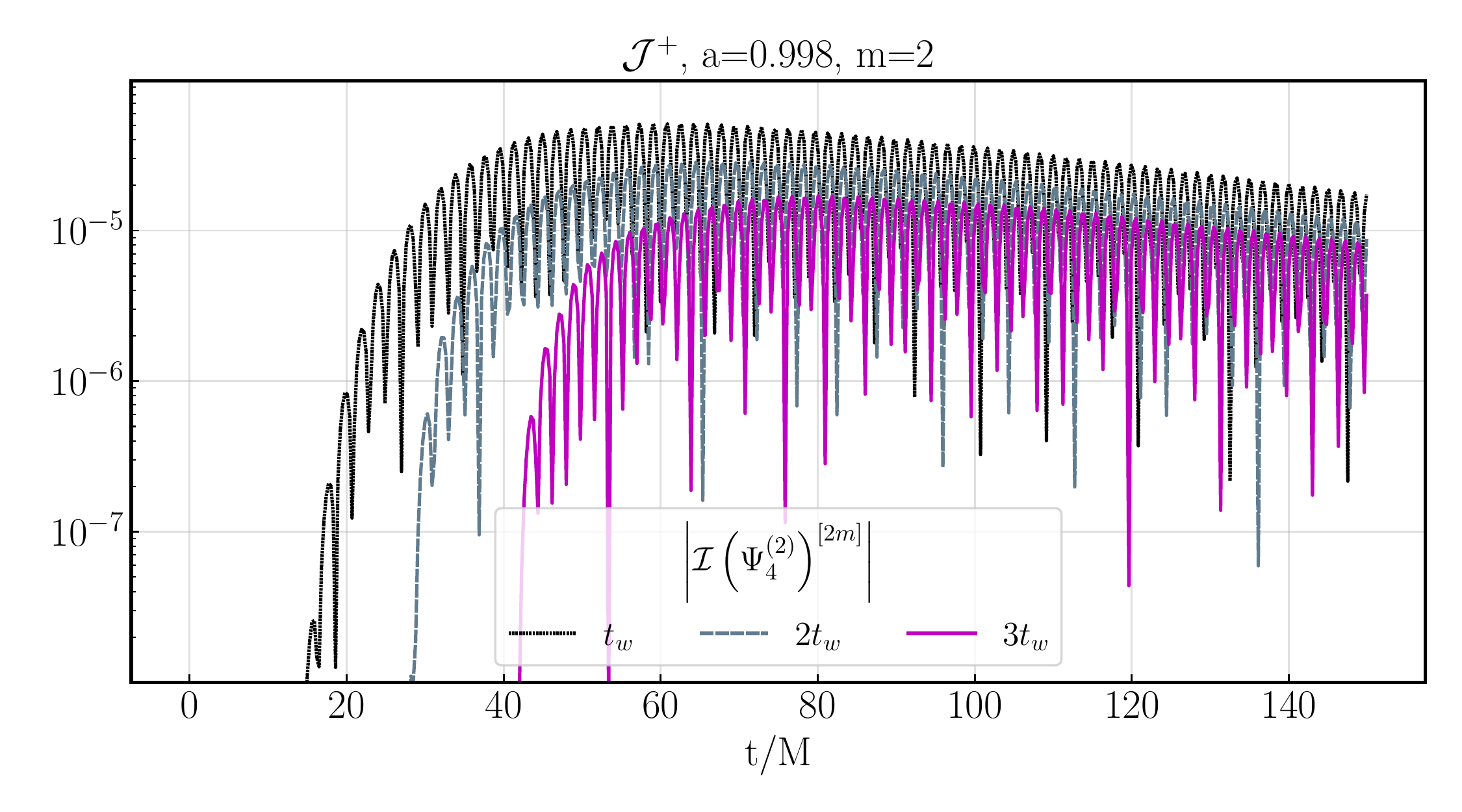}}}%
  \hfill
  \centering
  \subfloat[$\mathcal{R}\psi_4^{(2)[0]}$]{{\includegraphics[width=0.5\textwidth]{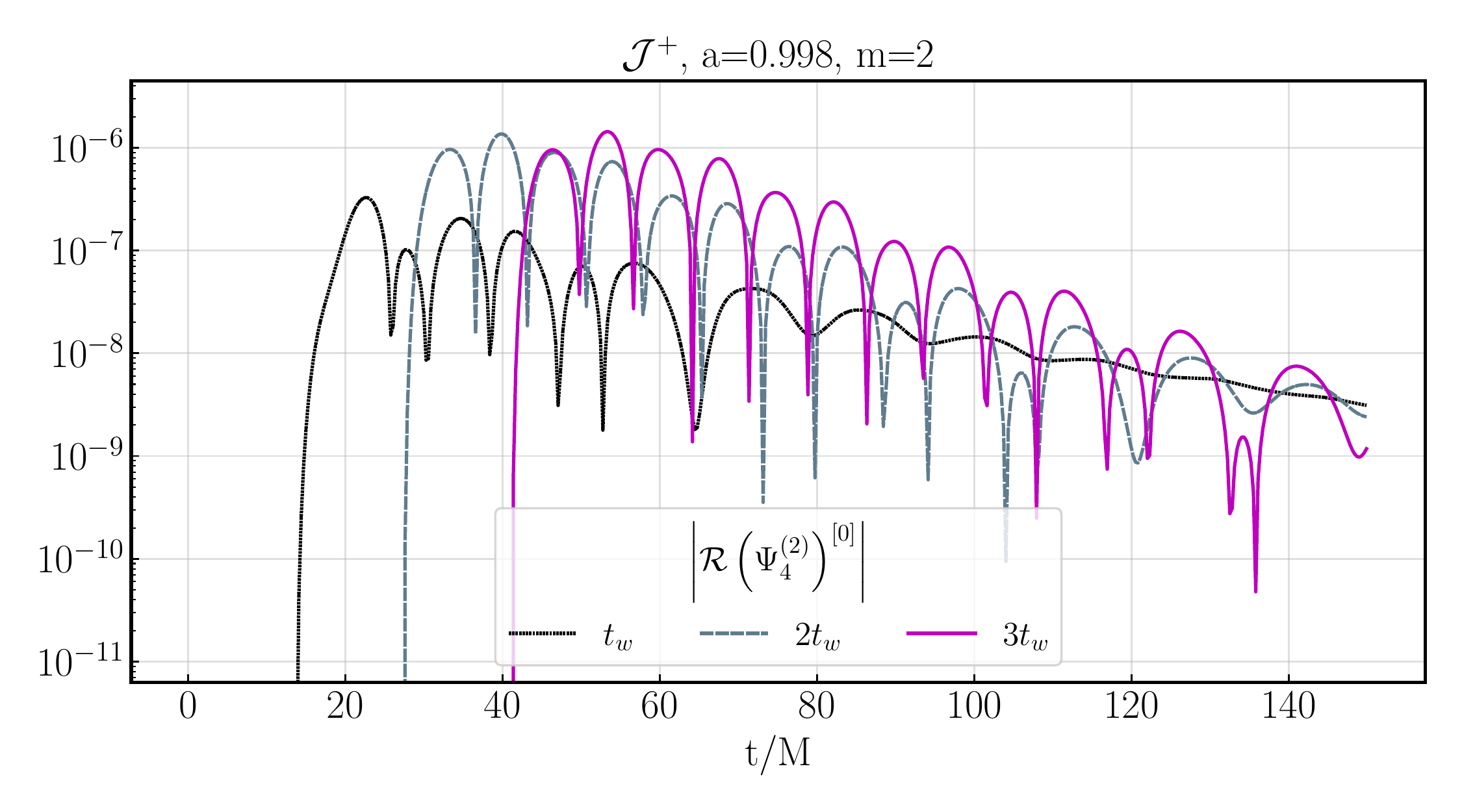}}}%
  \hfill
  \centering
  \subfloat[$\mathcal{I}\psi_4^{(2)[0]}$]{{\includegraphics[width=0.5\textwidth]{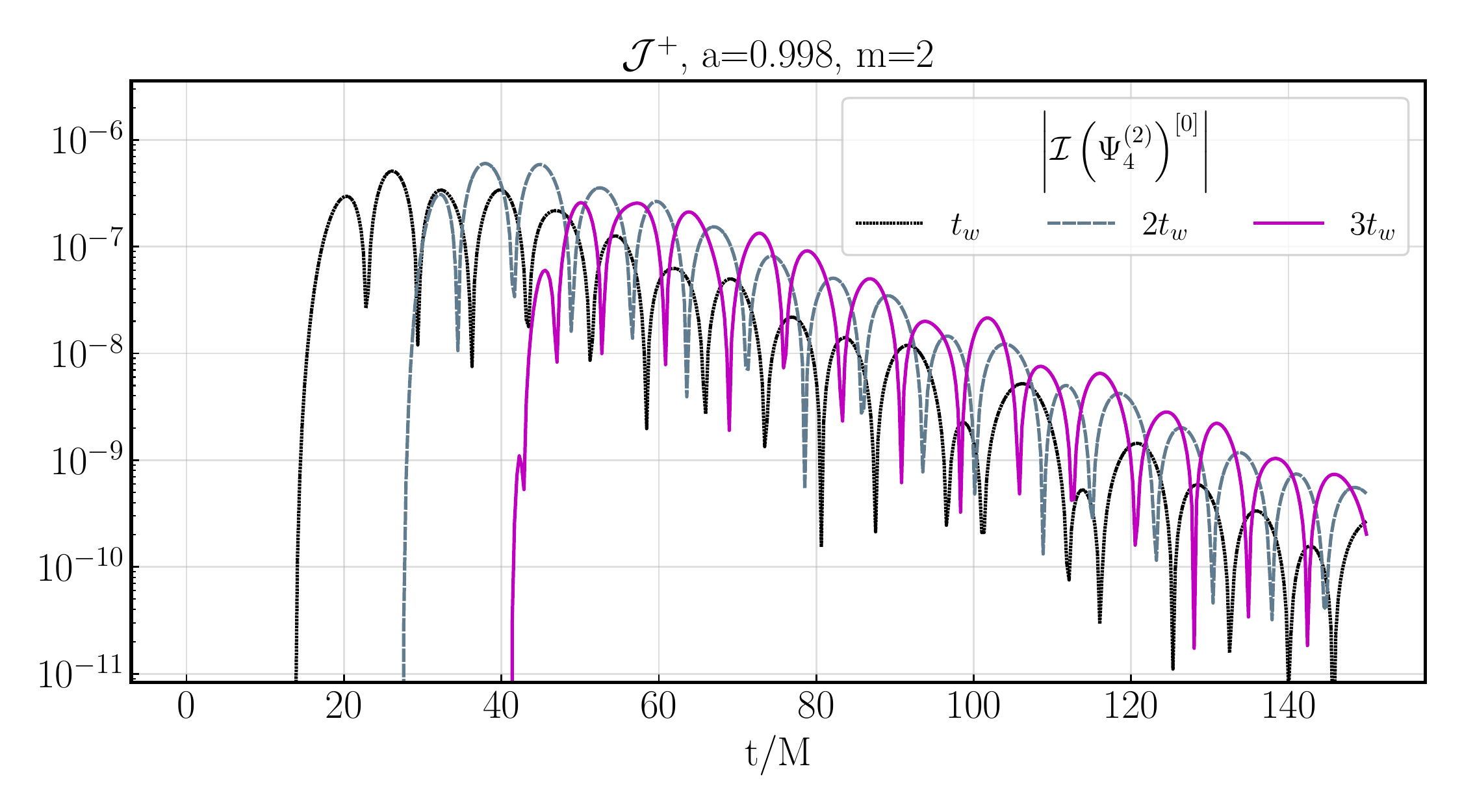}}}%
  \hfill
\caption{Comparison of the real (left) and imaginary (right)
components of the second order $\Psi_4^{(2),[4]}$ (top)
and $\Psi_4^{(2),[0]}$ (bottom) fields, from the same $a=0.998$ first
order perturbation depicted in Fig.~\ref{fig:spin_0.998_psi4}, as a function of when we begin
evolving the second order field. Three cases are shown, including for reference the $T_w$
case also shown in Fig.~\ref{fig:spin_0.998_psi4}.
}
\label{fig:a0998_m2_startStudy}
\end{figure*}

\begin{figure*}
   \centering
   \subfloat[$\mathcal{F}\mathcal{R}$, window $(41M,95M)$]{{\includegraphics[width=0.5\textwidth]{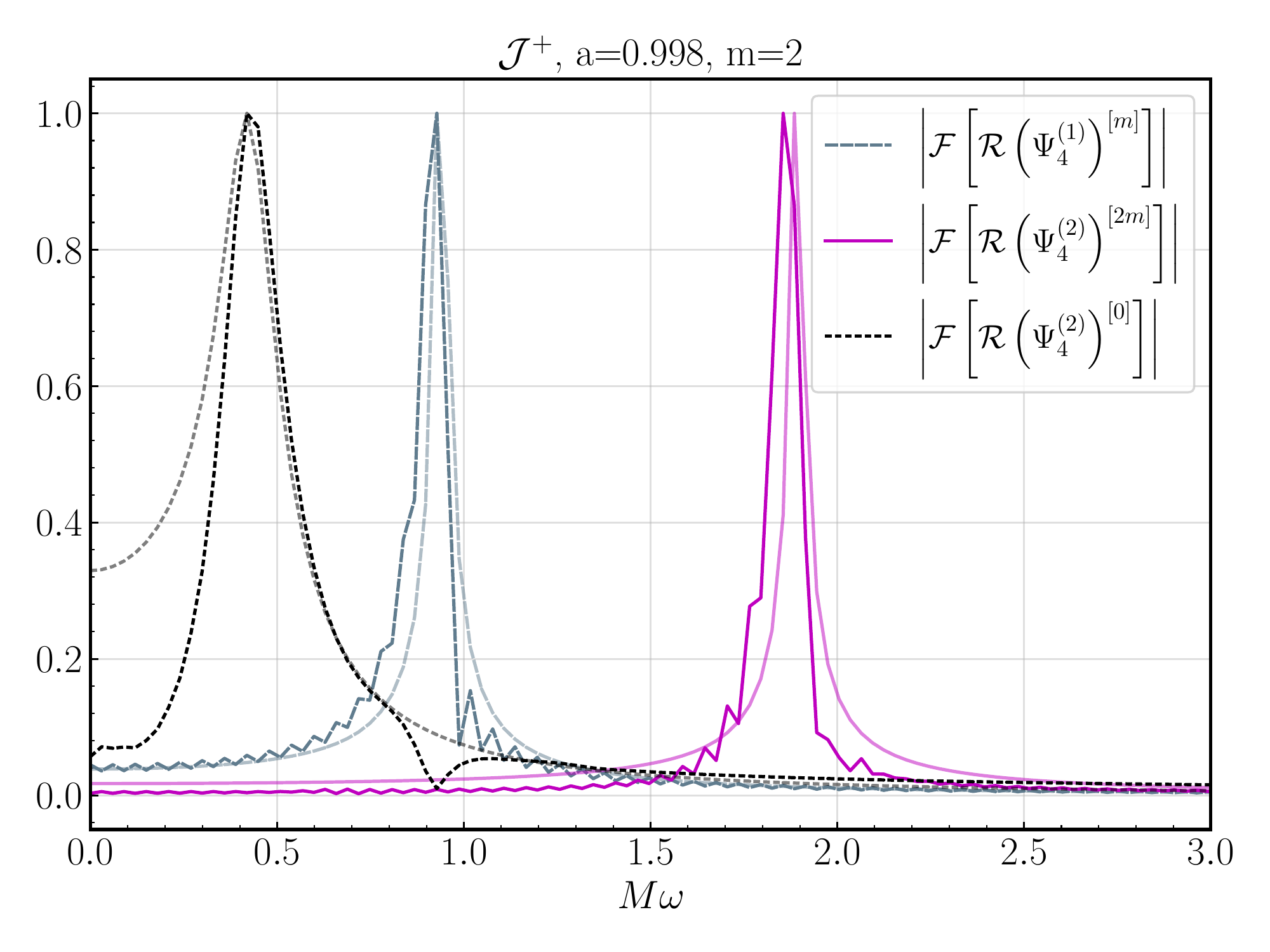}}}%
   \hfill
   \centering
   \subfloat[$\mathcal{F}\mathcal{I}$, window $(41M,95M)$]{{\includegraphics[width=0.5\textwidth]{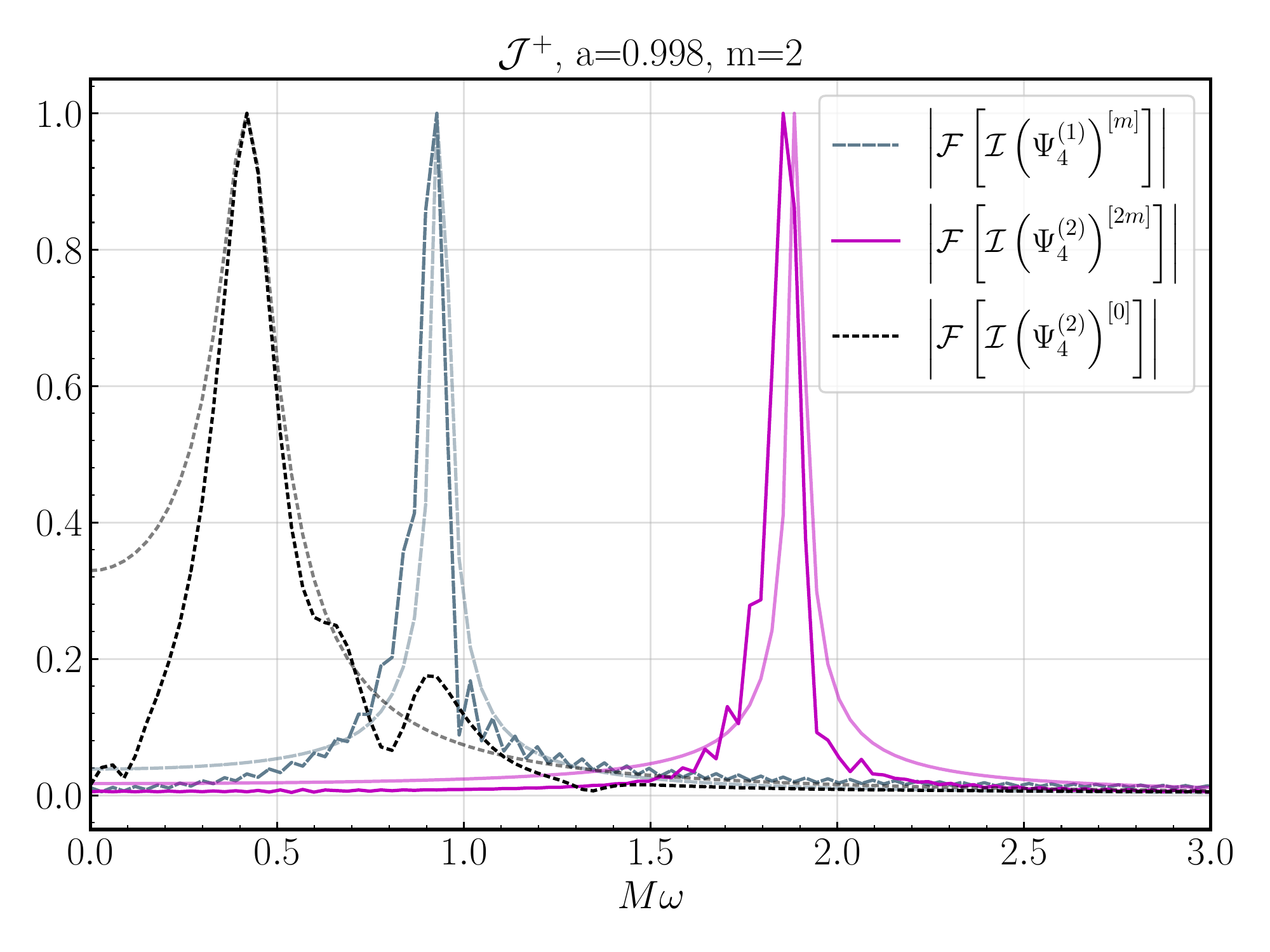}}}%
   \hfill
   \centering
   \subfloat[$\mathcal{F}\mathcal{R}$, window $(68M,150M)$]{{\includegraphics[width=0.5\textwidth]{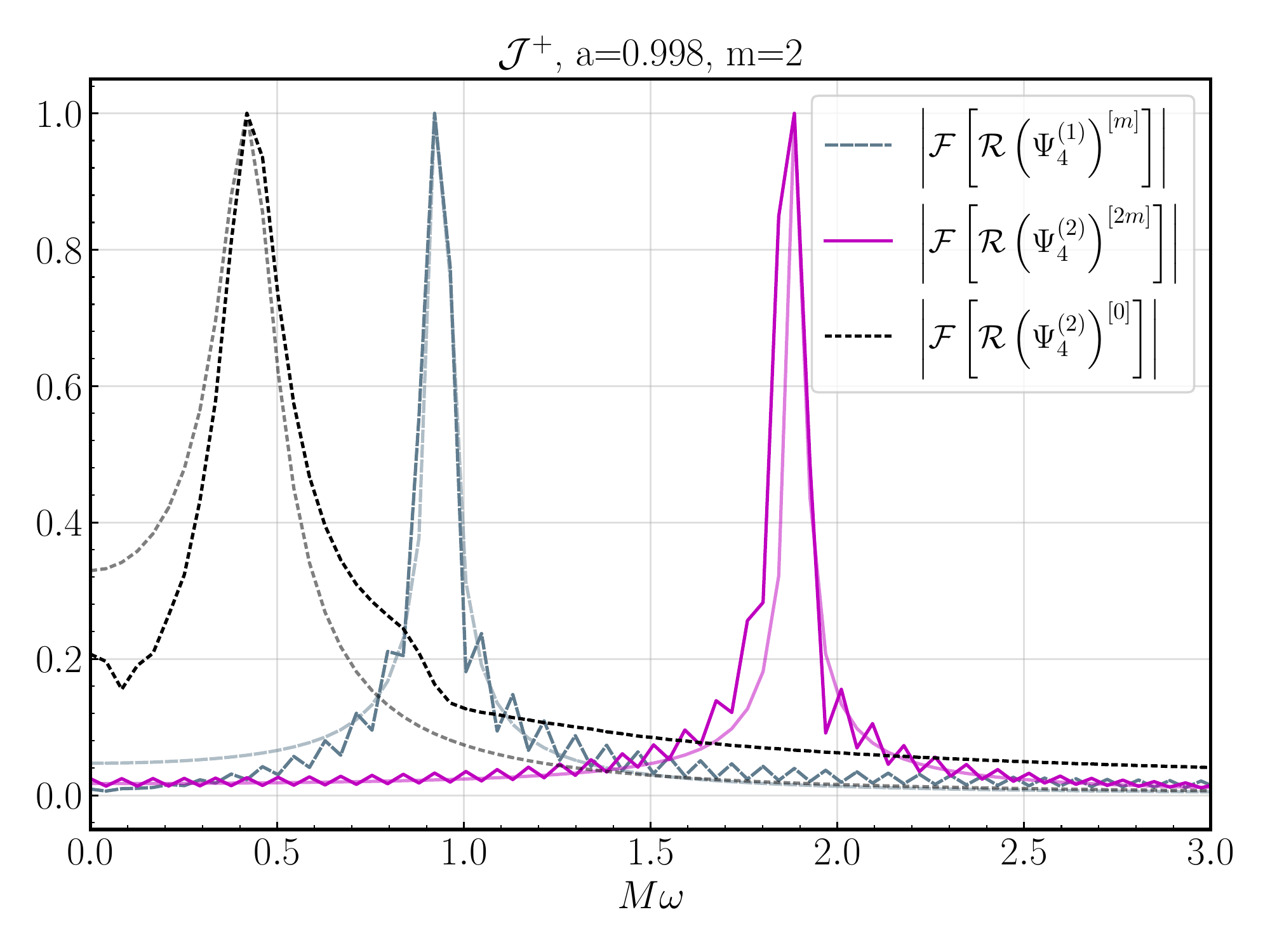}}}%
   \hfill
   \centering
   \subfloat[$\mathcal{F}\mathcal{I}$, window $(68M,150M)$]{{\includegraphics[width=0.5\textwidth]{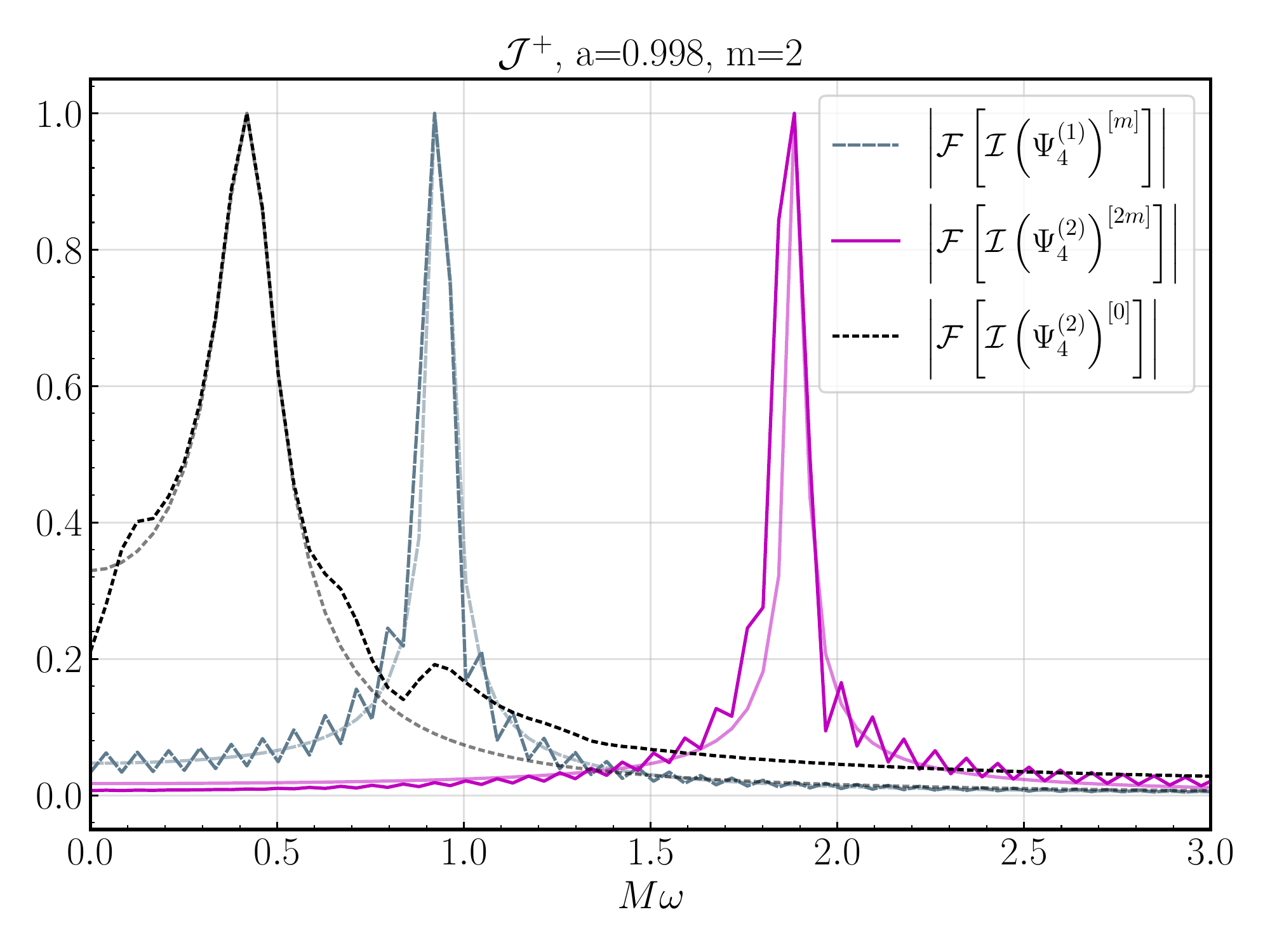}}}%
   \hfill
\caption{Normalized absolute value of the Fourier Transform (\ref{norm_ft}) of
the real (left) and imaginary (right) parts of
$\Psi_4^{(1),[2]}$, $\Psi_4^{(2),[4]}$, and $\Psi_4^{(2),[0]}$,
taken over two different windows, for the $a=0.998$ case (see Table~\ref{table:sim_params_a0.998}).
As in Fig.\ref{fig:spin_0pt7_FT_comparison}
for the $a=0.7$ case, the data for the second order
components come from the $3 T_w$ start time (see Fig.\ref{fig:a0998_m2_startStudy}),
the top (bottom) panels use a $[3 T_w,150M]$ ($[5 T_w,150M]$) window,
darker lines are from the numerical output,
and the lighter lines are the Fourier transforms of the corresponding
quasinormal modes of a spin $a=0.998$ black hole.
The small angular oscillations in the measured Fourier transforms
are due to our (Dirichlet) windowing of the measured waveform.
}
\label{fig:spin_0pt998_FT_comparison}
\end{figure*}

We show the same set of data as from the $a=0.7$ runs : the magnitudes
of $\Psi_4^{(1)}$ and $\Psi_4^{(2)}$ at future null infinity (Fig.\ref{fig:spin_0.998_psi4}),
the magnitudes of $\Psi_4^{(1)}$ and $\mathcal{S}^{(2)}$ at the horizon (Fig.\ref{fig:a0998_m2_horizon_psi4_source}),
a resolution study of the latter (Fig.~\ref{fig:a0998_m2_horizon_source2_source0_resolution}),
convergence of the metric reconstruction independent residuals (Fig.\ref{fig:convergence_indep_res_a0998}),
the second order response with varied start time (Fig.\ref{fig:a0998_m2_startStudy}), and
Fourier transforms of $\Psi_4^{(1)}$ and $\Psi_4^{(2)}$ at future null infinity (Fig.\ref{fig:spin_0pt998_FT_comparison}).

For the most part the interpretation of the results is similar to the $a=0.7$ case,
taking into account the shorter evolution time in terms of $T_d$ for the $a=0.998$ case.
A notable difference though is a significant non-oscillatory component
to the second order $m=0$ mode. One can roughly understand why such a component
might appear given our initial data for $\Psi_4^{(1),[m]}\propto e^{i\omega_R t - \omega_I t}$
(we follow the quasinormal mode convention 
where an exponential $e^{i \omega t}$ has
complex frequency $\omega\equiv\omega_R + i \omega_I$).
The $m=0$ source term largely comes form reconstructed fields of the form
$p^{[m]}\times\overline{q^{[m]}}$, where $p^{[m]},q^{[m]}\propto \Psi_4^{(1),[m]}$;
hence their oscillatory components can cancel, leaving a real exponential piece decaying
at roughly twice the rate of $\Psi_4^{(1),[m]}$. 
For near extremal spins (in contrast to the $a=0.7$
case), this driven component has
a decay rate quite close to the fundamental $m=0,l=2$ harmonic 
\footnote{See e.g. Table II of~\cite{Berti:2005ys} for their $a=0.98$ case,
the closest spin to our value that they list.}, which
is why it remains visible in the waveform at late times. We find that
how much of an oscillatory vs pure exponential piece is visible
in either of the real or imaginary parts 
of $\Psi_4^{(2),[0]}$ depends quite sensitively on the relative
amplitudes and phases of the real vs imaginary components
of $\Psi_4^{(1)}$ in the initial data.

\section{Conclusion and further extensions}
\label{sec:conclusion}
	We have presented a new numerical evolution scheme to reconstruct the
linear metric from the Weyl scalar $\Psi_4^{(1)}$ in Kerr spacetimes, along
with a numerical implementation of the equations of motion for the second
order perturbation $\Psi_4^{(2)}$ (a more detailed discussion
of the analytic framework we used is 
discussed in the Appendices of this paper, and in our first
paper \cite{Loutrel:2020wbw}).
This first implementation is limited in several respects, and in the
remainder of this section we outline possible extensions
that will allow more direct application to our desired goals of
studying second order effects in post-merger black hole ringdown,
investigating gravitational wave turbulence,
and other related issues for rapidly rotating Kerr black holes.

   In this study we only considered mode coupling from
a single mode of angular number $m$, $\Psi_4^{(1),[m]}$,
to produce the
frequency doubled second order components
$\Psi_4^{(2),[2m]}$ and $\Psi_4^{(2),[0]}$.
Astrophysically realistic
sets of initial data will include many different modes, and the
second order perturbations for an $m$ mode
will be a sum over all modes $(m_1,m_2)$
such that $m_1+m_2=m$. We leave exploring such
more complicated mode mixing to future work.

   Constructing astrophysically realistic initial data for
$\Psi_4^{(1)}$, and the $l=0,1$ components of the metric
and NP scalars that represent the changes in mass and spin
corresponding to the given $\Psi_4^{(1)}$,
remain unsolved problems
in black hole perturbation theory.
This will require specifying a $\Psi_4^{(1)}$ that matches
a desired scenario at $T=0$, and then solving a set 
of constraint equations to give consistent initial
conditions (perturbed metric and related NP quantities)
for the reconstruction transport equations. These constraint
equations are most naturally expressed in terms of geometric
quantities intrinsic to the spacelike hypersurface $T=0$,
giving rise to the familiar Hamiltonian and momentum contraint
equations. One approach to take could be based on solving these
equations using established approaches~\cite{2010nure.book.....B},
and then transforming the solutions to initial conditions for our scheme.
Following a merger, finding appropriate values of $\Psi_4^{(1)}$ (or
equivalently choosing the free initial data for $h_{\mu\nu}$ in a traditional
scheme) describing the perturbation of the remnant, is less well understood.
One possible approach to tackle this is to follow the lines
of earlier analytical studies, including the close limit approximation to 
comparable mass black hole mergers~\cite{Price:1994pm}, or related work
done for the EMRI problem~\cite{Apte:2019txp,Lim:2019xrb}.
Another approach might be to map the data from a constant time slice of
a full numerical simulation 
to our coordinates, and try to extract a 
perturbation relative to the late time Kerr solution.
(And we note that, as discussed in more detail in the companion
paper~\cite{Loutrel:2020wbw}, our goal is not to simply ``solve'' for the
post merger waveform to second order; numerical relativity
can already give us the full nonlinear solution as accurately
as computer resources allow. Rather, we want to be able to interpret ringdown
studies in terms of quasinormal modes,
which requires understanding the waveform
at a quantitative level that the full ``answer'', in terms of
a waveform by itself, cannot give.)

Another area of future work we mention is to investigate whether
one can adapt our scheme to a gauge condition that is
less restrictive on matter/effective matter in the spacetime
than the outgoing radiation gauge. For at present we cannot
study (for example) the EMRI problem, where there is a matter
source representing the small body, and we cannot reconstruct
the second order metric corresponding to the second order piece of
$\Psi_4$. To list two potential ways forward, it may be possible
to continue to work in a radiation gauge but evolve a ``corrector tensor''
to include matter fields as in \cite{Green:2019nam}, or to
directly reconstruct the metric in a different gauge, as is done
in \cite{Chandrasekhar_bh_book,Andersson:2019dwi}.

We emphasize that we have implemented a form of
``ordinary'' perturbation theory (e.g. \cite{bender2013advanced}),
based on expansion in the curvature (or equivalently the peturbed
metric)
\begin{align}
   \Psi_4
   =
   \Psi_4^{(1)}
+  \Psi_4^{(2)}
+  \cdots
   ,
\end{align} 
where the corrections are solved order-by-order, and at each
order the new correction is assumed smaller than the
prior sum. Up to second order then, the only non-linear
phenomena we can explore is mode-coupling. We mentioned
one of our goals was to understand whether Kerr black holes
could exhibit ``turbulence'', though exactly what turbulence means
in the case of black holes is a bit nebulous. Regardless, the authors
of \cite{2015PhRvL.114h1101Y,Green:2013zba} who first suggested turbulence
could be present in the 4D case argued it would be 
a more subtle effect than can be described
by ordinary perturbation theory, or at least would require
some ``resumming'' of many terms in the sum.
Instead then, they proposed an alternative perturbative
expansion to try to capture such effects at leading beyond-linear
order, showing in a scalar field model that a kind of parametric
instability resulted that might be able to drive a turbulent cascade.
It is still unknown whether such an effect occurs for gravitational perturbations
of Kerr, though mode coupling more along the lines of what we describe
here has been seen in full numerical solutions of both perturbed 4D
Kerr black holes~\cite{East:2013mfa} and black holes in 5D AdS spacetime~\cite{Bantilan:2012vu}.
We leave it to future work for a more thorough comparison with
full numerical results, as well as how close second order mode coupling,
augmented with energy transfer between modes guided by measurements
at null infinity mentioned in Sec.~\ref{sec:gauge_invariance_measurements},
could come to describing a turbulent-like cascade of energy 
for rapidly spinning black holes.

\section*{Acknowledgements}
We are grateful to Andrew Spiers and Adam Pound
for sharing results of their calculations which confirmed our calculation of 
Eq.~\eqref{eq:pert_Rici_rot} (and which confirmed several errors in
Eq. (A4) of \cite{Campanelli:1998jv}), 
to Stefan Hollands for discussions on the metric
reconstruction method described in \cite{Green:2019nam},
and to Scott Hughes and Richard Price for a helpful discussion on
our project and
earlier work on second order black hole perturbation theory.
N.L. \& F.P. acknowledge support from NSF grant PHY-1912171,
the Simons Foundation, 
and the Canadian Institute for Advanced Research (CIFAR).
E.G. acknowledges support from NSF grant DMS-2006741.
The simulations presented in this article were performed
on computational resources managed and supported by
Princeton Research Computing,
a consortium of groups including the
Princeton Institute for Computational Science and Engineering (PICSciE)
and the Office of Information Technology's High Performance Computing Center
and Visualization Laboratory at Princeton University.
\appendix

\section{Conventions for discrete norms and Fourier transforms}
\label{sec:symbols}
	As fields are typically complex in the NP
formalism, the discrete two norm
of a field $f$ at a time level $n$ is defined to be the sum
\begin{align}
\label{eq:def_two_norm}
   |f(t_n)|_2
   \equiv\nonumber\\
   \bigg(
      \frac{1}{N_xN_y}
      \sum_{i=1}^{N_x}\sum_{j=1}^{N_y}
      \big(
            &\left( \mathcal{R}f(t_n,r_i,\vartheta_j)\right)^2
            \nonumber\\&
      +	\left(\mathcal{I}f(t_n,r_i,\vartheta_j)\right)^2
      \big)
   \bigg)^{1/2}
   .
\end{align}
	Our conventions for the Fourier transform are
\begin{align}
	\hat{f}(\omega)
	=
	\int_{-\infty}^{\infty}dt e^{-i\omega t}f(t)
	,
        \nonumber\\
	f(t)
	=
	\int_{-\infty}^{\infty}
	\frac{d\omega}{2\pi} e^{i\omega t}\hat{f}(\omega)
	.
\end{align}
	And we define a normalized Fourier transform by
\begin{align}\label{norm_ft}
	\mathcal{N}\hat{f}(\omega)
	\equiv
	\frac{1}{\mathrm{max}\hat{f}}|\hat{f}(\omega)|
	.
\end{align}
	For reference the absolute value of the Fourier transform of
$f(t)=\Theta(t)e^{-\omega_I t}\mathrm{sin}\left(\omega_R t\right)$ is
\begin{align}
\label{eq:Fourier_transform_damped_sinusoid}
	\left|\hat{f}(\omega)\right|
	=
	\frac{
		\omega_R
	}{
		\sqrt{
			\left(\omega_R^2+\omega_I^2-\omega^2\right)^2
		+	4\omega_I^2\omega^2
		}
	}
	.
\end{align}
\section{Derivation of metric reconstruction equations and source term
	in the outgoing radiation gauge}

\label{sec:derivation_metric_recon_source}
\subsection{Tetrad, gauge, and first order spin coefficients}
	We assume the background is type D, that the background
tetrad has been chosen to set
$\Psi_0=\Psi_1=\Psi_3=\Psi_4=\kappa=\sigma=\nu=\lambda=0$, the
linearized tetrad is as in (\ref{lin_ops}), and we are using 
outgoing radiation gauge for the first order metric perturbation (\ref{org_def}).
For the Kerr spacetime we can also rotate the background tetrad to set $\gamma=0$
(see Sec.~\ref{sec:derivation_coordinates}), and we assume we have done this.

It is worth noting that the tetrad components of the
metric perturbation (\ref{h_projs}) are all scalars
of definite spin and boost weight, which we catalogue in
Table~\ref{table:spin_boost_hij} (for the angular tetrad
projections we list the spin and boost
of their complex conjugates, which is what we mostly use).

\begin{table}
\begin{tabular}{ c | c | c | c }
\hline
 scalar & weight &  spin weight &  boost weight \\ 
\hline
$h_{ll}$                 & $\{ 2, 2\}$ & $ 0$ & $ 2$ \\ 
$h_{l\bar{m}}$         & $\{ 0, 2\}$ & $-1$ & $ 1$ \\ 
$h_{\bar{m}\bar{m}}$ & $\{-2, 2\}$ & $-2$ & $ 0$ \\ 
\hline
\end{tabular}
\caption{Nonzero contractions of the perturbed metric.}
\label{table:spin_boost_hij}
\end{table}
	We can write the first order perturbed tetrad in
terms of the background tetrad.
Following \cite{PhysRevD.13.806}
(see also \cite{Campanelli:1998jv,Loutrel:2020wbw})
we have
\begin{subequations}
\begin{align}\label{tetrad_pert}
	l^{(1)}_{\mu}
	=&
	\frac{1}{2}h_{ll}n_{\mu}
	,\\
	n^{(1)}_{\mu}
	=&
	0
	,\\
	m^{(1)}_{\mu}
	=&
	h_{lm}n_{\mu}
-	\frac{1}{2}h_{mm}\bar{m}_{\mu}
	,
\end{align}
\end{subequations}
        which then immediately gives the expressions for the
perturbed derivative operators listed in (\ref{lin_ops}).

With the above choices for the tetrad/gauge, we find
the following first order corrections
to the spin coefficients (for the more general version
of these expressions, without the choice of
ingoing radiation gauge and $\gamma=0$,
see \cite{Loutrel:2020wbw}):
\begin{subequations}
\label{eq:pert_Rici_rot}
\begin{align}
\label{eq:pert_ka_1}
	\kappa^{(1)}
	=&
	\left(D - 2\epsilon - \bar{\rho}\right)h_{lm} 
         \nonumber\\& 
-	\frac{1}{2}\left(\delta - 2\bar{\alpha} - 2\beta + \bar{\pi} + \tau\right)h_{ll} 
	,\\
\label{eq:pert_la_1}
	\lambda^{(1)}
	=&
-	\frac{1}{2}\left(\Delta - \mu + \bar{\mu}\right)h_{\bar{m}\bar{m}} 
	,\\
\label{eq:pert_si_1}
	\sigma^{(1)}
	=&
	\frac{1}{2}\left(D - 2\epsilon + 2\bar{\epsilon} + \rho - \bar{\rho}\right)h_{mm} 
         \nonumber\\& 
-	\left(\bar{\pi} + \tau\right)h_{lm} 
	,\\
\label{eq:pert_ep_1}
	\epsilon^{(1)}
	=&
-	\frac{1}{4}\left(\Delta - \mu + \bar{\mu}\right)h_{ll} 
         \nonumber\\& 
-	\frac{1}{4}\left(\delta - 2\bar{\alpha} + \bar{\pi} + 2\tau\right)h_{l\bar{m}} 
         \nonumber\\& 
+ 	\frac{1}{4}\left(\bar{\delta} - 2\alpha - 3\pi - 2\bar{\tau}\right)h_{lm} 
	,\\
\label{eq:pert_rh_1}
	\rho^{(1)}
	&=
	\frac{1}{2}\mu h_{ll} 
+	\frac{1}{2}\left(\bar{\delta} - 2\alpha - \pi\right)h_{lm} 
         \nonumber\\& 
-	\frac{1}{2}\left(\delta - 2\bar{\alpha} + \bar{\pi} + 2\tau\right)h_{l\bar{m}}  
	,\\
\label{eq:pert_al_1}
	\alpha^{(1)}
	=&
-	\frac{1}{4}\left(\Delta - 2\mu + \bar{\mu}\right)h_{l\bar{m}} 
         \nonumber\\& 
-	\frac{1}{4}\left(\delta - 2\bar{\alpha} + \bar{\pi} + \tau\right)h_{\bar{m}\bar{m}}  
	,\\
\label{eq:pert_be_1}
	\beta^{(1)}
	=&
-	\frac{1}{4}\left(\Delta + \mu + 2\bar{\mu}\right)h_{lm}  
         \nonumber\\& 
+ 	\frac{1}{4}\left(\bar{\delta} + 2\bar{\beta} - \pi - \bar{\tau}\right)h_{mm}  
	,\\
\label{eq:pert_pi_1}
	\pi^{(1)}
	=&
-	\frac{1}{2}\left(\Delta + \bar{\mu}\right)h_{l\bar{m}}
- 	\frac{1}{2}\tau h_{\bar{m}\bar{m}}
	,\\
\label{eq:pert_ta_1}
	\tau^{(1)}
	=&
	\frac{1}{2}\left(\Delta+\mu\right)h_{lm}
- 	\frac{1}{2}\pi h_{mm} 
	.
\end{align}
\end{subequations}
	The following perturbed NP scalars are zero
\begin{align}\label{gauge-NP-scalars-zero}
	\nu^{(1)}=\gamma^{(1)}=\mu^{(1)}=0
	.
\end{align}
	Notice that
\begin{align}
\label{eq:pi_ta_relation_pert}
	\pi^{(1)}+\bar{\tau}^{(1)}
	=&
	-\frac{1}{2}\left(\bar{\pi} + \tau\right)h_{\bar{m}\bar{m}}
	.
\end{align}
	so it is straightforward to find, e.g. $\bar{\tau}^{(1)}$ once
we know $\pi^{(1)}$ and $h_{\bar{m}\bar{m}}$.

\subsection{Reconstructing the metric from $\Psi_4^{(1)}$}

Here we list the sequence of step we use to reconstruct all the metric coefficients $h_{ll}$,
$h_{l\bar{m}}$, and $h_{\bar{m}\bar{m}}$ from the curvature component $\Psi_4^{(1)}$.

\begin{enumerate}[leftmargin=*]
\item 
	With $\Psi_4^{(1)}$
one can find $\Psi_3^{(1)}$ and $\lambda^{(1)}$.
We begin with the following 
Bianchi identity
(Eq. (1.321.h) in \cite{Chandrasekhar_bh_book}):
	\begin{align}
	\ \ \ &	3\nu\Psi_{2} 
	-	2\left(\gamma+2\mu\right) \Psi_{3} 
        \nonumber\\&
	+ 	\left(4\beta - \tau\right) \Psi_{4} 
	+ 	\delta\left(\Psi_{4}\right) 
	- 	\Delta\left(\Psi_{3}\right)
	= 
	0; \label{Bianchi-Delta-Psi3}
	\end{align}
linearizing this gives
	\begin{align}
	\ \ \ &	3\nu^{(1)}\Psi_{2} 
	-	4\mu  \Psi_{3}^{(1)} 
        \nonumber\\
	&+ 	\left(4\beta - \tau\right) \Psi_{4}^{(1)} 
	+ 	\delta\Psi_{4}^{(1)} 
	- 	\Delta \Psi_{3}^{(1)}
	= 
	0.
	\end{align}
	Using the gauge condition \eqref{gauge-NP-scalars-zero}
and writing out the NP $\{\delta,\bar{\delta}\}$ derivatives
in terms of the GHP derivatives $\{\edth,\edth^{\prime}\}$,
i.e. $\edth \Psi_4^{(1)}= (\delta + 4 \beta) \Psi_4^{(1)}$, we obtain
\begin{align}
	\left(\edth-\tau\right)\Psi_{4}^{(1)} 
-	\left(\Delta+4\mu\right)\Psi_{3}^{(1)}
	=&0.
\end{align}
The above is a first order differential equation for $\Psi_3^{(1)}$
in terms of the known $\Psi_4^{(1)}$.
Similarly, the linearization of
\begin{align}
\ \ \ &	\lambda\left(3\gamma - \bar{\gamma} + \mu + \bar{\mu}\right)
- 	\nu\left(3\alpha + \bar{\beta} + \pi - \bar{\tau}\right) 
        \nonumber\\&
+ 	\Psi_{4} 
+ 	\Delta\left(\lambda\right)
- 	\bar{\delta}\left(\nu\right)
	= 
	0 \label{Ricci-Delta-lambda}
\end{align}
(Eq.~(1.310.j) in \cite{Chandrasekhar_bh_book})
gives
\begin{align}
-	\Psi_{4}^{(1)} 
-	\left(\Delta+\mu+\bar{\mu}\right)\lambda^{(1)}
	=&0,
\end{align}	
a differential equation for $\lambda^{(1)}$. 

\item With $\Psi_3^{(1)}$ we can find $\Psi_2^{(1)}$.
The linearization of
\begin{align}
\ \ \ &	2\nu\Psi_{1} 
- 	3\mu\Psi_{2} 
+ 	2\beta\Psi_{3} 
- 	2\tau\Psi_{3} 
        \nonumber\\&
+ 	\sigma\Psi_{4} 
+ 	\delta\left(\Psi_{3}\right) 
- 	\Delta\left(\Psi_{2}\right)
	= 
	0 \label{Bianchi-Delta-Psi2}
\end{align}
(Eq.~(1.321.g) of \cite{Chandrasekhar_bh_book}) 
gives
\begin{align}
\ \ \ &- 	3\mu\Psi_{2}^{(1)} 
+ 	2\beta\Psi_{3}^{(1)} 
- 	2\tau\Psi_{3}^{(1)} 
        \nonumber\\&
+ 	\sigma\Psi_{4}^{(1)} 
+ 	\delta \Psi_{3}^{(1)} 
- 	\Delta \Psi_{2}^{(1)}
	= 
	0.
\end{align}
By using
$\edth \Psi_3^{(1)}= (\delta + 2 \beta) \Psi_3^{(1)}$, we obtain
the desired differential equation for $\Psi_2^{(1)}$:
\begin{align}
	\left(\edth- 2\tau \right) \Psi_{3}^{(1)} 
- 	\left(\Delta+3\mu\right) \Psi_{2}^{(1)}
	= &
	0.
\end{align}
\item With $\lambda^{(1)}$ we can now solve for $h_{\bar{m}\bar{m}}$
using \eqref{eq:pert_la_1}.
\item With $\lambda^{(1)}$, $h_{\bar{m}\bar{m}}$, and $\Psi_3^{(1)}$
	we can find $\pi^{(1)}$. Using the linearization of
	\begin{align}
\ \ \ &	\left(3\epsilon + \bar{\epsilon}\right)\nu 
- 	\gamma\pi 
+ 	\bar{\gamma}\pi 
- 	\lambda\left(\bar{\pi} + \tau\right) 
        \nonumber\\&
- 	\mu\left(\pi + \bar{\tau}\right) 
- 	\Psi_{3} 
+ 	D\left(\nu\right)
- 	\Delta\left(\pi\right)
	= 
	0, \label{Ricci-Delta-pi}
	\end{align}
(Eq.~(1.310.i) of \cite{Chandrasekhar_bh_book})
	namely,
	\begin{align}
\ \ \ -&\lambda^{(1)}\left(\bar{\pi} + \tau\right) 
- 	\mu\left(\pi + \bar{\tau}\right)^{(1)}  
        \nonumber\\&
- 	\Psi_{3}^{(1)} 
- 	\Delta \pi^{(1)}
	= 
	0, 
	\end{align}
	and using \eqref{eq:pi_ta_relation_pert} gives the
        differential equation to solve for $\pi^{(1)}$:
\begin{align}
\ \ \ -&\lambda^{(1)}\left(\bar{\pi} + \tau\right) 
+	\frac{1}{2}\mu\left(\bar{\pi}+\tau\right) h_{\bar{m}\bar{m}}
        \nonumber\\&
- 	\Psi_{3}^{(1)} 
- 	\Delta \pi^{(1)}
	=0
	.
\end{align}
\item With $\pi^{(1)}$ and $h_{\bar{m}\bar{m}}$ we can solve
for $h_{l\bar{m}}$ using \eqref{eq:pert_pi_1}.
\item With $\Psi_2^{(1)}$, $h_{lm}$ and $h_{mm}$ we can find $h_{ll}$.
The linearization of
\begin{align}
	\Psi_2
	= &
	\epsilon\mu 
+ 	\bar{\epsilon}\mu 
+ 	\kappa\nu 
+ 	\bar{\alpha}\pi 
- 	\beta\pi 
- 	\pi\bar{\pi} 
        \nonumber\\&
- 	\mu\bar{\rho} 
- 	\lambda\sigma 
+ 	D\left(\mu\right) 
- 	\delta\left(\pi\right)
	, \label{Ricci-identity-Psi_2}
\end{align}
(Eq.~(1.310.h) of \cite{Chandrasekhar_bh_book})
	gives
\begin{align}\label{expression-Psi-2-inter}
\Psi_2^{(1)}
	= &
	( \epsilon^{(1)} + \bar{\epsilon}^{(1)}) \mu
+ 	\left( \bar{\alpha}^{(1)} - \beta^{(1)}\right) \pi
        \nonumber\\&
+ 	\left(- \delta+\bar{\alpha}- \beta-\bar{\pi}\right) \pi^{(1)} 
- 	\pi\bar{\pi}^{(1)}
	\nonumber\\ & 
- 	\mu\bar{\rho}^{(1)} 
-	\frac{1}{2}h_{ll}\Delta\left(\mu\right)
	\nonumber\\ & 
+	\left(	
		h_{lm}\Delta
	-	\frac{1}{2}h_{mm}\bar{\delta}
	\right)\left(\pi\right) 
	,
\end{align}
where we used \eqref{lin_ops}.
From \eqref{eq:pert_ep_1}, we deduce
\begin{align}
	\epsilon^{(1)}+\bar{\epsilon}^{(1)}
	&=
-	\frac{1}{2} \Delta h_{ll} 
-	\left(  \bar{\pi} + \tau\right)h_{l\bar{m}} 
	\nonumber\\ & 
- 	\left(  \pi + \bar{\tau}\right)h_{lm} 
\end{align}
and from \eqref{eq:pert_rh_1}, we find
\begin{align}\label{rho_1_eqn}
	\bar{\rho}^{(1)}
	&=
	\frac{1}{2}\bar{\mu} h_{ll} 
+	\frac{1}{2}\left(\edth - \bar{\pi}\right)h_{l \bar{m}} 
	\nonumber\\ & 
-	\frac{1}{2}\left(\edth^{\prime} + \pi + 2\bar{\tau}\right)h_{lm},
\end{align}
where we have used
$\edth h_{l\bar{m}}=(\delta- 2 \bar{\alpha}) h_{l \bar{m}}$ and
$\edth^{\prime} h_{l m} = (\bar{\delta}- 2 \alpha) h_{lm}$. 

From  \eqref{eq:pert_al_1} and \eqref{eq:pert_be_1}, we obtain
\begin{align}\label{alpha-beta}
	\bar{\alpha}^{(1)}-\beta^{(1)}
	&=
	  \bar{\mu} h_{lm} 
	\nonumber\\ & 
-	\frac{1}{2}\left(\edth^{\prime} +\alpha-  \bar{\beta} \right)h_{mm},
\end{align}
where we have used $\edth^{\prime} h_{mm}=(\bar{\delta}+ 2 \bar{\beta} - 2\alpha) h_{mm}$. 

Substituting the above in \eqref{expression-Psi-2-inter} gives
\begin{align}
\ \ \ \   \Psi_2^{(1)}
&	= 
	\bigg(-	\frac{1}{2} \Delta h_{ll} 
-	\left(  \bar{\pi} + \tau\right)h_{l\bar{m}} 
- 	\left(  \pi + \bar{\tau}\right)h_{lm} \bigg) \mu \nonumber \\
&
+ 	\left(
		\bar{\mu} h_{lm} 
	-	\frac{1}{2}
		\left(\edth^{\prime} +\alpha-  \bar{\beta} \right)h_{mm}
	\right)\pi
	\nonumber \\
&
+ 	\left(- \edth -\bar{\pi}\right) \pi^{(1)}   - \pi\bar{\pi}^{(1)}
	\nonumber\\
	& - \mu\bigg(	
		\frac{1}{2}\bar{\mu} h_{ll} 
	+	\frac{1}{2}\left(\edth - \bar{\pi}\right)h_{l \bar{m}} 
	\nonumber\\ & 
	-	\frac{1}{2}\left(\edth^{\prime} + \pi + 2\bar{\tau}\bigg)h_{lm}
	\right) 
-	\frac{1}{2}h_{ll}\Delta\left(\mu\right)
	\nonumber\\ & 
+	\left(
		h_{lm}\Delta
	-	\frac{1}{2}h_{mm}
		\left(\edth^{\prime} - \alpha + \bar{\beta}\right)
	\right)\left(\pi\right) 
\end{align}
which we rewrite as
\begin{align}
 &
-	\frac{1}{2} \left( \Delta+\bar{\mu}\right) \left(\mu h_{ll}\right)
-	\frac{1}{2}\mu\left(\edth + \bar{\pi} +2 \tau\right)h_{l\bar{m}} 
	\nonumber\\ & 
+ 	\frac{1}{2} \left( 
		\mu \left(\edth^{\prime} - \pi \right)
	+ 	2\bar{\mu}\pi  
	\right) h_{lm} 
	\nonumber \\
&
-	\frac{1}{2}\pi \edth^{\prime}  h_{mm} 
- 	\left( \edth +\bar{\pi}\right) \pi^{(1)} 
- 	\pi\bar{\pi}^{(1)}  
	\nonumber\\ & 
+	h_{lm}\Delta\pi
-	\frac{1}{2}h_{mm} \edth^{\prime} \pi 
-	\Psi_2^{(1)}
	=0.
\end{align}
Using \eqref{Ricci-Delta-pi}, and
Eq. (1.310.g) in \cite{Chandrasekhar_bh_book},
\begin{align}
\ \ \ &	3\epsilon\lambda 
+ 	\bar{\kappa}\nu 
- 	\pi\left(\alpha - \bar{\beta} + \pi\right) 
- 	\lambda\left(\bar{\epsilon} + \rho\right) 
	\nonumber\\ & 
- 	\mu\bar{\sigma}  
+ 	D\left(\lambda\right) 
- 	\bar{\delta}\left(\pi\right)
	= 
	0, \label{Ricci-D-lambda} 
\end{align}
	evaluated on a type D background ($\gamma=0$)
to write 
\begin{align}
\ \ \  	\Delta\left(\pi\right)
	= 
- 	\mu\left(\pi + \bar{\tau}\right) 
	, \, \, \, \,
	\edth^{\prime}\left(\pi\right)
	= 
- 	\pi  \pi,
	\end{align}
we finally obtain the transport equation for $h_{ll}$
\begin{align}
 & -	\left( \Delta+\bar{\mu}\right) \left(\mu h_{ll}\right) 
-	\mu\left(\edth + \bar{\pi} +2 \tau\right)h_{l\bar{m}} 
  + 	
	\nonumber\\ & 
 	\left(
		\mu \left(\edth^{\prime} 
	- 	3\pi 
	-	2 \bar{\tau}\right)+ 2\bar{\mu}\pi
	\right) h_{lm}
	\nonumber \\
&
-	\pi \left(\edth^{\prime} -\pi\right) h_{mm} 
-	2 \left( \edth +\bar{\pi}\right) \pi^{(1)}
	\nonumber \\&
- 	2\pi\bar{\pi}^{(1)}
-	2\Psi_2^{(1)}
	=
	0.
\end{align}

\end{enumerate}
	With $\{h_{ll},h_{l\bar{m}},h_{\bar{m}\bar{m}}\}$
one can readily compute the rest of the NP scalars in outgoing radiation gauge,
and we are now able to compute the second order source term.

\subsection{Source term for $\Psi_4^{(2)}$}

	In this section we rewrite the vacuum source term $\mathcal{S}^{(2)}$
for the Teukolsky equation for $\Psi_4^{(2)}$ \cite{Campanelli:1998jv},
\begin{align}
\label{eq:second_order_Teukolsky_source}
& \ \ \	\mathcal{S}^{(2)} 
	\equiv 
	\nonumber \\ &
-	\bigg[
		d_4\left(D+4\epsilon-\rho\right)^{(1)} 
	-	d_3\left(\delta+4\beta-\tau\right)^{(1)} 
	\bigg]
	\Psi_4^{(1)}
	\nonumber \\
	&
+	\bigg[
		d_4\left(\bar{\delta}+2\alpha+4\pi\right)^{(1)} 
	-	d_3\left(\Delta+2\gamma+4\mu\right)^{(1)} 
	\bigg]
	\Psi_3^{(1)}
	\nonumber \\
	&
-	3\left[
		d_4\lambda^{(1)}
	-	d_3\nu^{(1)}
	\right]
	\Psi^{(1)}_2
	\nonumber \\
	&
+	3\Psi^{(0)}_2\bigg[
		\left(d^{(1)}_4-3\mu^{(1)}\right)\lambda^{(1)}
	-	\left(d^{(1)}_3-3\pi^{(1)}\right)\nu^{(1)}
	\bigg]
	,
\end{align}
in outgoing radiation gauge, with a tetrad chosen so that  $\gamma=0$.
In the above we have introduced the background operators
\begin{subequations}
\begin{align}
	d_3
	\equiv&
	\bar{\delta}+3\alpha+\bar{\beta}+4\pi-\bar{\tau}
	,\\
	d_4
	\equiv&
	\Delta+4\mu+\bar{\mu},
\end{align}
and their first order corrections
\begin{align}
	d_3^{(1)}
	\equiv&
	\bar{\delta}^{(1)}+3\alpha^{(1)}+\bar{\beta}^{(1)}+4\pi^{(1)}-\bar{\tau}^{(1)}
	,\\
	d_4^{(1)}
	\equiv&
	0
	.	
\end{align}
\end{subequations}

	We now consider the expansion of $\mathcal{S}^{(2)}$ line by line.
\begin{enumerate}[leftmargin=*]
\item	The first line is
\begin{align}
&-	\left[
		d_4\left(D+4\epsilon-\rho\right)^{(1)} 
	-	d_3\left(\delta+4\beta-\tau\right)^{(1)} 
	\right]
	\Psi_4^{(1)}.
\end{align}
By using \eqref{eq:pert_ep_1}, we expand 
\begin{align}
&\left(D+4\epsilon-\rho\right)^{(1)}  \Psi_4^{(1)} \nonumber \\
&= \Big(
-	\frac{1}{2}h_{ll}\Delta-\left(\Delta - \mu + \bar{\mu}\right)h_{ll} 
	\nonumber\\&
-	\left(\delta - 2\bar{\alpha} + \bar{\pi} + 2\tau\right)h_{l\bar{m}} 
	\nonumber\\&
+ \left(\bar{\delta} - 2\alpha - 3\pi - 2\bar{\tau}\right)h_{lm} -\rho^{(1)}
	\Big)  \Psi_4^{(1)}
	\nonumber \\
&= -	\frac{1}{2}h_{ll}\Delta\Psi_4^{(1)}
	-\Psi_4^{(1)}\left(\Delta - \mu + \bar{\mu}\right)h_{ll} 
	\nonumber\\&
-	\Psi_4^{(1)}\left(\edth + \bar{\pi} + 2\tau\right)h_{l\bar{m}} 
	\nonumber\\&
	+\Psi_4^{(1)} \left(\edth^{\prime} - 3\pi - 2\bar{\tau}\right)h_{lm} 
	-\Psi_4^{(1)}\rho^{(1)}.
\end{align}
By using \eqref{eq:pert_be_1} and \eqref{eq:pi_ta_relation_pert}, we expand
\begin{align}
\ \ \ &	\left(\delta+4\beta-\tau\right)^{(1)} \Psi_4^{(1)} 
	\nonumber \\
&=	\Big(
-	h_{lm}\Delta
+	\frac{1}{2}h_{mm}\bar{\delta}
-	\left(\Delta + \mu + 2\bar{\mu}\right)h_{lm}  
	\nonumber\\&
+ 	\left(\bar{\delta} + 2\bar{\beta} - \pi - \bar{\tau}\right)h_{mm}
+	\bar{\pi}^{(1)}+\frac{1}{2} (\pi+\bar{\tau})h_{mm}
	\Big) \Psi_4^{(1)}
	\nonumber \\
&=-	h_{lm}\Delta\Psi_4^{(1)}
+	\frac{1}{2}h_{mm}\bar{\delta}\Psi_4^{(1)}
	\nonumber \\&
+        \Psi_4^{(1)}\bigg[
-	\left(\Delta + \mu + 2\bar{\mu}\right)h_{lm}  
+	\left(
		\bar{\delta} + 2\bar{\beta} - \pi - \bar{\tau}
	\right)h_{mm}
	\nonumber \\&
+	\bar{\pi}^{(1)}+\frac{1}{2} 
        (\pi+\bar{\tau})h_{mm}
        \bigg]
	\nonumber \\
&=-	h_{lm} \left(\Delta + \mu + 2\bar{\mu}\right) \Psi_4^{(1)}
+	\frac{1}{2}h_{mm}\edth^{\prime}\Psi_4^{(1)}
	\nonumber\\&
+        \Psi_4^{(1)}\bigg[
-	\Delta h_{lm}  
+	\left(\edth^{\prime} 
	-	\frac{1}{2}  \pi 
	-	\frac{1}{2}  \bar{\tau}
	\right)h_{mm}
+	\bar{\pi}^{(1)}
	\bigg].
\end{align}
The above quantity has GHP weight $\{p,q\}=\{-3, -1\}$,
and therefore $d_3$ can be written as
$d_3
=\bar{\delta}+3\alpha+\bar{\beta}+4\pi-\bar{\tau}
=\edth^{\prime}+4\pi-\bar{\tau}$.
This finally gives
\begin{align}\label{line_1_eqn}
\ \ \	&-\left[
		d_4\left(D+4\epsilon-\rho\right)^{(1)} 
	-	d_3\left(\delta+4\beta-\tau\right)^{(1)} 
	\right]
	\Psi_4^{(1)}
	\nonumber\\&
	=
-	\left(\Delta+4\mu+\bar{\mu}\right)
	\Bigg[
        \Psi_4^{(1)}\bigg(
	-	\left(\edth+\bar{\pi}+2\tau\right)h_{l\bar{m}}
	-	\rho^{(1)}
	\nonumber\\&
	+	\left(\edth^{\prime}-3\pi-2\bar{\tau}\right)h_{lm}
	-	\left(\Delta-\mu+\bar{\mu}\right)h_{ll}
                \bigg)
	\nonumber\\&
	-	\frac{1}{2}h_{ll}\Delta\Psi_4^{(1)}
	\Bigg]	
	\nonumber\\&
+	\left(\edth^{\prime}+4\pi-\bar{\tau}\right)
	\Bigg[
		\frac{1}{2}h_{mm}\edth^{\prime}\Psi_4^{(1)}
	\nonumber\\&
	-	h_{lm}\left(\Delta+\mu+2\bar{\mu}\right)\Psi_4^{(1)}
	\nonumber\\&
        +\Psi_4^{(1)}\bigg(
		\bar{\pi}^{(1)}
	-	\Delta h_{lm}
	+	\left(
			\edth^{\prime}
		-	\frac{1}{2}\pi
		-	\frac{1}{2}\bar{\tau}
		\right)h_{mm}
                \bigg)
	\Bigg]
	.
\end{align} 
\item	The second line is
\begin{align}
&	\left[
		d_4\left(\bar{\delta}+2\alpha+4\pi\right)^{(1)} 
	-	d_3\left(\Delta+2\gamma+4\mu\right)^{(1)} 
	\right]
	\Psi_3^{(1)} \nonumber \\
	&=d_4 \left(\bar{\delta}+2\alpha+4\pi\right)^{(1)} \Psi_3^{(1)},
	\end{align}
	where we used that $\left(\Delta+2\gamma+4\mu\right)^{(1)} =0$.
By using equation \eqref{eq:pert_al_1},
we rewrite the expression follow $d_4$ in the following way:
	\begin{align}
	&\left(\bar{\delta}^{(1)}+2\alpha^{(1)}+4\pi^{(1)}\right)\Psi_3^{(1)}
	\nonumber \\&
	=\Big(
-	h_{l \bar{m}}\Delta
+	\frac{1}{2}h_{\bar{m}\bar{m}}\delta
-	\frac{1}{2}\left(\Delta - 2\mu + \bar{\mu}\right)h_{l\bar{m}} 
	\nonumber\\&
-	\frac{1}{2}\left(
		\delta - 2\bar{\alpha} + \bar{\pi} + \tau
	\right)h_{\bar{m}\bar{m}}
+	4\pi^{(1)}
	\Big)\Psi_3^{(1)}
	\nonumber \\
&=\left(
-	h_{l \bar{m}}\Delta
+	\frac{1}{2}h_{\bar{m}\bar{m}}\delta+4\pi^{(1)} \right)\Psi_3^{(1)}
	\nonumber\\
& -	\frac{1}{2}\Psi_3^{(1)}\bigg[
        \left(
		\Delta - 2\mu + \bar{\mu}
	\right)h_{l\bar{m}} 
	\nonumber \\&
+	\left(
		\delta - 2\bar{\alpha} + \bar{\pi} + \tau
	\right)h_{\bar{m}\bar{m}}
        \bigg]
	\nonumber\\
&=\left(
-	h_{l \bar{m}}\Delta
+	\frac{1}{2}h_{\bar{m}\bar{m}}\edth+4\pi^{(1)} \right)\Psi_3^{(1)}
	\nonumber\\&
-	\frac{1}{2}\Psi_3^{(1)}\bigg[
        \left(
		\Delta - 2\mu + \bar{\mu}
	\right)h_{l\bar{m}} 
+	 \left(\edth + \bar{\pi} + \tau
	\right)h_{\bar{m}\bar{m}} 
        \bigg]
	,
\end{align}
	where we used
$\edth h_{\bar{m}\bar{m}}=(\delta+ 2 \beta - 2\bar{\alpha}) h_{\bar{m}\bar{m}}$ and
$\edth \Psi_3^{(1)} =(\delta+ 2 \beta) \Psi_3^{(1)} $. This finally gives
\begin{align}
	&d_4\left(\bar{\delta}+2\alpha+4\pi\right)^{(1)}\Psi_3^{(1)}
	\nonumber\\&
	=
	\left(\Delta+4\mu+\bar{\mu}\right)
	\Bigg[
	\left(
	-	h_{l\bar{m}}\Delta
	+	\frac{1}{2}h_{\bar{m}\bar{m}}\edth
	+	4\pi^{(1)}
	\right)\Psi_3^{(1)}
	\nonumber\\&
-	\frac{1}{2}\Psi_3^{(1)}\bigg(
        \left(
		\edth+\bar{\pi}+\tau
	\right)h_{\bar{m}\bar{m}}
+       \left(
		\Delta
	-	2\mu
	+	\bar{\mu}
	\right)h_{l\bar{m}}
        \bigg)
	\Bigg]
	.
\end{align} 
\item The third line is given by 
\begin{align}
&-	3\left[
		d_4\lambda^{(1)}
	-	d_3\nu^{(1)}
	\right]
	\Psi^{(1)}_2=
	\nonumber\\&
-	3\left(\Delta+4\mu+\bar{\mu}\right)
	\left(\lambda^{(1)}\Psi_2^{(1)}\right)
\end{align}
since $\nu^{(1)}=0$. 
\item The fourth line
\begin{align}
3\Psi_2\left[
		\left(d^{(1)}_4-3\mu^{(1)}\right)\lambda^{(1)}
	-	\left(d^{(1)}_3-3\pi^{(1)}\right)\nu^{(1)}
	\right]=0
	\end{align}
	since $d^{(1)}_4=\mu^{(1)}=\nu^{(1)}=0$. 
\end{enumerate}
	We have thus rewritten the second order source term entirely in terms
of the variables reconstructed from $\Psi_4^{(1)}$ (though
in the form listed in Eq.~\eqref{eq:def_mathfrak_s} we have additionally
replaced $\rho^{(1)}$ in line 1 above (\ref{line_1_eqn}) using (\ref{rho_1_eqn})).
\section{Derivation of horizon penetrating hyperboloidally compactified
coordinates for Kerr spacetime}
\label{sec:derivation_coordinates}
   A Mathematica notebook that contains some of the algebraic manipulations
we undertook to derive the coordinates and tetrad we used
can be accessed at \cite{notebook_online}.
\subsection{
	Starting point:
	Kerr in Boyer-Lindquist coordinates and the Kinnersley tetrad
}
	We begin with the Kerr metric in Boyer-Lindquist coordinates
(e.g. \cite{Chandrasekhar_bh_book})
\begin{align}
\label{eq:Kerr_BL_coords}
	ds^2
	=&
	\left(1-\frac{2Mr}{\Sigma_{BL}}\right)dt^2
+	2\left(\frac{2Mar\mathrm{sin}^2\vartheta}{\Sigma_{BL}}\right)dtd\varphi
	\nonumber \\&
-	\frac{\Sigma_{BL}}{\Delta_{BL}}dr^2
-	\Sigma_{BL}d\vartheta^2
	\nonumber \\&
-	\mathrm{sin}^2\vartheta\left(
		r^2+a^2+2Ma^2r\frac{\mathrm{sin}^2\vartheta}{\Sigma_{BL}}
	\right)
	d\varphi^2
	,
\end{align}
	where
\begin{subequations}
\begin{align}
\label{eq:def_SigmaBL}
	\Sigma_{BL}\equiv & r^2+a^2\mathrm{cos}^2\vartheta 
	, \\
\label{eq:def_DeltaBL}
	\Delta_{BL}\equiv & r^2-2Mr+a^2
	.
\end{align}
\end{subequations}
	The outer and inner horizons are at $\Delta(r_{\pm})=0$.

The Kinnersley tetrad \cite{1969JMP....10.1195K}
in Boyer-Lindquist coordinates is
\begin{subequations}
\label{eq:Kinnersley_tetrad_BL}
\begin{align}
	l_{Kin}^{\mu}
	= &
	\left(\frac{r^2+a^2}{\Delta_{BL}},1,0,\frac{a}{\Delta_{BL}}\right)
	, \\
	n_{Kin}^{\mu}
	= &
	\frac{1}{2\Sigma_{BL}}\left(r^2+a^2,-\Delta_{BL},0,a\right)
	, \\
	m_{Kin}^{\mu}
	= &
	\frac{1}{2^{1/2}\left(r+ia\mathrm{cos}\vartheta\right)}
	\left(
		ia\mathrm{sin}\vartheta,0,1,\frac{i}{\mathrm{sin}\vartheta}
	\right)
	.
\end{align}
\end{subequations}
	The Kinnersley tetrad vectors $l^{\mu}_{Kin}$ and $n^{\mu}_{Kin}$ are
aligned with the outgoing and ingoing principle null directions of Kerr.
The Kinnersley tetrad sets the maximal number of
NP scalars to zero for a general type-D spacetime, and sets
$\epsilon=0$ as well. 
\subsection{
	Intermediate step: Kerr in ingoing Eddington-Finkelstein coordinates
}
	We transform to ingoing Eddington-Finkelstein
coordinates, which renders the metric 
nonsingular on the black hole horizon, via
\begin{subequations}
\begin{align}
	dv 
	\equiv & 
	dt + dr_* - dr 
	, \\
	d\phi 
	\equiv & 
	d\varphi + \frac{a}{r^2+a^2}dr_*
	.
\end{align}
\end{subequations}
	where
\begin{align}
	\frac{dr_*}{dr} 
	\equiv
	\frac{r^2+a^2}{\Delta_{BL}}
	.
\end{align}
	The results are
\begin{align}
	ds^2
	= &
	\left(1-\frac{2Mr}{\Sigma_{BL}}\right)dv^2
-	\frac{4Mr}{\Sigma_{BL}}\left(
		dr
	-	a\mathrm{sin}^2\vartheta d\phi
	\right)
	dv
	\nonumber \\ &
-	\left(1+\frac{2Mr}{\Sigma_{BL}}\right)
	\left(
		dr^2
	-	2a\mathrm{sin}^2\vartheta drd\phi
	\right)
	\nonumber \\ &
-	\Sigma d\vartheta^2
-	\left(
		a^2+r^2+2Mr\frac{a^2}{\Sigma_{BL}}\mathrm{sin}^2\vartheta
	\right)
	d\phi^2
	.
\end{align}
	and
\begin{subequations}
\label{eq:Kinnersley_tetrad_IEF}
\begin{align}
	l_{Kin}^{\mu}
	= &
	\left(
		1+\frac{4Mr}{\Delta_{BL}},
		1,
		0,
		\frac{2a}{\Delta_{BL}}
	\right)
	, \\
	n_{Kin}^{\mu}
	= &
	\frac{1}{2\Sigma_{BL}}\left(
		\Delta_{BL},
		-\Delta_{BL},
		0,
		0
	\right)
	, \\
	m_{Kin}^{\mu}
	= &
	\frac{1}{2^{1/2}\left(r+ia\mathrm{cos}\vartheta\right)}
	\left(
		ia\mathrm{sin}\vartheta,0,1,\frac{i}{\mathrm{sin}\vartheta}
	\right)
	.
\end{align}
\end{subequations}
	This tetrad is still singular on the horizons.
Furthermore, it is 
more useful for metric reconstruction in outgoing radiation gauge to have
\begin{align}
	\epsilon
	\neq
	0
	,
	\qquad
	\gamma
	=
	0
\end{align}
(the Kinnersley tetrad has the opposite property). 
Therefore, we rescale and rotate the tetrad to
obtain one that is regular on the horizon,
and has $\gamma=0,\epsilon\neq 0$:
\begin{subequations}
\begin{align}
	l^{\mu}
	\to &
	\frac{\Delta_{BL}}{2\Sigma_{BL}}l^{\mu}
	, \\
	n^{\mu}
	\to &
	\frac{2\Sigma_{BL}}{\Delta_{BL}}n^{\mu}
	, \\
	m^{\mu}
	\to &
	\mathrm{exp}\left[
		-2i\mathrm{arctan}\left[\frac{r}{a\mathrm{sin}\vartheta}\right]
	\right]m^{\mu}
	,
\end{align}
\end{subequations}
	giving
\begin{subequations}
\label{eq:new_tetrad_EF}
\begin{align}
	l^{\mu}
	= &
	\left(
		\frac{r^2+2Mr+a^2}{2\Sigma_{BL}},
		\frac{\Delta_{BL}}{2\Sigma_{BL}},
		0,
		\frac{a}{\Sigma_{BL}}
	\right)
	, \\
	n^{\mu}
	= &
	\left(
		1,-1,0,0
	\right)
	, \\
	m^{\mu}
	= &
	\frac{1}{2^{1/2}\left(r-ia\mathrm{cos}\vartheta\right)}
	\left(
		-ia\mathrm{sin}\vartheta,0,-1,-\frac{i}{\mathrm{sin}\vartheta}
	\right)
	.
\end{align}
\end{subequations}
\subsection{Coordinates used in our code:
Kerr in horizon penetrating hyperboloidally compactified coordinates}
\label{eq:coordinates_used_in_code}
	Now we give the final step,
hyperboloidal compactification (see~\cite{Zenginoglu_2008} for a more
general description of this)
to arrive at the coordinates and tetrad we use in our code.

The ingoing/outgoing radial null characteristic speeds\footnote{We
do not need to consider angular characteristic speeds as those
die off more quickly as we go to future null infinity.
} for Kerr in ingoing Eddington-Finkelstein coordinates are found by solving
for the characteristics of the eikonal equation for the metric
\begin{align}
	g^{\mu\nu}\xi_{\mu}\xi_{\nu}
	=&
	0
	,
\end{align}
	setting $\xi_{\vartheta}=\xi_{\phi}=0$, and then computing
\begin{align}
	c_{\pm}
	\equiv&
	\mp\frac{\xi_v}{\xi_r}
	.
\end{align}
	We obtain
\begin{subequations}
\begin{align}
	c_+
	= &
	1
-	\frac{4Mr}{2Mr+\Sigma_{BL}}
	, \\
	c_-
	= &
	-1
	.
\end{align}
\end{subequations}
	We define a new radial coordinate $R(r)$ and
time coordinate $T(v,r)$.
The ingoing/outgoing radial null characteristic speeds are now
\begin{align}
	\tilde{c}_{\pm}
	=
	\frac{dR/dr}{\frac{1}{c_{\pm}}\partial_vT+\partial_rT}
	.
\end{align}
	We want to choose a time coordinate that sets
$\tilde{c}_-|_{r=\infty}=0$ while keeping $0<\tilde{c}_+|_{r=\infty}<\infty$.
We choose the time coordinate to be of the form
\begin{align}
	T(v,r) 
	= 
	v + h(r) 
	. 
\end{align}
	We compactify the radial coordinate by choosing
\begin{align}
	R(r) \equiv \frac{L^2}{r}
	,
\end{align}
	where $L$ is a constant length scale (we set $L=1$ in numerical code).
Series expanding about $r=\infty$, we have
\begin{subequations}
\begin{align}
	\tilde{c}_+
	= &
	\left(
		1+\frac{4M}{r}+\frac{8M^2}{r^2}
	+	\mathcal{O}\left(\frac{1}{r^3}\right)
	+	\frac{dh}{dr}
	\right)^{-1}
	\left(-\frac{L^2}{r^2}\right)
	, \\
	\tilde{c}_-
	= &
	\left(-1+\frac{dh}{dr}\right)^{-1}
	\left(-\frac{L^2}{r^2}\right)	
	.
\end{align}
\end{subequations}
	We see that the choice
\begin{align}
	\frac{dh}{dr} = -1 - \frac{4M}{r}
	,
\end{align}
	sets
$\tilde{c}_-|_{R=0}=0$ while keeping
$0>\tilde{c}_+|_{R=0}>-\infty$ (our choice of compactification
flips the signs of the ingoing and outgoing characteristics, and
$r=\infty$ is mapped to $R=0$). 

	In summary we choose
\begin{subequations}
\begin{align}
	R(r)
	\equiv&
	\frac{L^2}{r}
	,\\
	T(v,r)
	\equiv&
	v-r-4M\mathrm{ln}r
	.
\end{align}
\end{subequations}
	In these coordinates future null infinity is located
at $R=0$, and the black hole horizon is located at $R(r_+)$.

	We apply this set of coordinate transformations
to the tetrad Eq.~\eqref{eq:new_tetrad_EF} to obtain
\begin{subequations}
\label{eq:tetrad_IEF_HC}
\begin{align}
	l^{\mu} 
	= &
	\frac{R^2}{L^4+a^2R^2\mathrm{cos}^2\vartheta}\Bigg(
		2M\left(2M-\left(\frac{a}{L}\right)^2R\right),
        \nonumber\\&
	-	\frac{1}{2}\left(L^2-2MR+\left(\frac{a}{L}\right)^2R^2\Bigg),
		0,
		a
	\right)
	, \\
	n^{\mu} 
	= &
	\left(
		2+\frac{4MR}{L^2},\frac{R^2}{L^2},0,0
	\right)
	, \\
	m^{\mu}
	= &
	\frac{R}{2^{1/2}\left(L^2-iaR\mathrm{cos}\vartheta\right)}
	\left(
		-ia\mathrm{sin}\vartheta,
		0,
		-1,
		-\frac{i}{\mathrm{sin}\vartheta}
	\right)
	.
\end{align}
\end{subequations}
	We list the nonzero Ricci rotation coefficients in
this coordinate system in
Eqs.~\eqref{eq:NP_IEF_HC}.
\section{Spin-weighted spherical harmonics}
\label{sec:swaL}
	Here we collect relevant properties of the
spin-weighted spherical harmonics
for reference, as we found them to be useful in evaluating
the $\edth$ and $\edth^{\prime}$ operators.
For further discussion see, e.g.
\cite{1967JMP.....8.2155G,1977RSPSA.358...71B,2018arXiv180410320V}. 
\subsection{Basic properties}

	The spin-weighted spherical harmonics are
eigenfunctions of the spin-weighted Laplace-Beltrami operator on the
two-sphere
\begin{align}
\label{eq:swLB_S2}
	\spindersphere Y^m_l
	\equiv &
	\frac{1}{\mathrm{sin}\vartheta}\partial_{\vartheta}
	\left(
		\mathrm{sin}\vartheta\;\partial_{\vartheta}{}_sY^m_l
	\right)
        \nonumber \\&
+	\left(
		s
	-	\frac{
			\left(
				-i\partial_{\varphi}+s\mathrm{cos}\vartheta
			\right)^2
		}{
			\mathrm{sin}^2\vartheta
		}
	\right)
	{}_sY^m_l	
	\nonumber \\
	= &
-	\left(l-s\right)\left(l+s+1\right){}_sY^m_l
	.
\end{align}
	We write ${}_sY^m_l(\vartheta,\varphi)$ as
\begin{align}
	{}_sY^m_l(\vartheta,\varphi)
	\equiv
	e^{im\varphi}{}_sP^m_l(\vartheta)
	,
\end{align}
	where the spin-weighted associated Legendre (swaL) functions
${}_sP^m_l(\vartheta)$ satisfy (setting 
$y\equiv-\mathrm{cos}\vartheta$)
\begin{align}
\label{eq:swaL_equation}
	&\frac{d}{dy}\left(\left(1-y^2\right)\frac{d{}_sP^m_l(y)}{dy}\right)
        \nonumber\\&
+	\left(
		\left(l-s\right)\left(l+s+1\right)
	+	s
	-	\frac{(m-sy)^2}{1-y^2}
	\right){}_sP^m_l(y)
	=
	0
	.
\end{align}
	There exist explicit formulas for the swaL functions.
	For a function of spin-weight $s$, ${}_sf$,
it is convenient to define\footnote{Note that unlike the standard references,
we use $\Edth$ (capital $\edth$), to avoid confusion with the
GHP $\edth$.} 
\begin{subequations}
\begin{align}
	\Edth {}_sf
	\equiv&
	\left(
-	\partial_{\vartheta}
-	\frac{i}{\mathrm{sin}\vartheta}\partial_{\varphi}
+	s\mathrm{cot}\vartheta
	\right)
	{}_sf
	,\\
	\Edth^{\prime} {}_sf
	\equiv&
	\left(
-	\partial_{\vartheta}
+	\frac{i}{\mathrm{sin}\vartheta}\partial_{\varphi}
-	s\mathrm{cot}\vartheta
	\right)
	{}_sf
	.
\end{align}
\end{subequations}
	One can show that
\begin{align}
	{}_sY^m_l
	=
	\left[\frac{(l-s)!}{(l+s)!}\right]^{1/2}\Edth^sY^m_l
	,
\end{align}
	and moreover that
\begin{align}
	\spindersphere=\Edth^{\prime}\Edth
	.
\end{align}
	We also have
\begin{align}
	{}_s\bar{Y}^m_l
	= &
	\left(-1\right)^{m+s}{}_{-s}Y^{-m}_l
	, \\
	\Edth {}_sY^m_l
	= &
        \label{Dplus}
	\left[\left(l-s\right)\left(l+s+1\right)\right]^{1/2}{}_{s+1}Y^m_l
	, \\
        \label{Dminus}
	\Edth^{\prime}{}_sY^m_l
	= &
-	\left[\left(l+s\right)\left(l-s+1\right)\right]^{1/2}{}_{s-1}Y^m_l
	.
\end{align}
\subsection{Relation between spin-weighted spherical harmonics
and the Jacobi polynomials}

   To evaluate the spin-weighted spherical harmonics in our code,
we write them in terms of Jacobi polynomials, which can be straightforwardly
computed using well-known numerical packages (such as {\bf mpmath} \cite{mpmath}).
Here we review the steps that establish
how those two classes of functions are related to one another.

	We rearrange Eq.~\eqref{eq:swaL_equation} to obtain the
``generalized associated Legendre equation''
(e.g. \cite{virchenko2001generalized})
\begin{align}
	&\frac{d}{dy}\left(\left(1-y^2\right)\frac{d{}_sP^m_l(y)}{dy}\right)
        \nonumber\\&
+	\left(
		l(l+1)
	-	\frac{\mu_1^2}{2(1-y)}
	-	\frac{\mu_2^2}{2(1+y)}
	\right){}_sP^m_l(y)
	=
	0
	,
\end{align}
	where
\begin{align}
	\mu_1^2
	\equiv &
	(m-s)^2
	, \\
	\mu_2^2
	\equiv &
	(m+s)^2
	.
\end{align}
	This equation has regular singular points at $\{\pm1,\infty\}$,
so we can reduce it to a hypergeometric equation. In fact 
we can also reduce it to a Jacobi equation, and thus write the
${}_sP^m_l$ in terms of Jacobi polynomials\footnote{We follow
the conventions of \cite{NIST:DLMF}.}. We define the transformation
\begin{align}
	{}_sP^m_l(y)
	\equiv
	\left(1-y\right)^{|\mu_1|/2}
	\left(1+y\right)^{|\mu_2|/2}
	f(y)
	,
\end{align} 
	and obtain the Jacobi differential equation 
\begin{align}
	\left(1-y^2\right)
	\frac{d^2f}{dy^2}
+	\left(
		\beta-\alpha-\left(\alpha+\beta+2\right)y
	\right)
	\frac{df}{dy}
        \nonumber\\
+	n\left(
		n+\alpha+\beta+1
	\right)
	f
	=
	0
	,
\end{align}
	where
\begin{align}
	\alpha
	= &
	|\mu_1|
	=
	|m-s|
	, \\
	\beta
	= &
	|\mu_2|
	=
	|m+s|
	, \\
	n
	= &
	l-\frac{\alpha+\beta}{2}
	.
\end{align}

	The solutions $f$ are the Jacobi polynomials $f=P^{(\alpha,\beta)}_n$. 
	The variables $\alpha$, $\beta$, and $n$ are all positive integers
(note as well that for the Jacobi polynomials
we need $n$ to be a positive integer). 
We see that the orthonormal swaL functions 
are
\begin{align}
	{}_s\hat{P}^m_l(y)
	=
	{}_s\mathcal{N}^m_l
	\left(1-y\right)^{\alpha/2}\left(1+y\right)^{\beta/2}
	P^{(\alpha,\beta)}_n(y)
	,
\end{align}
	where ${}_s\mathcal{N}^m_l$ is a normalization constant
to make the functions orthonormal (see also \cite{2018arXiv180410320V}).
We can compute ${}_s\mathcal{N}^m_l$ with the identity
\begin{align}
	\int_{-1}^1dx
	(1-x)^{\alpha}(1+x)^{\beta}
	P_m^{(\alpha,\beta)}(x)P_n^{(\alpha,\beta)}(x)
	=\nonumber\\
	\frac{2^{\alpha+\beta+1}}{2n+\alpha+\beta+1}
	\frac{
		\Gamma(n+\alpha+1)\Gamma(n+\beta+1)
	}{
		n!\Gamma(n+\alpha+\beta+1)
	}
	\delta_{mn}
	,
\end{align}
   so that the ${}_sP^m_l$ are orthonormal: 
$\int_{-1}^1 dx {}_sP^m_l(x) {}_sP^m_{l^{\prime}}(x)=\delta_{ll^{\prime}}$.

	As $\alpha$ and $\beta$ are positive integers for us, 
we can replace the Gamma functions with factorials.
We conclude the normalization factor is
\begin{align}
	{}_s\mathcal{N}^m_l
	= &
	\left(-1\right)^{\mathrm{max}\left(m,-s\right)} 
        \times\nonumber\\&
        \left(
	\frac{
		2n+\alpha+\beta+1
	}{
		2^{\alpha+\beta+1}
	}
	\frac{
		n!(n+\alpha+\beta)!
	}{
		(n+\alpha)!(n+\beta)!
	}
	\right)^{1/2}
	\nonumber \\
	= &
	\left(-1\right)^{\mathrm{max}\left(m,-s\right)}
        \times\nonumber\\&
        \left(
	\frac{
		2l+1
	}{
		2^{2l_{min}+1}
	}
	\frac{
		\left(l-l_{min}\right)!
		\left(l+l_{min}\right)!
	}{
		\left(l-l_{min}+\alpha\right)!
		\left(l-l_{min}+\beta\right)!
	}
	\right)^{1/2}
	,
\end{align}
	where we have defined
\begin{align}
	l_{min}
	\equiv
	\frac{\alpha+\beta}{2}
	.
\end{align}
\bibliography{../references}

\end{document}